\documentclass[preprints,article,accept,moreauthors,pdftex]{Definitions/mdpi} 

\firstpage{1} 
\pubvolume{1}
\issuenum{1}
\articlenumber{0}
\doinum{}
\pubyear{}
\copyrightyear{2020}
\datereceived{} 
\dateaccepted{} 
\datepublished{} 
\hreflink{https://doi.org/} 

\usepackage{amssymb}
\usepackage{latexsym}

\Title{Hybrid electrostatic-atomic accelerometer for future space gravity missions}

\Author{Nassim Zahzam $^{1}$, Bruno Christophe $^{1}$, Vincent Lebat $^{1}$, Emilie Hardy $^{1}$, Phuong-Anh Huynh $^{1}$, No\'emie Marquet $^{1}$, C\'edric Blanchard $^{1}$, Yannick Bidel $^{1}$, Alexandre Bresson $^{1}$, Petro Abrykosov $^{2}$, Thomas Gruber $^{2}$, Roland Pail $^{2}$, Ilias Daras $^{3}$, Olivier Carraz $^{4}$}

\address{%
$^{1}$ \quad DPHY, ONERA, Université Paris-Saclay, Chemin de la Hunière - BP80100, F-91123 Palaiseau, France; bruno.christophe@onera.fr (B.C.); vincent.lebat@onera.fr (V.L.); emilie.hardy@onera.fr (E.H.); phuong-anh.huynh@onera.fr (P.-A.H.); noemie.marquet@onera.fr (N.M.); cedric.blanchard@onera.fr (C.B.); yannick.bidel@onera.fr (Y.B.); alexandre.bresson@onera.fr (A.B.)\\
$^{2}$ \quad Lehrstuhl für Astronomische und Physikalische Geodäsie, Technische Universität München, Arcisstraße 21, 80333 München, Germany; petro.abrykosov@tum.de (P.A.); thomas.gruber@tum.de (T.G.); roland.pail@tum.de (R.P.)\\
$^{3}$ \quad European Space Agency, Keplerlaan 1, P.O. Box 299, 2200 AG Noordwijk, the Netherlands; ilias.daras@esa.int (I.D.)\\
$^{4}$ \quad RHEA for ESA – European Space Agency, Keplerlaan 1, P.O. Box 299, 2200 AG Noordwijk, the Netherlands; olivier.carraz@esa.int (O.C.)}

\corres{Correspondence: nassim.zahzam@onera.fr (N.Z.)}

\abstract{Long term observation of temporal Earth's gravity field with enhanced temporal and spatial resolution is a major objective for future satellite gravity missions. Improving the performance of the accelerometers present in such missions is one of the main path to explore. In this context, we propose to study an original concept of a hybrid accelerometer associating a state-of-the-art electrostatic accelerometer (EA) and a promising quantum sensor based on cold atom interferometry. To assess the performance potential of such instrument, numerical simulations have been performed to determine its impact in term of gravity field retrieval. Taking advantage of the long term stability of the cold atom interferometer (CAI), it has been shown that the reduced drift of the hybrid sensor could lead to improved gravity field retrieval. Nevertheless this gain vanishes once temporal variations of the gravity field and related aliasing effects are taken into account. Improved de-aliasing models or some specific satellite constellations are then required to maximize the impact of the accelerometer performance gain. To evaluate the achievable acceleration performance in orbit, a numerical simulator of the hybrid accelerometer has been developed and preliminary results are given. The instrument simulator has been in part validated by reproducing the performance achieved with a hybrid lab prototype operating on ground. The problematic of satellite rotation impact on the CAI has been also investigated both with instrument performance simulations and experimental demonstrations. It has been shown that the proposed configuration, where the EA's proof-mass acts as the reference mirror for the CAI, seems a promising approach to allow the mitigation of satellite rotation. To evaluate the feasibility of such instrument for space applications, a preliminary design has been elaborated along with a preliminary error, mass, volume and electrical power consumption budget.}

\keyword{Cold atom interferometer; Electrostatic accelerometer; Hybrid accelerometer; Gravity mission; Quantum space gravity; Satellite geodesy} 

\begin{document}
\section{Introduction}
\subsection{General context of space gravity missions and emergence of cold atom interferometry}
Sustained observations from dedicated satellite gravity missions (e.g. GOCE , GRACE and GRACE-FO) are key in monitoring the Earth system dynamic processes related to mass transport and in understanding their coupling mechanisms. Satellite gravimetry is a unique technique that enables the detection of sub-surface storage variations of groundwater or sub-glacial water mass exchanges that are generally difficult to access from other remote sensing techniques. It also contributes to the quantification of all relevant processes of the global water cycle, allowing to directly estimate their contribution to sea level rise. Future satellite gravity missions are expected to provide enhanced sustained observations with increased spatio-temporal resolution, higher accuracies and new products that could directly be incorporated in operational services such as early warning of hydrological extremes and monitoring of geo-hazards. Moreover, the enhanced measurements shall contribute to the improvement of Essential Climate Variables (ECV) with unprecedented quality for ground water, as well as unique measurements on climate applications, Earth energy balance closure, sea level change, mass balance of ice sheets and glaciers, heat and mass transport,…\\
The largest error contributor of state-of-the-art missions that monitor the time-variable gravity field (e.g., GRACE and GRACE-FO) is the effect of aliasing that results from the poor observation geometry and the insufficient spatiotemporal sampling of the gravity signal. The ESA/NASA MAGIC mission concept \citep{massotti_next_2021} is a well-designed satellite constellation that aims to tackle this limitation. Moreover, accelerometers used so far in gravity missions exhibit relatively high noise at low frequency and are the next largest error contributor after the aliasing effects. Quantum Spaceborne Gravimetry (QSG) measurement techniques hold the promise of substantially improving on current technologies since they can provide absolute measurements with impressive performance in term of long term stability. The key innovative technique, the cold-atom interferometry-based quantum sensor, should exploit their full potential in space by the ability to achieve long interrogation times. This could fill the technological gap and help to increase the spatial and temporal resolution of mass transport products.\\
Several space projects involving the operation of cold atom sensors have emerged overs the past thirty years mostly for fundamental physics exploration \citep{jentsch_hyper_2004, tino_atom_2007, cacciapuoti_space_2009, tino_sage_2019} such as for testing the weak equivalence principle \citep{aguilera_ste-questtest_2014, williams_quantum_2016}, detecting gravitational waves \citep{chiow_laser-ranging_2015, hogan_atom-interferometric_2016} or dark matter \citep{el-neaj_aedge_2020} but also for studying their potential in the context of satellite geodesy \citep{carraz_spaceborne_2014, chiow_laser-ranging_2015, douch_simulation-based_2018, trimeche_concept_2019, migliaccio_mocass_2019, leveque_gravity_2021}. The first accomplishments of cold atom experiments in space occurred recently in 2017-2018 with the in-orbit operation of a cold atomic clock (CACES) onboard China's Tiangong-2 space laboratory \citep{liu_-orbit_2018} and the first realizations of a Bose-Einstein Condensate (BEC) in space, on a sounding rocket with the MAIUS-1 project \citep{becker_space-borne_2018} and on the ISS with the CAL project \citep{aveline_observation_2020}. Neverthless, up to now, no demonstration of inertial measurements with a cold atom interferometer (CAI) has been yet demonstrated in space. 

\subsection{Cold atom interferometry}
\label{CAI_Intro}
This new generation of instruments, relying on the manipulation of matter waves through atom interferometry \citep{kasevich_atomic_1991}, appears nowadays as one of the most promising candidates for highly precise and accurate inertial measurements \citep{geiger_high-accuracy_2020}. Cold atom interferometers have indeed proven on ground to be very high performance sensors with the development in recent decades of cold atom gravimeters \citep{peters_high-precision_2001,louchet-chauvet_influence_2011,hu_demonstration_2013,freier_mobile_2016,karcher_improving_2018}, gravity gradiometers \citep{mcguirk_sensitive_2002,sorrentino_sensitivity_2014} and gyroscopes \citep{gustavson_precision_1997,savoie_interleaved_2018}. In addition to the undeniable contribution they could bring in practical applications such as inertial navigation \citep{jekeli_navigation_2005,cheiney_navigation-compatible_2018}, they also appear very promising for exploring fundamental physics such as for the determination of the fine-structure constant \citep{parker_measurement_2018,morel_determination_2020}, the gravitational constant \citep{fixler_atom_2007,rosi_precision_2014} and for testing Einstein's theory of general relativity with quantum objects \citep{tino_testing_2021}. In that field, atom interferometers seem notably promising for detecting gravitational waves \citep{dimopoulos_atomic_2008} and testing the weak equivalence principle \citep{asenbaum_atom-interferometric_2020,tino_precision_2020}. This technology has demonstrated impressive results in laboratory environment and begin also to prove its interest for onboard applications with the recent demonstrations of gravity field mapping during shipborne \citep{bidel_absolute_2018} and airborne \citep{bidel_absolute_2020} campaigns with a cold atom gravimeter. Progressively, proposals for using cold atom interferometers in space have emerged targeting at the beginning more particularly fundamental physics applications \citep{dimopoulos_atomic_2008,aguilera_ste-questtest_2014} but recently opened up to Earth observation objectives \citep{carraz_spaceborne_2014,chiow_laser-ranging_2015,trimeche_concept_2019,abrykosov_impact_2019,migliaccio_mocass_2019,leveque_gravity_2021}. Space offers indeed a micro-gravity environment that should allow in principle to increase the interrogation time and thus the scale factor of the atom interferometer compared to ground based atomic instruments where the size of the instrument limits the duration of atoms free-fall and thus the sensitivity of the instrument.\\ 

\begin{figure}[H]
  \centering
  \includegraphics[scale=0.3]{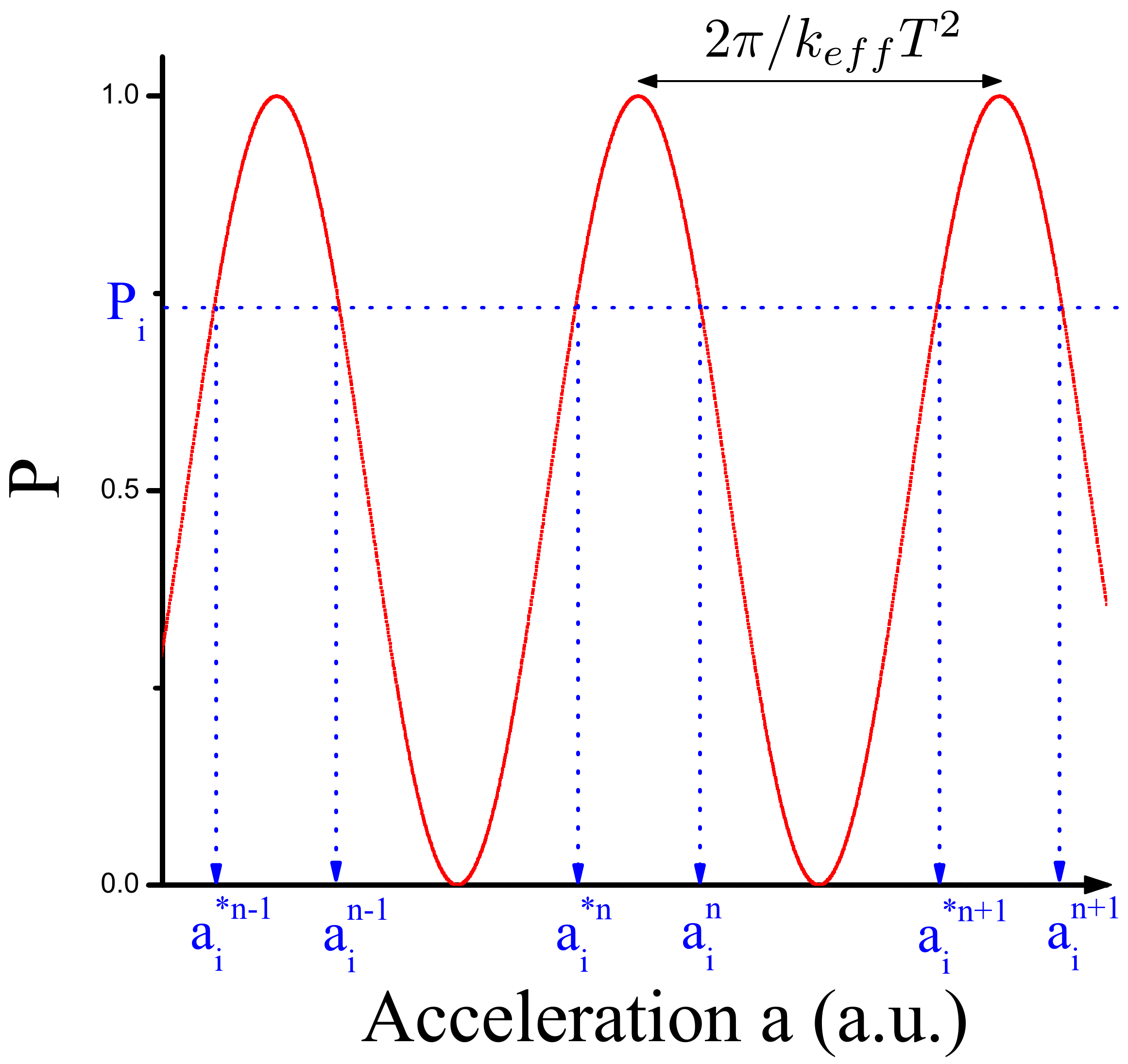}
  \caption{Output signal $P$ of a cold atom accelerometer. One measurement $P_i$ corresponds to several possible acceleration values $a_i^n$ , $a_i^{*n}$. This ambiguity can be resolved thanks to hybridization with a classical accelerometer. Here $P_{\mathrm{0}} = 0.5$ and $C=1$.}
  \label{fig:FringeAI}
\end{figure}

\begin{figure}[H]
  \centering
  \includegraphics[scale=0.5]{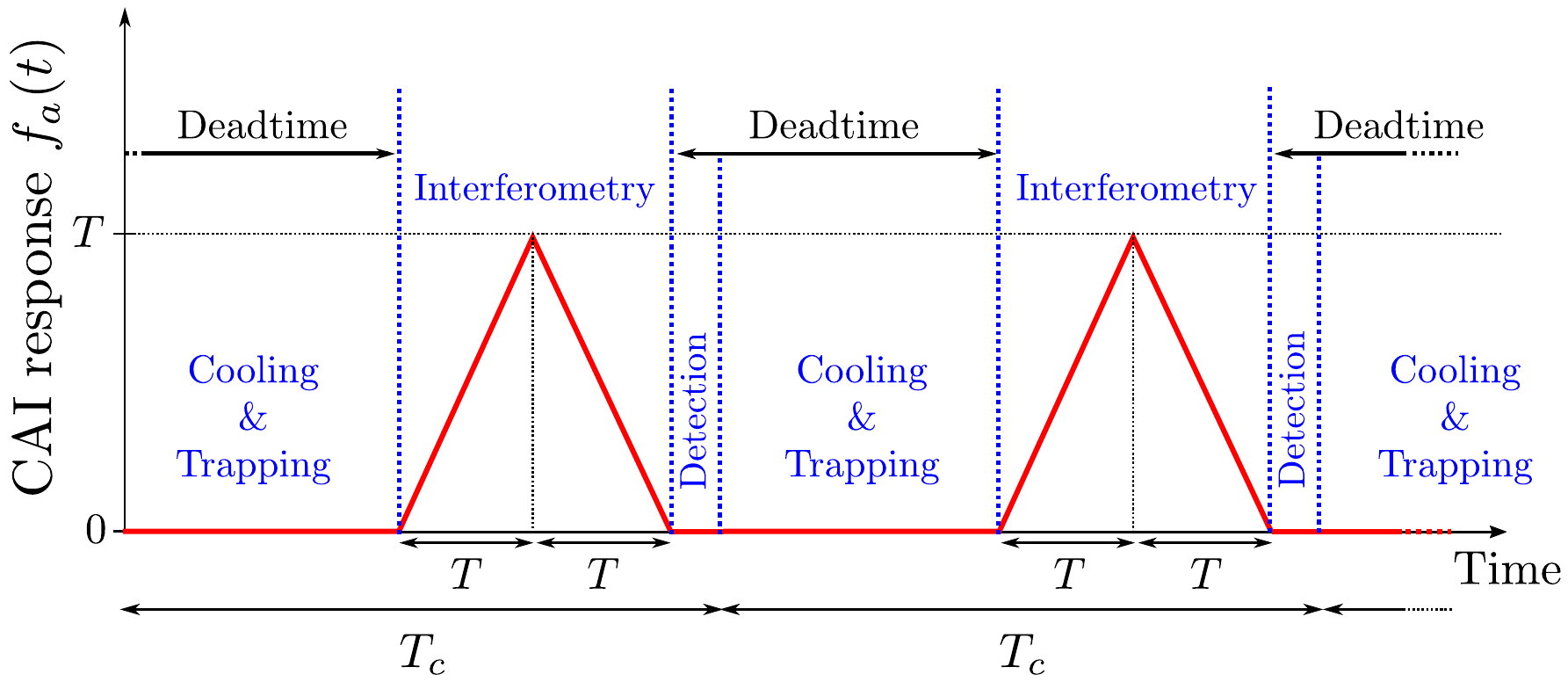}
  \caption{Typical atom interferometer sequence. After cooling and trapping the atoms, the interferometer is realized with a total duration time $2T$. The phase at the output of the interferometer is measured during the subsequent detection phase. The measurement is given with a cycling time $T_c$. The atomic instrument is only sensitive to acceleration during the interferometer period.}
  \label{fig:SequenceAI}
\end{figure}

In the rest of the document, the study deals with an atom interferometer dedicated only to acceleration measurements which is the objective of the proposed space instrument. In a cold atom accelerometer, the test mass is a gas of cold atoms obtained by laser cooling and trapping techniques \citep{metcalf_laser_2003}. This cloud of cold atoms is released from a trap, evolving in free-fall in a vacuum chamber, and its acceleration is measured by an atom interferometry technique. Typically, a Mach-Zehnder type interferometer is realized, consisting in a sequence of three equally spaced laser pulses which drive transitions between two atomic stable states. During each light pulse, the atoms interact with the laser beam and have a probability to absorb a photon out of the laser beam. In this way, the recoil of a photon can be transferred to the atom. This absorption probability can be adjusted by controlling the duration and intensity of the light pulses such that the atomic wave can be either equally split, deflected or re-combined so to form an atom interferometer. At the end of the interferometer, the proportion $P$ of atoms in one of the two stable states depends sinusoidally on the phase $\Delta \Phi$ of the interferometer (see Figure~\ref{fig:FringeAI}) which is proportional to the acceleration $\vec{a}$ of the atoms, assumed constant here, along the laser direction of propagation defined by its wave vector $\vec{k}_{\mathrm{eff}}$. The expression of the output signal is given at first order by:
\begin{eqnarray}
P & = & P_{\mathrm{0}}-\frac{C}{2} \cos(\Delta \Phi) \label{Proba_CAI}\\
\textnormal{with } \Delta \Phi & \approx &\vec{k}_{\mathrm{eff}} \cdot \vec{a} \cdot T^{2} \textnormal{.}
\end{eqnarray}
Here $T$ is the time between the three laser pulses constituting the Mach-Zehnder interferometer, $P_{\mathrm{0}}$ is the fringe offset et $C$ the fringe contrast. The output signal $P$ of the atomic sensor, corresponding to the probability for the atom to be in one of the two output ports of the interferometer, could be measured typically thanks to fluorescence or absorption imaging techniques.\\
In the case where the acceleration to be measured is not subjected to large acceleration variations, i.e. $\Delta a\ll \frac{\pi}{2k_{\mathrm{eff}}T^2}$, it's possible thanks to a specific technique \citep{farah_underground_2014} to identify the fringe index corresponding to the measurement and to retrieve unambiguously the true acceleration value. Otherwise, for large shot-to-shot acceleration variations ($\Delta a > \frac{\pi}{2k_{\mathrm{eff}}T^2}$), the true acceleration value could not anymore be recovered. This ambiguity concerning the acceleration determination reduces consequently the shot-to-shot measurement range of cold atom sensors. In regards to this limitation, it's typical to associate a classical accelerometer or seismometer to the cold atom instrument \citep{le_gouet_limits_2008,merlet_operating_2009,lautier_hybridizing_2014,geiger_detecting_2011,bidel_absolute_2018,bidel_absolute_2020}. This classical accelerometer allows the identification of the fringe index corresponding to the atomic acceleration measurement (see Figure~\ref{fig:FringeAI}). Moreover, the interferometer phase duration ($2T$) being smaller than the cycling time ($T_c$), due mainly to the loading time of the cold atomic cloud (cooling \& trapping phase) and to the detection period, the atomic instrument presents measurement dead times (see Figure~ \ref{fig:SequenceAI}) leading to potential errors \citep{dick_local_1987,santarelli_frequency_1998}. The atom interferometer is thus sensitive to acceleration only during the interferometer phase. Note that the measurement rate is typically of few Hz for this kind of instruments on ground. The interferometer temporal acceleration response function $f_{\mathrm{a}}(t)$ \citep{cheinet_measurement_2008,geiger_detecting_2011} follows a triangle shape (see Figure~\ref{fig:SequenceAI}) and highlights that the CAI, for each cycle, gives an average of the acceleration between the first and third pulse with a peak sensitivity during the second laser pulse.
In first approximation, this acceleration response function $f_{\mathrm{a}}(t)$ is related to the output interferometer phase $\Delta \Phi$ as follows \citep{cheinet_measurement_2008, geiger_detecting_2011}:
\begin{eqnarray}
\Delta \Phi &=& \vec{k}_{\mathrm{eff}} \cdot \int f_{\mathrm{a}}(t) \cdot \vec{a}(t) dt
\label{Phi_respFunc}
\end{eqnarray}
\begin{eqnarray}
\textnormal{with } f_{\mathrm{a}}(t) &\approx&  (t+T) \cdot \Theta(t+T)-2t \cdot \Theta(t)
 + (t-T) \cdot \Theta(t-T)
\label{Equ_CAI-response}
\end{eqnarray}
\indent where $\Theta(t)$ is the Heaviside function:
$$\left\lbrace
\begin{array}{l}
\Theta(t) = 0 \quad \textnormal{for} \quad t < 0\\
\Theta(t) = 1 \quad \textnormal{for} \quad t \geq 0
\end{array}
\right.
$$
This acceleration response function corresponds, in the frequency domain, to a low pass filter transfer function of $1/2T$ cut-off frequency \citep{cheinet_measurement_2008}.
\subsection{Electrostatic accelerometer}
\label{subsection_EA}
\begin{figure}[H]
  \centering
  \includegraphics[scale=0.38]{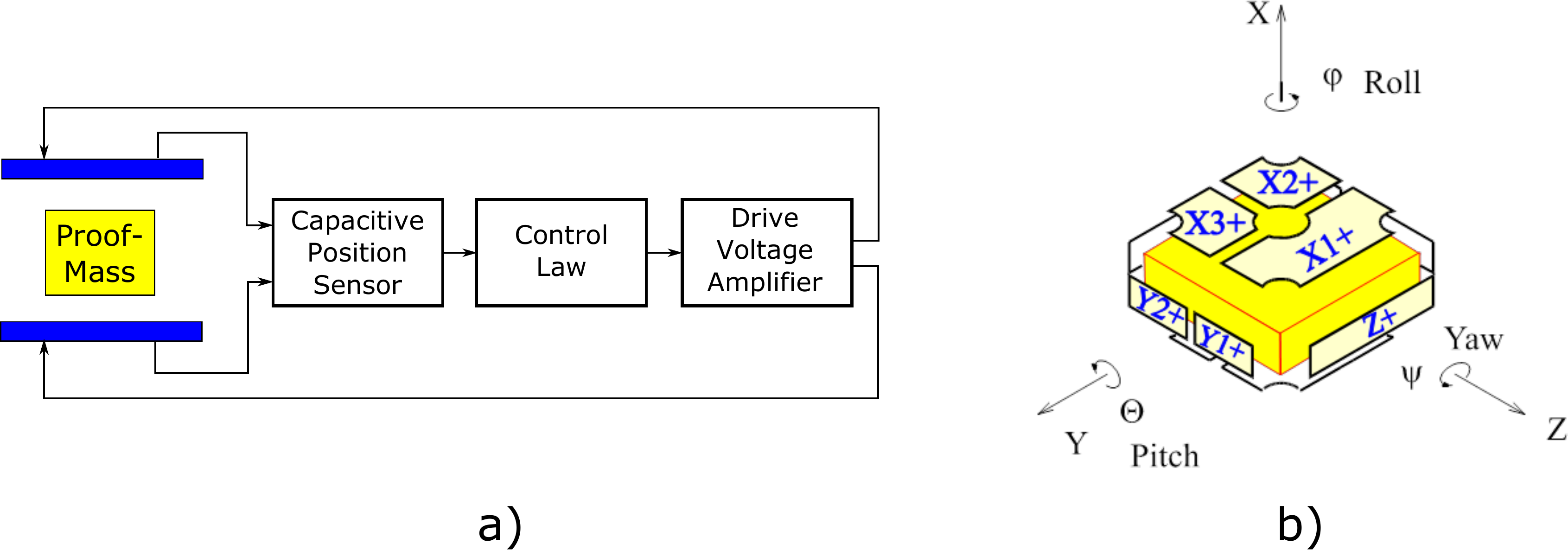}
  \caption{a) Principle of the servo-control loop for one accelerometer's axis. b) Configuration of the electrodes around the proof-mass for the six degrees of freedom control.}
  \label{fig:EA_ServoLoop}
\end{figure}

\begin{figure}[H]
  \centering
  \includegraphics[scale=0.42]{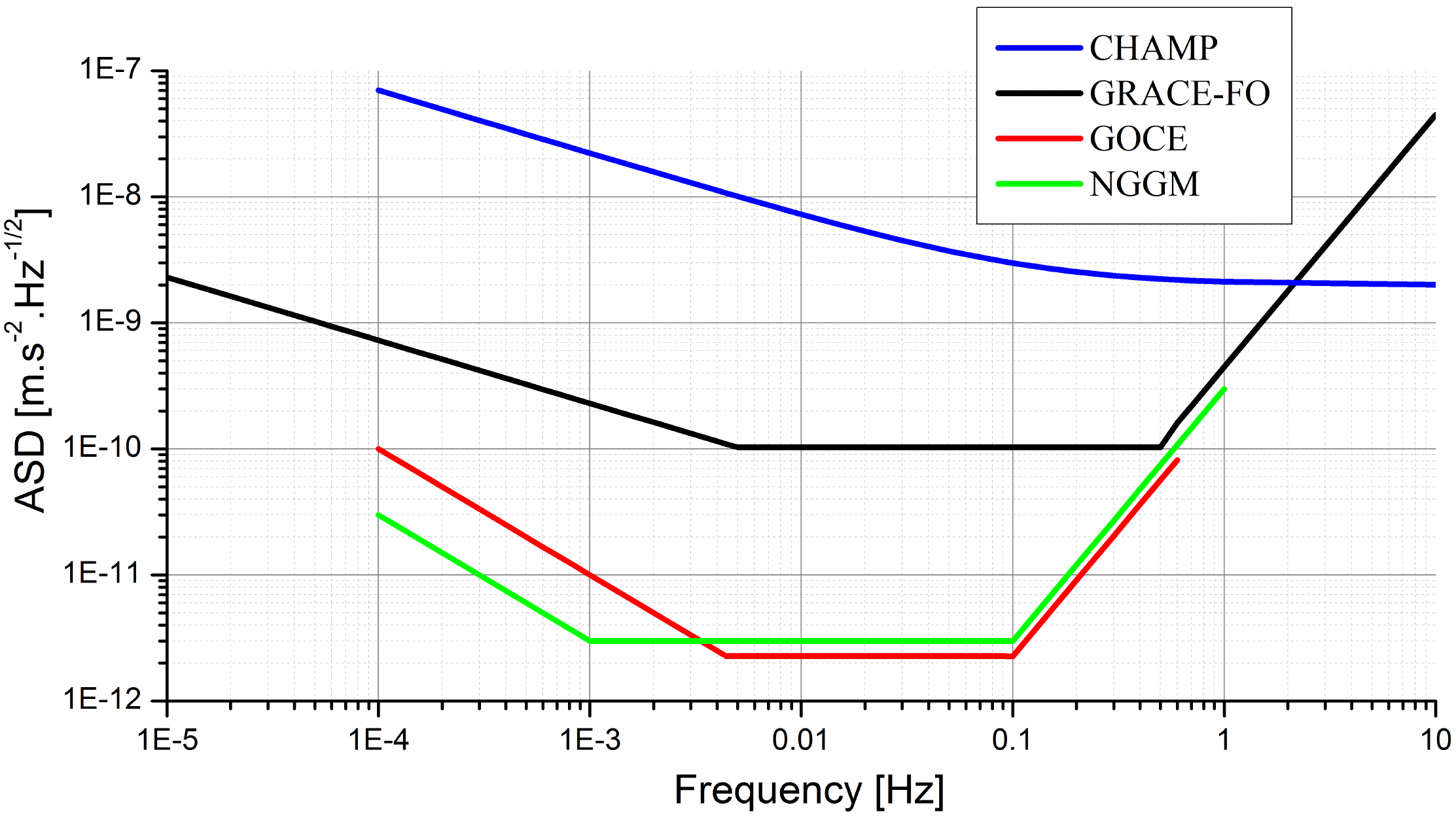}
  \caption{Noise performance specifications on the Ultra Sensitive Axis (USA) of the EAs used in CHAMP, GRACE-FO, GOCE and foreseen for NGGM.}
  \label{fig:EA_Noise}
\end{figure}

The present space electrostatic accelerometers \citep{touboul_orbit_2004, touboul_champ_2012}, used for geodesy missions, are based on the electrostatic suspension of an inertial proof-mass (PM) which is controlled to remain motionless at the centre of a cage by applying adequate voltages on the internal walls of a cage. The electrostatic forces applied on the PM compensate its relative acceleration with respect to the cage. The control voltages are therefore representative of the PM acceleration. As a standalone instrument, when placed at the spacecraft centre of gravity, the voltages provide the measurement of the non-gravitational acceleration. This concept of electrostatic servo-controlled accelerometer (see Figure~\ref{fig:EA_ServoLoop}-a) is well suited for space applications since it relies on electrostatic forces that give the possibility to generate very weak but accurate accelerations. The capacitive sensing offers also a high position resolution with negligible back-action. The accelerometer proof-mass is fully suspended with six servo-control loops acting along its six degrees of freedom, suppressing mechanical contact to the benefit of the resolution and leading to a six-axis accelerometer with three linear accelerations along X, Y, Z and three angular accelerations around $\varphi$, $\Theta$, $\psi$ (see Figure~\ref{fig:EA_ServoLoop}-b).
Such electrostatic accelerometers based on this concept have already flown on various space geodesy missions like CHAMP, GRACE or GOCE \citep{touboul_orbit_2004}, and are even present in-orbit on GRACE-FO satellites. They were also used for fundamental physics in the MICROSCOPE mission to test the Einstein Equivalence Principle \citep{touboul_microscope_2017}.
We summarize in Table \ref{EA_Specs} the main characteristics of the onboard electrostatic accelerometers for each of these space missions. We can see that the EAs used for each mission don't have the same characteristics and performance depending  on the mission objectives. The specification on the Amplitude Spectral Density (ASD) of EA noise for the three different space missions CHAMP, GRACE-FO and GOCE are represented on Figure~\ref{fig:EA_Noise}. This figure highlights the potential to improve significantly the EA noise depending on the mission objectives but it comes at the expense of a reduced measurement range (see Table \ref{EA_Specs}). 
We see in any case in the low frequency part of the spectrum an increase of noise that is mainly due to environment temperature stability that impacts the EA signal. In the worst case, this increase could scale as $f^{-3}$. Ultimately, the noise increase would be limited by the gold wire damping attached to the proof-mass that would set a $f^{-1/2}$ slope on the low frequency part of the accelerometer noise. On the high frequency part, the EA noise is mainly limited by the capacitive detector noise.
\begin{specialtable}[H]
\small
\caption {Main characteristics of last flying space EAs in CHAMP, GOCE, GRACE-FO and the foreseen EAs for NGGM, the future ESA mission. For CHAMP, the mass and power consumption don't take into account the Interface and Control Unit (ICU). USA and LSA stand respectively for Ultra Sensitive Axis and Less Sensitive Axis. For instance, in GRACE-FO, the USA are the along-track and radial axis and the LSA is the cross-track axis.}
\label{EA_Specs}
\begin{tabular}{c||cccc}
\hline\hline
EA & \multicolumn {4}{c}{Space Missions} \\
characteristics&CHAMP&GOCE&GRACE-FO&NGGM\\
\hline\hline
PM mass [g] & 72 & 320 & 72 & 507\\
GAP USA [$\mu$m] & 75 & 299 & 175 & 300\\
GAP LSA [$\mu$m] & 60 &  32 &  60 & 300\\
Meas. range USA $\lbrack$m.s$^{-2}$$\rbrack$ & $\pm10^{-4}$ & $\pm6\times10^{-6}$ & $\pm7\times10^{-5}$ & $\pm 10^{-5}$\\
Meas. range LSA $\lbrack$m.s$^{-2}$$\rbrack$ & $\pm10^{-4}$ & $\pm5\times10^{-4}$ & $\pm6\times10^{-4}$ & $\pm 10^{-5}$\\
Noise floor USA $\lbrack$m.s$^{-2}$.Hz$^{-1/2}$$\rbrack$ & $2\times10^{-9}$ & $2\times10^{-12}$ & $10^{-10}$ & $ 3\times10^{-12}$\\
Noise floor LSA $\lbrack$m.s$^{-2}$.Hz$^{-1/2}$$\rbrack$ & $2\times10^{-9}$ & $2\times10^{-10}$ & $10^{-9}$ & $ 3\times10^{-12}$\\
Total Mass $\lbrack$kg$\rbrack$ (w. elec.) & 10 (w/o ICU) & 9 & 11 & 15 \\
Total Elec. Cons. $\lbrack$W$\rbrack$ & 2 (w/o ICU) & 9 & 11 & 15\\
Total Volume $\lbrack$l$\rbrack$ & 13 (w/o ICU) & 11 & 14 & 16\\
\hline\hline
\end{tabular}
\end{specialtable}
\subsection{Hybrid atomic-electrostatic accelerometer concept}
Both atomic and electrostatic technologies have their own drawbacks and advantages in terms of performance. Typically, electrostatic accelerometers offer in space state-of-the-art sensitivity, a bandwidth of typically 1 Hz ($<$ 10 Hz), a measurement range of $10^{6}$ relatively to the achievable sensitivity and continuous measurements. They suffer nevertheless from bias instabilities mainly due to temperature dependencies which degrade the long-term stability of the measurement. In cold atom interferometers, the measurement is based on a very well-known and controlled scale factor, determined at first order only by the inter-pulse timing $T$ and the interferometer laser wavelength and relies on the well-defined energy separation between two atomic states. Atomic interferometers are therefore potentially excellent in terms of accuracy and stability. They suffer nevertheless from a low measurement rate ($1/T_{c}$), errors due to dead times measurements ($T_{c}-2T$), and a relatively low measurement range (shot-to-shot acceleration variations should stay smaller than $\approx \frac{\pi}{2k_{eff}T^{2}}$ corresponding to an interfringe variation for the atom interferometer, see Figure~\ref{fig:FringeAI}). 
Hybridizing an atom accelerometer and an electrostatic accelerometer should offer the possibility to take advantage of both technologies interests and mitigate their respective weaknesses. The atom interferometer could be used to correct the drift of the EA and simultaneously the EA could be used to fill the measurement dead times of the atomic sensor. If necessary, the EA could help to determine the fringe index of the interferometer instrument. Such hybridization scheme is reminiscent of methods already implemented for ground cold atom gravimeters where the atomic sensor is coupled to a classical accelerometer or seismometer to compensate for the same lacks of atomic instruments \citep{le_gouet_limits_2008,merlet_operating_2009,lautier_hybridizing_2014,geiger_detecting_2011,bidel_absolute_2018}. Note that the previous demonstrated hybrid instruments on Earth never involved electrostatic instruments similar to the one operating in space since this technology is not in principle adapted to ground operations, the electrostatic forces to compensate for the gravity acceleration being too large. One original aspect of the proposed atomic-electrostatic hybridization is the ability to create a close link between both accelerometers. This link is materialized by the interferometer laser beam of the atomic instrument that could be retro-reflected directly on the electrostatic accelerometer proof-mass. In this configuration, the EA proof-mass becomes the reference for the atomic measurement. The output signal from the atomic sensor gives the differential acceleration between the free-falling atomic cloud and the EA proof-mass. It offers the possibility to rotate dynamically the CAI reference mirror during the measurement in orbit and to compensate for the detrimental impact of satellite rotation on the atomic signal \citep{lan_influence_2012,barrett_dual_2016,trimeche_concept_2019, duan_suppression_2020, zhao_extension_2021}. This method of operation is currently at a very exploratory stage to assess its potential for space applications. One has for instance to guarantee that, during rotation of the proof-mass, the EA is still measuring the 3-axis accelerations with the same performance. A more conservative approach to atomic-electrostatic hybridization would be to operate two independent accelerometers whose output signals are merged through an adequate hybridization algorithm. In that case, the atom interferometer would dispose of its own mirror mounted on a piezo driven tip-tilt stage \citep{hogan_light-pulse_2009,lan_influence_2012,hauth_first_2013,zhao_extension_2021} to rotate the mirror and compensate the satellite rotation impact mainly due to Coriolis effect. In this article, we will only consider the more exploratory concept where the EA proof-mass is used as the atom interferometer mirror.

\subsection{Satellite mission scenario}
To figure out more concretely what could bring such kind of hybrid atomic-electrostatic instrument, we consider one particular scenario that has been clearly identified for a future space geodesy mission. This scenario is based on a Low-Low Satellite-to-Satellite Tracking (SST-LL) configuration \citep{wolff_direct_1969,jekeli_determination_1999}. SST-LL missions involve the measurement of the differences in satellite orbital perturbations over baselines of a few hundred kilometers. These measurements are typically implemented with two satellites on the same orbit in the Earth's gravity field, one satellite tracking the other. The gravity field is recovered by determining the inter-satellite distance variation produced exclusively by the geopotential. The distance variation between the two satellites is measured either by a microwave link, adopted on GRACE and GRACE-FO, or by laser interferometry, which has been also adopted on GRACE-FO as a technology demonstrator \citep{abich_-orbit_2019} and is the foreseen method for NGGM mission since it should allow higher resolution than a microwave link. The contribution of non-gravitational effects, produced for example by atmospheric drag, are separately measured by accelerometers on each satellite and accounted for in data processing. We will thus consider a payload allowing to retrieve the gravity field that is mainly composed of a ranging system and accelerometers.

\subsection{Overview of the study}
In this article, first are presented the numerical simulations conducted to investigate the impact of this hybrid accelerometer for gravity field retrieval \citep{abrykosov_impact_2019}. These simulations were carried out in the context of GRACE-type and Bender-type missions where different level of noise specifications of the cold atom accelerometer are considered. In this study, the hybrid accelerometer takes advantage of the long term stability of the CAI that is assumed to benefit from a white acceleration noise. The impact of the improved scale factor from which the hybrid instrument benefits is also investigated.\\
In a second step, we report some experimental demonstrations of atomic-electrostatic hybridization that have been conducted on ground, in a lab, with a dedicated electrostatic accelerometer whose proof-mass plays the role of the reference mirror for a cold atom gravimeter. The possibility to compensate satellite rotation by controlling the electrostatic PM has been also investigated experimentally and is here briefly reported.\\
Then, we present a numerical modeling of the hybrid accelerometer that has been validated by the previously described experimental demonstrations. This simulator would be of great importance to extrapolate the performance of such a hybrid instrument in the context of a space missions for which prior ground performance validations are hardly achievable.\\
Finally, a preliminary design of a space hybrid instrument will be also presented along with first estimations of achievable performances taking into account the detrimental effect of satellite rotation on the atom instrument signal. 
\section{Materials and Methods}
\subsection{Numerical gravity performance simulations}
\begin{figure}[h]
\includegraphics[scale=1.45]{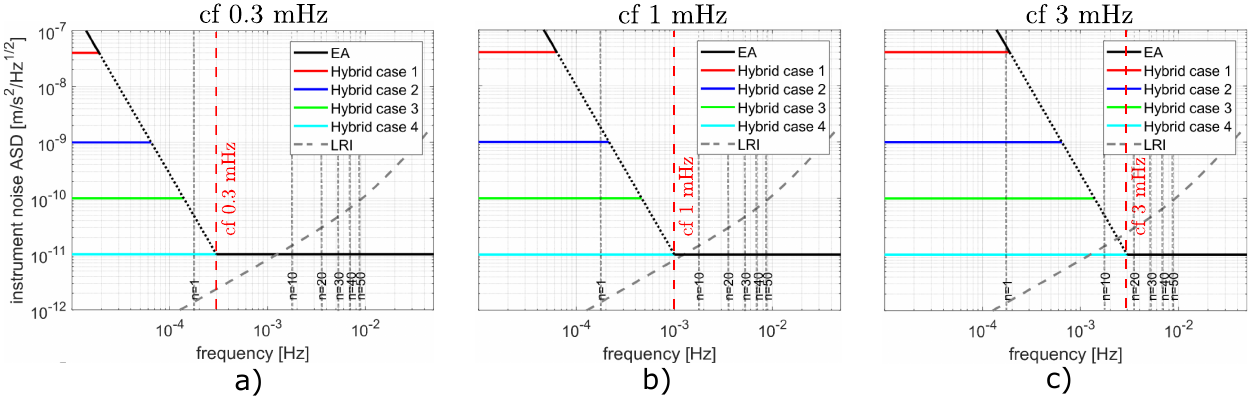}
\caption{ASD of EA/CAI hybridization scenario and a regular EA assuming a $1/f^3$ noise slope and corner frequencies  (cf) of 0.3 mHz (a), 1 mHz (b) and 3 mHz (c). The dotted line depicts the spectrum common to the respective hybridization scenario and the stand-alone EA. Vertical dashed grey lines represents the maximal contributing observation frequency to spherical harmonic coefficients of the given degree. The thick dashed grey line represents the Laser Ranging Interferometer noise in terms of range accelerations.}
  \label{fig:Hybrid_Noise_param}
\end{figure}

\begin{figure*}
\centering
  \includegraphics[scale=0.8]{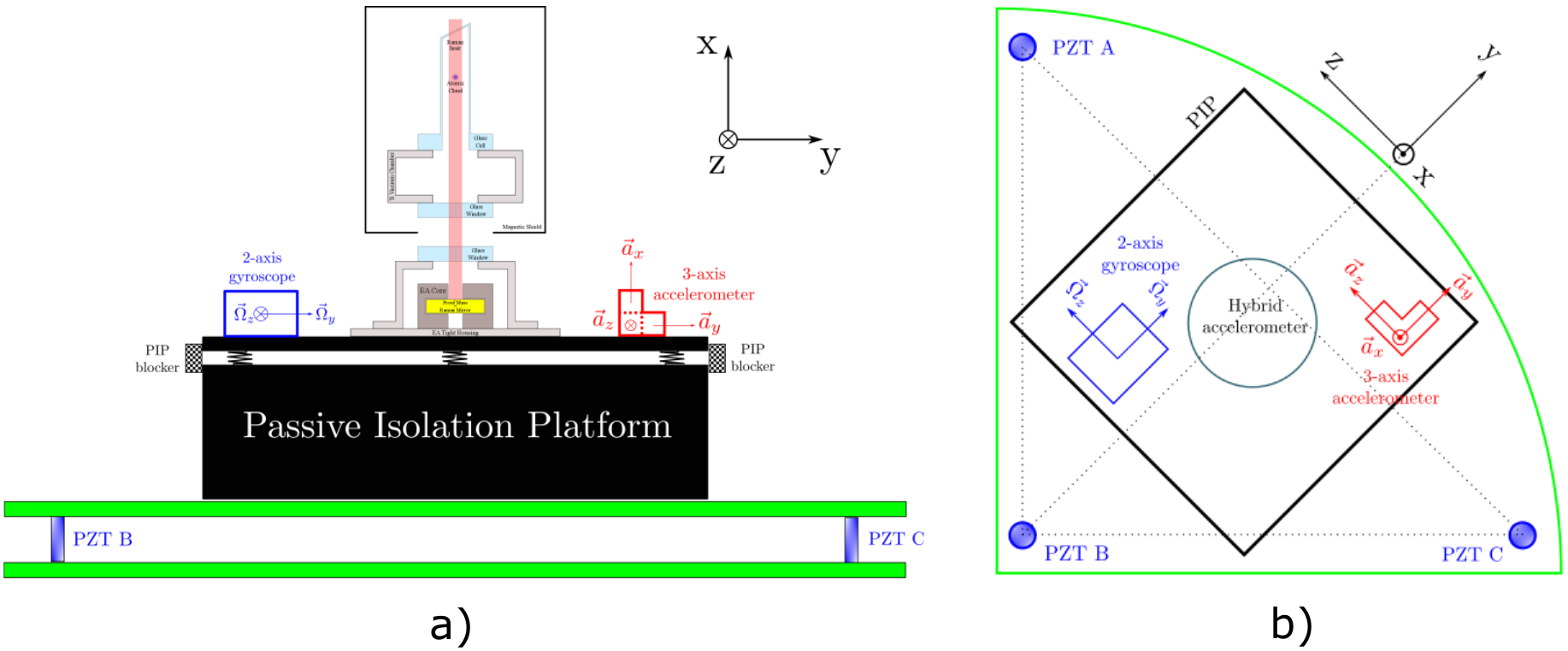}
  \caption{Representation of the experimental setup (a- side view and b- top view) used to study rotation impact on the CAI and to demonstrate rotation compensation through EA proof-mass actuation. The dimensions are not to scale. The distance between PZT A – PZT B and PZT B - PZT C is of 56 cm.}
	\label{fig:Exp_Rot_Setup}
\end{figure*}

\begin{figure}[h]
\centering
  \includegraphics[scale=0.190]{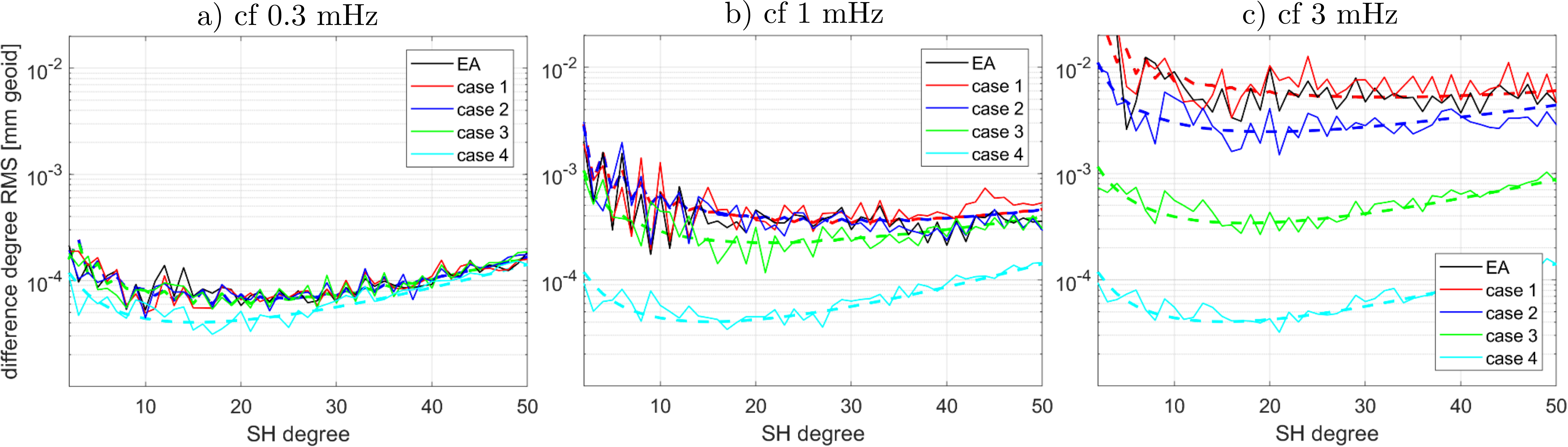}
  \caption{Degree RMS of residual coefficients for a \textbf{GRACE-type mission} under consideration of the static gravity field signal and a hybrid accelerometer with a $1/f^3$ EA noise slope for an EA corner frequency a) of 0.3 mHz (left), b) of 1 mHz (center) and c) of 3 mHz (right). The respective formal errors are shown as dashed lines of the corresponding colour.}
	\label{fig:Simu_cf_GRACE}
\end{figure}

\begin{figure}[h]
\centering
  \includegraphics[scale=0.190]{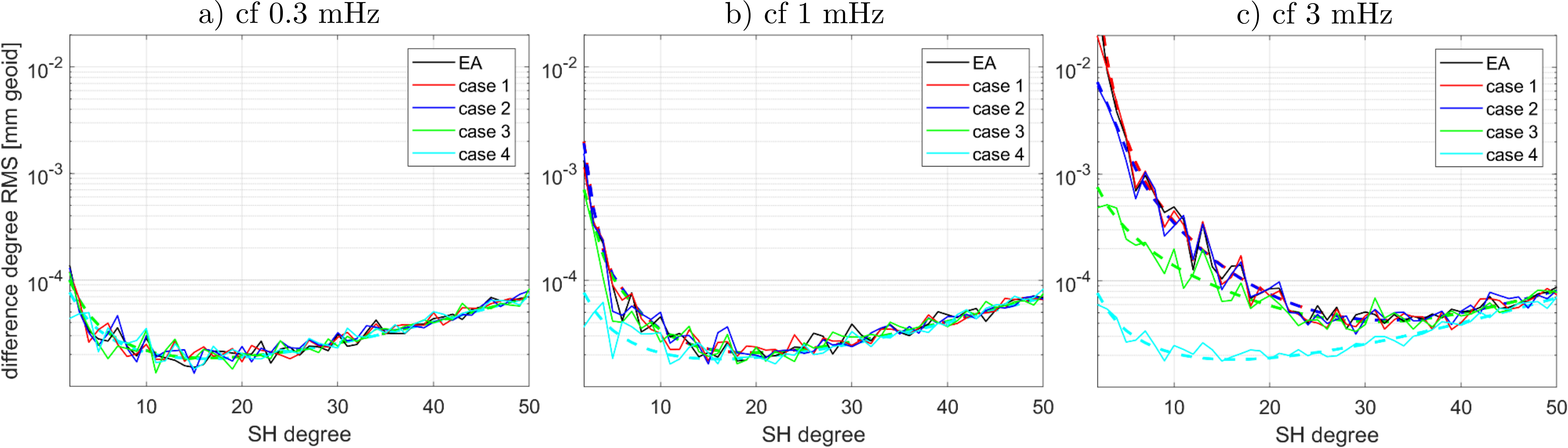}
  \caption{Degree RMS of residual coefficients for a \textbf{Bender-type mission} under consideration of the static gravity field signal and a hybrid accelerometer with a $1/f^3$ EA noise slope for an EA corner frequency a) of 0.3 mHz, b) of 1 mHz and c) of 3 mHz. The respective formal errors are shown as dashed lines of the corresponding colour.}
  \label{fig:Simu_cf_Bender}
\end{figure}

Gravity performance simulations were performed aiming to assess the added value of such hybrid accelerometer for future geodesy missions. Some additional details concerning the performed simulations can be found in \citep{abrykosov_impact_2019}. In brief, series of simulations have been conducted on IAPG's closed-loop reduced-scale simulation software, described in \citep{murbock_optimal_2014}, to study the hybrid accelerometer’s impact on the performance of gravity field retrieval. Due to several simplifications implemented in these reduced-scale simulations as a trade-off for improved computation time, their results have been validated by a comparison to IAPG's full-scale simulation tool \citep{daras_treatment_2017} where all relevant aspects of a real gravity field mission simulation and the subsequent gravity retrieval can be taken into account. The reduced-scale results exhibit a highly consistent behavior with the full-scale ones \citep{abrykosov_impact_2019}. Although the total amplitude of the residuals differ slightly between full- and reduced-scale simulations, the relative behavior among the solutions is very similar and proves the validity of the reduced-scale simulation results. All the results that will be presented in the following derive from reduced-scale simulations in case of a GRACE-type and a Bender-type satellite flight formation. The GRACE-type constellation consists of a pair of identical satellites which follow each other on a near-polar orbit, while Bender features an additional GRACE-type satellite pair on an inclined orbit. Here, both the orbit altitudes and the initial mean anomalies (resp. mean anomaly differences of the satellites within one pair) are chosen to closely match the corresponding average values of the actual GRACE- resp. GRACE-FO missions (450 – 500 km orbit altitude, 200 km inter-satellite distance). The retrieval period is set to a total of 30 days, enabling one full repeat cycle of the polar pair and nearly three full repeat cycles of the inclined one. For the instrument noise, only the two most dominant contributors to the error budget are taken into account, which are on one hand the accelerometer and on the other hand the inter-satellite ranging instrument. For the latter, the performance specifications of the laser raging interferometer (LRI) implemented on GRACE-FO is assumed \citep{Iran_assessment_2015}. For the hybrid accelerometer a number of possible performance scenarii were predefined, based on the CAI's overall noise floor (4 cases, cf. Table \ref{Simu_Param}), the EA's noise increase in the low observation frequencies (stemming primarily from thermal fluctuations), the extent of the EA's designated measurement bandwidth defined by the corner frequency at which the temperature dependent effects in the low frequencies become the major error source. The model specifications are presented in Table \ref{Simu_Param} and a selection is visualized in Figure~\ref{fig:Hybrid_Noise_param} where, in this case, the EA is assumed to present a $1/f^3$ noise slope at low frequency and a corner frequency of 1 mHz. For the CAI, we assumed different noise levels ranging from \textit{case 1}, corresponding to achieved on-ground sensitivities with state-of-the-art cold atom gravimeters \citep{hu_demonstration_2013,farah_underground_2014,freier_mobile_2016-1}, to \textit{case 4}, corresponding roughly to performance foreseen in future challenging space projects with cold atom inertial sensors \citep{aguilera_ste-questtest_2014,trimeche_concept_2019}.
\begin{specialtable}[H]
\small
\caption {Accelerometer parameters for simulations.}
\label{Simu_Param}
\begin{tabular}{cc|cc}
\hline\hline
\multicolumn {2}{c}{Parameter} & Value & Unit\\
\hline\hline
\multirow{4}{*}{CAI Noise level} & case 1 & 4$\times10^{-8}$  &\multirow{4}{*}{m.s$^{-2}$.Hz$^{-1/2}$}\\
                                 & case 2 & 1$\times10^{-9}$  & \\
			                           & case 3 & 1$\times10^{-10}$ & \\													
																 & case 4 & 1$\times10^{-11}$ & \\
\hline														
\multirow{3}{*}{EA corner frequency} &     & 0.3             &\multirow{3}{*}{mHz}\\
                                    &     & 1               & \\
																		&     & 3			          & \\
\hline
\multirow{3}{3cm}{EA low-frequency noise slope} & & $1/f$      & \multirow{3}{*}{-}\\
                                                & & $1/{f^2}$  & \\
	 															                & & $1/{f^3}$	& \\
\hline\hline
\end{tabular}
\end{specialtable}
\end{paracol}
\begin{paracol}{2}
\switchcolumn
Note that, in the following, the results assume a hybridization in just one axis that can be considered as aligned along the line-of-sight between the two satellites, pointing from one satellite's centre of mass to the other's \citep{abrykosov_impact_2019}. A selection of simulation results is presented based on specific parameters to show the impact of hybridization in gravity retrieval. In the first group of simulation results, presented in Figure~\ref{fig:Simu_cf_GRACE}, Figure~\ref{fig:Simu_cf_Bender} and Figure~\ref{fig:Simu_Stripes}, we disregard temporal gravity signals and thus temporal aliasing errors and only take into account instrument errors of accelerometers and the LRI. Afterwards, temporal gravity is included to compare the error contribution related to temporal aliasing with the instrument errors.
\subsection{Experimental demonstrations}
Experimental activities have been conducted to explore with the help of a cold atom gravimeter and a ground electrostatic accelerometer prototype the concept of space acceleration measurements with a hybrid instrument. The main objectives were here more particularly to perform some preliminary experimental demonstrations of hybridization and to address experimentally the satellite rotation issue that would induce a complete loss of contrast of the atomic signal.
\subsubsection{Hybrid lab-prototype}
To perform the experimental tests of hybridization, a cold atom gravimeter, described in detail in \citep{bidel_compact_2013}, has been used. For the electrostatic technology, a dedicated ground prototype has been assembled by reusing parts of previously developed space accelerometers. The EA has been mounted just below the atomic gravimeter, replacing the Raman mirror in a standard configuration (see Figure~\ref{fig:Exp_Setup_Hybrid}). Both instruments have been mounted on a passive vibration isolation system (see Figure~\ref{fig:Exp_Rot_Setup}), allowing tests performed in high or low vibration conditions ($10^{-6}$ m.s$^{-2}$ < $\sigma_{vib}$ <$10^{-2}$ m.s$^{-2}$, where $\sigma_{vib}$ is the standard deviation related to ground vibrations).
\begin{figure}[H]
  \centering
  \includegraphics[scale=0.25]{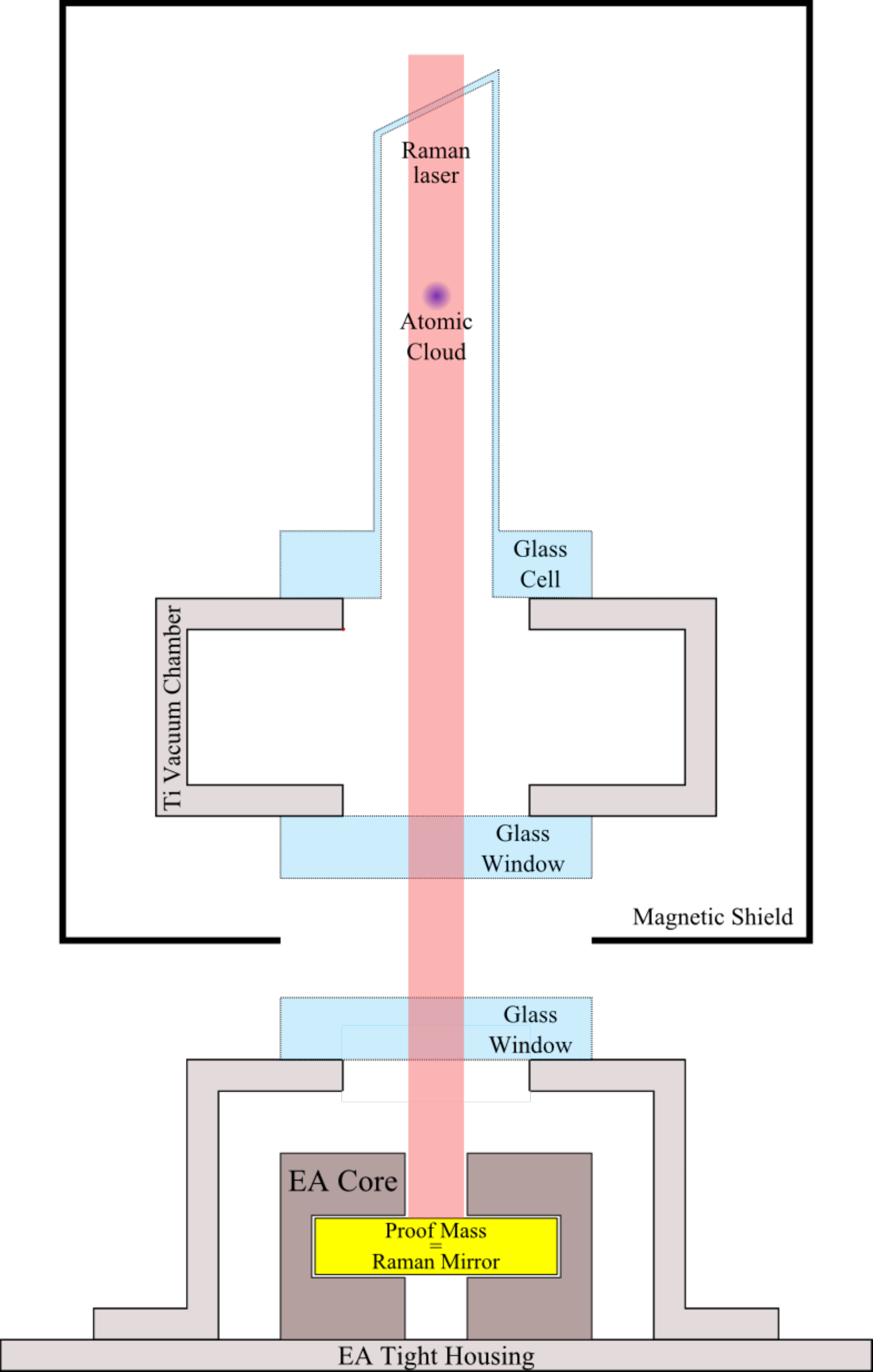}
  \caption{Scheme of the hybrid lab-prototype comprising a cold atom gravimeter and an electrostatic accelerometer. In this configuration, the proof-mass of the EA acts as the retro-reflecting Raman mirror, the reference mirror for the CAI measurement.}
  \label{fig:Exp_Setup_Hybrid}
\end{figure}
In all the experiments having demonstrated the coupling of an atomic sensor with a classic one, the Raman mirror, acting as the reference for the atomic measurement, is fixed rigidly to the classic accelerometer housing, reducing the separation between the two measuring points. Here, the association of the atom accelerometer with the electrostatic sensor allows to use the proof-mass of the latter as the Raman reference mirror. To our knowledge, no experiment has reported hybridization by directly linking the atoms and the proof-mass through the Raman laser.\\
As can be seen on Figure~\ref{fig:Exp_Rot_Setup}, the whole setup is put on an excitation platform actuated by three piezoelectric actuators allowing vertical translation and rotation around the two horizontal axes. 
\paragraph{Cold atom gravimeter}
The cold atom gravimeter used for the hybridization tests is described in detail in \citep{bidel_compact_2013}. In brief, the cold atoms are produced in a vacuum chamber based on a glass cell located inside a magnetic shield consisting of four layers of mu-metal. The sensor head containing the vacuum chamber, the magnetic shield, the magnetic coils, and the optics for shaping the laser beams and collecting the fluorescence has a height of 40 cm and a diameter of 33 cm. The maximum allowed interaction time $T$ is of 46 ms and the cycling time is $T_{c}$ = 250 ms. The Raman pulse duration is typically of 3.5 $\mu$s for a $\pi/2$ transition. The achieved contrast is typically $C = 0.39$. In standard conditions, the achieved sensitivity is of $8\times10^{-7}$ m.s$^{-2}$/Hz$^{1/2}$ and is limited by the residual vibration noise on the vibration isolation platform.
\paragraph{Ground electrostatic accelerometer}
The atom interferometer is only used in one direction, along the gravity, corresponding to the $X$ axis in the EA frame (see Figure~\ref{fig:EA_ServoLoop}). The EA proof-mass is used as the Raman mirror along the vertical axis. On ground, along this direction, the performance of the EA is strongly degraded due to the high voltage necessary to sustain the proof-mass. The servo-control parameters have been adapted to this original configuration to ensure a smooth capture of the proof-mass at switch on, and stability of its levitation with sufficient margins once it is at the center of the cage. The accelerometer cage comprises six pairs of electrodes (see Figure~\ref{fig:EA_ServoLoop}) which are used by both capacitive sensors (for proof-mass position and attitude detection) and electrostatic forces and torques. Electrode pairs, defined by two capacitor plates located in the core, are arranged around the proof-mass in such a way that the 6 degrees of freedom are controllable. Secondary entries for all position axes have been accommodated in the servo-control loops to allow for control of the proof-mass position by an external signal. Using the proof-mass of the EA for retro-reflecting the Raman laser offers an original mean to control the Raman mirror of the CAI.
\paragraph{Rotation issues for the CAI}
\label{Rot-comp_method}
Characterization of the CAI contrast behaviour is of great importance since it impacts directly the sensitivity of the accelerometer. Several effects can impact the contrast of the interferometer and we can classify these effects into two categories, those that make imperfect the Raman beamsplitters/mirrors and those that impact the atomic phase and that bring it position and/or velocity dependent. For this second category of effects, we can understand their impact on the contrast as phenomena that render less efficient the spatial/momentum overlap of the two wavepackets related to the two interferometer arms at the output of the interferometer. Another vision is to consider that these phenomena impact differently each atoms, according to their position or velocity, conducting to an averaged signal which is the contribution of dephased sine functions with a global contrast lower than the one of an individual atom signal. These two visions lead to the same evaluation of the contrast. Note that in our experimental setup, the detection scheme does not allow us to discriminate position/velocity dependent phases \cite{dickerson_multiaxis_2013}. 
The main issue of this kind of instrument in orbit that impacts the interferometer contrast is the satellite rotation and we will report here some experimental demonstrations dealing with this problematic.\\
The hybrid prototype, composed of the cold atom gravimeter and the electrostatic accelerometer (see Figure~\ref{fig:Exp_Setup_Hybrid}), is mounted on a Passive Isolation Platform (PIP) that is placed on a homemade excitation table driven by three linear piezoelectric actuators (PZT A, B, C). For rotation experiments (details will be given in a forthcoming article \citep{Marquet_Rotation}), the isolation platform is blocked so that its top plate stays fixed relative to the bottom. Consequently, no ground vibration isolation is provided anymore. The dynamic excitation is made through actuation of only PZT B, generating a rotation around the $Z$ axis. A two-axis gyroscope is used to measure the generated rotation $\Omega_{z}$ and unwanted residual rotation $\Omega_{y}$. Parasitic linear accelerations can also be measured thanks to 3 classical linear accelerometers.\\
\end{paracol}
\begin{paracol}{2}
\switchcolumn

To generate rotation of the whole setup , the PZT has been driven with a sine function to reduce parasitic excitations coming from numerous resonances of the mechanical structure. The same kind of excitation signals are provided to the secondary entries of the EA to control the EA's proof-mass for rotation compensation or for studying the impact of mirror rotation on the CAI contrast. These type of sine excitations are also useful for the proof-mass control to ensure a smooth and accurate control.The phase $\phi_{exc}$, frequency $\nu_{exc}$ and amplitude $\theta^{0}_{exc}$ of the excitation are controllable. Since the PZT excitation is a sine function, the hybrid instrument is submitted to angle variations $\theta(t)$, angular velocities $\Omega(t)$ and angular accelerations $\dot{\Omega}(t)$ with respective amplitude  $\theta^{0}_{exc}$, $\Omega^{0}_{exc}=\theta^{0}_{exc}\cdot\left(2\pi\nu_{exc}\right)$ and $\dot{\Omega}^{0}_{exc}=\theta^{0}_{exc}\cdot\left(2\pi\nu_{exc}\right)^{2}$.\\
The resulting acceleration coming from rotation can be written:
\begin{eqnarray}
\vec{a}_{rot}(t) = -\vec{\Omega}(t) \times \left[\vec{\Omega}(t) \times \vec{r}_{at}(t) \right] - 2\vec{\Omega}(t) \times \vec{v}_{at}(t) - \vec{\dot{\Omega}}(t)\times \vec{r}_{at}(t)
\label{Rot_accel}
\end{eqnarray}
\indent where $\vec{r}_{at}(t)$ and $\vec{v}_{at}(t)$ are respectively the time-dependent position, respectively velocity, of the considered atom relatively to the reference mirror frame. The first term corresponds to the centrifugal acceleration, the second one, to the Coriolis acceleration, and the third one, to the angular acceleration term.\\
To evaluate the impact of rotation on the CAI output, the interferometer phase has to be calculated by combining Equations \ref{Proba_CAI}, \ref{Phi_respFunc} and \ref{Rot_accel}. This resulting output would correspond to the contribution of only one atom. To take into account all the detected atoms, this one-atom probability has to be averaged over the position/velocity distribution of the detected atomic cloud. In our calculations, we assumed Gaussian distributions. For sake of simplicity and to give a first idea of the rotation impact on the CAI contrast, we give in the following a simplified expression of the atomic contrast considering constant, on the timescale of the interferometer phase, the angular velocity and the angular acceleration. The rotation is considered around the $Z$ axis (cf. Figure \ref{fig:Exp_Rot_Setup}), corresponding to a $\Psi$ rotation for the EA proof-mass (cf. Figure \ref{fig:EA_ServoLoop}), and the acceleration measurement along the $X$ axis. This leads to:
\begin{eqnarray}
C = C_{0} \cdot \exp \left[-2 \cdot k_{eff}^{2} T^{4} \cdot \left( \sigma_{y}^{2} \cdot \dot{\Omega}_{z}^{2} + \sigma_{x}^{2} \cdot \Omega_{z}^{4} + 4 \cdot \sigma_{v_{y}}^{2} \cdot \Omega_{z}^{2} \right) \right]
\label{Contrast_Decay_Formula}
\end{eqnarray}
\indent where $C_{0}$ refers to the CAI contrast without rotation. $\sigma_{x}$ and $\sigma_{v_{y}}$ correspond to the position and velocity dispersion of the atomic cloud respectively along the $X$ axis and the $Y$ axis.\\
The more general and complex expression of the CAI contrast, considering a non-constant angular velocity, will be given in \cite{Marquet_Rotation}.

\subsection{Hybrid instrument simulator}
The developpment of a simulation tool has been initiated with the objective to provide realistic measurements coming from an EA and a CAI both onboard a satellite in orbit around Earth. This simulator allows, depending on the chosen architecture, to hybridized both accelerometers with or without a shared EA proof-mass so this proof-mass can or does not play the role of the reference mirror for the CAI.\\
The software being developed recreates more particularly the dynamics of the satellite in its environment, the electrostatic accelerometer and the atomic interferometer as well as the hybridization of both measurements as a post-processing. The simulator is composed of several blocks modeling several subsystems (see Figure~\ref{fig:Simulator_Global}) which are described in the next paragraphs.
\begin{figure}[H]
  \centering
  \includegraphics[scale=0.5]{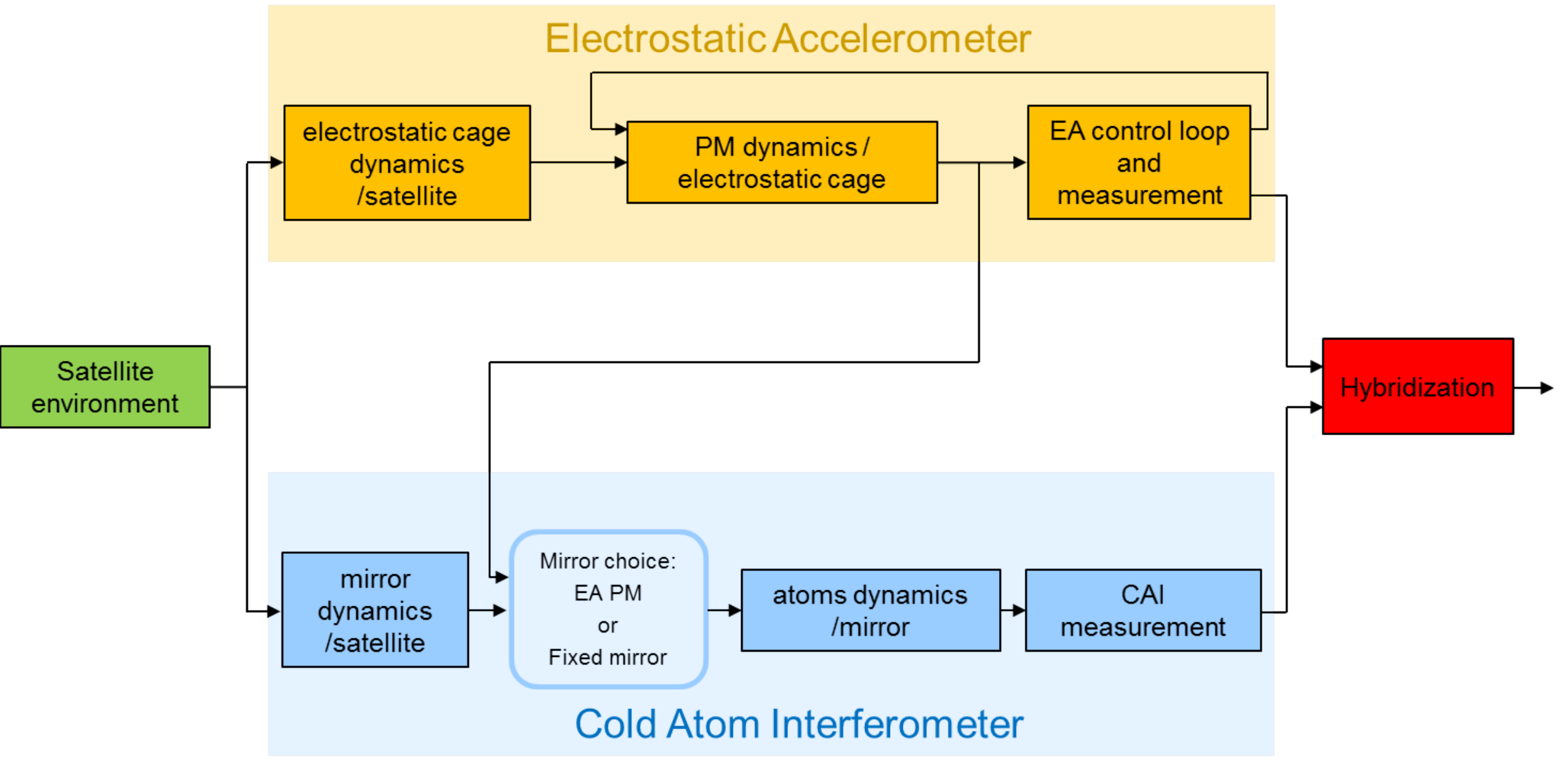}
  \caption{Global view of the software simulator for the in-orbit operation of the hybrid accelerometer.}
  \label{fig:Simulator_Global}
\end{figure}
\subsubsection{Satellite environment}
The first block of the simulator provides a simplified model of the spacecraft motion, following a perfect Keplerian orbit around the Earth. The satellite attitude is Earth pointing. Although the attitude has been considered ideal in our simulations, it is possible to add perturbation (noise or deterministic signal) in order to take into account the imperfections of the AOCS (Attitude and Orbit Control System). For the non-gravitational external perturbation, a simple constant model of the residual drag acceleration has been applied. The simulator also takes into account the gravity gradient which introduces acceleration in the measurements due to the offset between the center-of-mass of the satellite and of the electrostatic instruments on the one hand, and of the atomic instrument on the other hand.

\subsubsection{Electrostatic accelerometer}
The principle of operation of an EA has already been described in section \ref{subsection_EA}. The chosen HybridSTAR instrument (see section \ref{Design_EA} for details on the design) uses a cubic proof-mass involving that, by construction, all the three axes should have the same performance in orbit. The software simulator models the acceleration measured by the EA, that is to say, the dynamics of the proof-mass relatively to the center of the electrostatic cage. The dynamics is computed in two parts: a first block  computes the dynamics of the electrostatic cage in the reference frame of the satellite and a second one the dynamics  of the proof-mass in the reference frame of the electrostatic cage.\\
The proof-mass of the electrostatic accelerometer is surrounded by several sets of electrodes pairs. Its motion induces a variation of the gaps between the mass and the electrodes in regard, and therefore a variation of the corresponding capacity, which is measured by the capacitive detector (see Figure~\ref{fig:EA_ServoLoop}-a). In the simulator, the capacity is computed taking into account the position of the proof-mass, and the detector is modeled by a gain, a first order transfer function and a saturation, as well as its noise. By combining the outputs of the different electrodes pairs, the PM position is deduced along the six degrees of freedom (three translations and three rotations). The PID-type corrector computes the voltage to be applied to the electrodes in order to compensate for the PM motion and to keep it centered. The command is amplified by the DVA (Drive Voltage Amplifier), whose transfer function, saturation and noise are taken into account. This corrective voltage is applied to the PM by the same electrodes used for the position detection. The electrostatic stiffness of the control loop is taken into account, resulting in a perturbative term proportional to the PM position in the actuation. The resulting electrostatic actuation keeps the PM motionless in the satellite, and therefore provides a measurement of the acceleration of the PM relatively to the satellite. The simulation includes a secondary input for commanding the proof-mass position. This input is used when the proof-mass of the electrostatic instrument acts as the CAI reference mirror for the atom interferometer in order to compensate for the satellite rotation, as discussed in the following section \ref{Simulator_CAI} describing the CAI part of the simulator.
The main contributors to the EA noise have been included in the simulation: detectors noise, ADC (Analog-to-Digital Converter) measurement noise, thermal fluctuation of the bias and wire damping.

\subsubsection{Cold atom interferometer}
\label{Simulator_CAI}
In a cold atom interferometer, the test-mass is a gas of cold atoms obtained by laser cooling and trapping techniques. This cloud of atoms is released from a trap and its acceleration is measured by an atom interferometry technique (see section \ref{CAI_Intro} for more details). The simulator models the motion of each individual atom of the cloud along the three axes. Due to computation time, the number of atoms is limited to a few hundreds. Each one is assigned with a random initial position and velocity, corresponding to the atomic cloud spatial dispersion and the atomic cloud temperature.
Contrary to the electrostatic accelerometer with its high measurement rate of 1 kHz, the atomic interferometer presents a low sample frequency, limited by the interferometer cycle duration $T_{c}=4$ s in our simulations. The cycle begins with the trapping of the atomic cloud. The atoms are then released and the interferometer phase runs for 2 seconds, with the atoms interacting with the laser beams. Finally the proportion of atoms in the two atomic states is detected. The simulator therefore delivers the measurement for the atoms acceleration once per cycle.\\
For the CAI, a Mach-Zehnder type interferometer is simulated, consisting in a sequence of three laser pulses. A very simple model has been used where the measurement of the atom interferometer can be seen as a combination of three distance measurements between the atoms and the reference mirror. The simulator therefore models the dynamics of the atoms relatively to the mirror. As developed in the next part, this mirror can either be a fixed mirror or the proof-mass of the electrostatic accelerometer.\\
For simplification sake, the laser pulses durations of the CAI are considered infinitely short compared to the interrogation time $T$ and the intrinsic effects that modify the energy level of the two involved atomic states between the first and third laser pulse are not taking into consideration for the moment in our simulation. The interferometer laser beams are considered with a uniform intensity distribution, perfectly retro-reflected on the reference mirror and directed along the $X$ axis. It is possible to add in the simulation a phase noise and a phase bias.\\
The probability for one atom to be in one particular atomic state is deduced from the phase considering an ideal contrast $C=1$ and offset $P_{0}=\frac{1}{2}$, and the proportion of atoms measured in each atomic state is computed as the mean value of this probability for all the atoms composing the cloud, in addition to a white noise corresponding to the quantum projection noise (see section \ref{CAI_Perf}). The extraction of the measured acceleration is computed at each cycle $i$ from this mean probability $\overline{P}_{i}$ according to:
\begin{eqnarray}
a_{CAI,i}=\frac{\epsilon_{i}}{k_{eff}T^{2}}\arccos\left[\frac{2\left(\overline{P}_{0,i}-\overline{P}_{i}\right)}{\overline{C}_{i}}\right]+n_{i}\frac{2\pi}{k_{eff}T^{2}}
\end{eqnarray}
\indent where $\epsilon_{i}=\pm1$ and  $n_{i}\in\mathbb{Z}$\\

$\epsilon_{i}$ and $n_{i}$ result from the fringe indetermination due to the cosine function linking the interferometric phase and the acceleration. In the hybridization scheme, they are determined thanks to the EA. At this stage of simulator development, $\overline{C}_{i}$ and $\overline{P}_{0,i}$, resulting from the averaging of the signal over all the atoms, are considered perfectly known.

\subsubsection{Hybridization algorithm}
\label{section_Algo}
The objective of the hybridization algorithm is to combine both electrostatic and atomic measurements so that the EA can benefit from the long-term stability of the CAI measurement.\\
The principle of the algorithm is described in Figure~\ref{fig:Software_Hybridization}. The two inputs are the acceleration measurement of the EA ($a_{EA}$) at 1 kHz and the signal of the CAI ($\overline{P}_{i}$) at $1/T_{c}$, that is to say 0.25 Hz for a cycle time $T_{c}=4$ s.
\begin{figure}[H]
  \centering
  \includegraphics[scale=0.3]{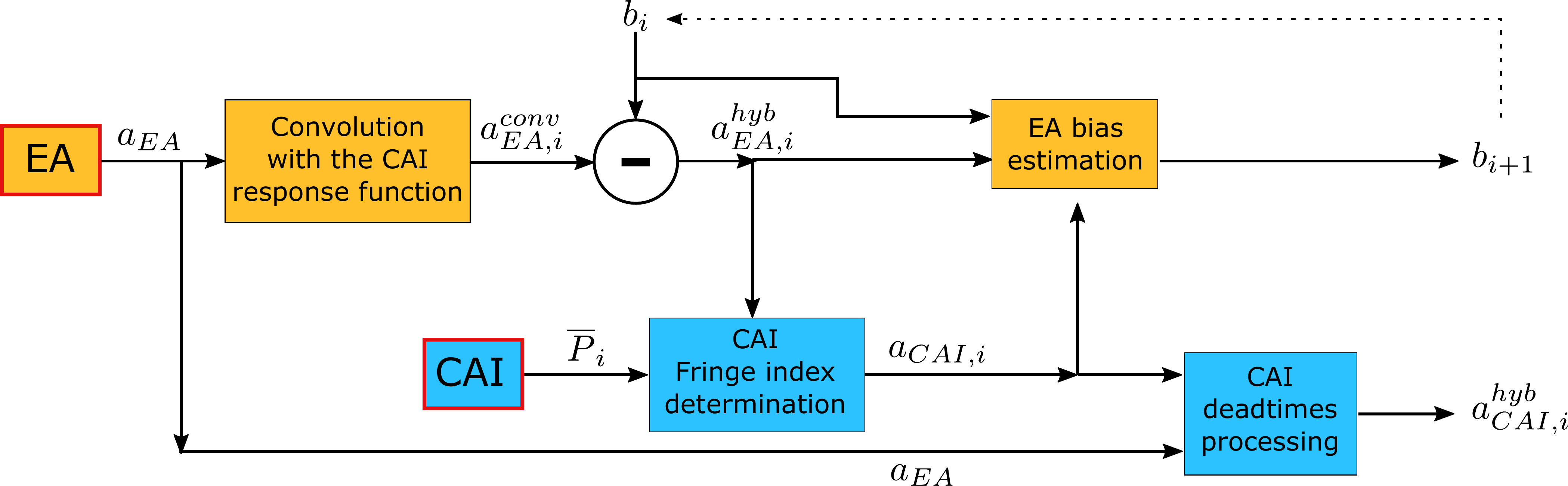}
  \caption{Global structure of the hybridization algorithm.}
  \label{fig:Software_Hybridization}
\end{figure}
The fringe index of the CAI is determined thanks to the EA measurement. For this purpose, the EA measurement is convoluted by the response of the atom interferometer (cf. Equation~\ref{Equ_CAI-response}) and corrected from the EA bias giving a debiased EA measurement ($a^{hyb}_{EA,i}$).\\
The estimation of the EA bias is done iteratively by comparing the value of acceleration provided by the CAI ($a_{CAI,i}$) and the value of acceleration provided by the EA for the current cycle after convolution with the atomic response function, and corrected by the bias estimated at the previous iteration ($a^{hyb}_{EA,i}$).\\
For the interferometric cycle $i$, the estimation $b_i$ of the EA bias is:
\begin{equation}
b_{i+1} = b_{i} + G_{b} \times \left(a^{hyb}_{EA,i}-a_{CAI,i}\right)\\
\end{equation}
\indent where $G_b$ is a gain to be adjusted to give more or less weight to the current measurement. It is linked to the hybridization characteristic time.\\
In the CAI, the atoms acceleration is measured once per cycle. The interaction phase is surrounded by two dead times (see Figure~\ref{fig:SequenceAI}), dedicated to the trapping and detection of the atoms. Because of these measurement dead times, the atom interferometer is subjected to errors. To avoid this effect, the EA measurements are used to fill the dead times. The hybrid acceleration measurement, after this processing filter, is for cycle i:
\begin{eqnarray}
a_{Hyb,i} = a_{CAI,i} + \left[a_{EA} \ast \left(f_{a,T_c}-f_{a,T}\right)/T^2\right]\left(t_i\right)
\end{eqnarray}
\indent where $f_{a,T}$ is the acceleration response function of the CAI for an interrogation time $T$ and $f_{a,T_c}$ for an interrogation time $T_c$.

The two possible outputs of interest are the continuous electrostatic acceleration measurement corrected with its bias estimation at each cycle, that is $a_{EA}-b_{i}$, and the interferometric measurement after determination of the fringe index and removal of dead times, $a_{CAI,i}^{hyb}$. The former is the one that we mainly consider in this paper and that we use for the numerical gravity performance simulations. The latter could be useful in the case where the CAI offers better noise performance compared to the EA and when the mission allows a lower measurement sampling rate.

\section{Results and discussions}
\subsection{Numerical gravity performance simulations}
\subsubsection{Instrument noise impact}
It is clearly observed on Figure~\ref{fig:Simu_cf_GRACE} that a higher EA corner frequency allows for a larger impact of the CAI, but simultaneously lowers the overall performance of gravity field retrieval. This is not observable for \textit{case 4} where the CAI performance is equal to the EA noise floor. Not all hybridization scenarii lead to an improvement of the solution quality compared to a stand-alone EA. Indeed, looking for instance at Figure~\ref{fig:Simu_cf_GRACE}, no significant differences can be observed between the coefficient residuals of the reference EA scenario and \textit{case 1} and only a slight improvement can be seen for \textit{case 2}, in the disadvantageous scenario, for the EA, to present a 3 mHz corner frequency. This behavior can be attributed to the fact that the obtained accelerometer improvement lies below the frequency bandwidth that is relevant for gravity field retrieval. The lower boundary of this frequency band is defined by the satellites orbital frequency which is approximately $1.8 \times 10^{-4}$ Hz (equivalent to complete revolution in about 90 min) for both the polar and the inclined pair. In \textit{case 2}, for an EA corner frequency of 1 mHz, the hybridization of the EA occurs at a frequency that is very close to the edge of the retrieval band and therefore no significant improvement could be achieved. This is not anymore the situation if now an EA corner frequency of 3 mHz is assumed. Considering a 1 mHz EA corner frequency, we can observe only significant improvements with respect to a stand-alone EA scenario when assuming a hybrid accelerometer compatible with \textit{case 3} and \textit{case 4}. The degree RMS of the residuals improves by up to a factor 3 in the SH degrees below $n=10$ for \textit{case 3} and, in the same spectrum area, by roughly one order of magnitude for \textit{case 4}.\\
Considering now a Bender-type configuration (cf. Figure~\ref{fig:Simu_cf_Bender}), we can see that the overall level of the coefficient residuals is significantly lower than for a GRACE-type one. This highlights one the interest of a Bender-type mission that improve the observation geometry and that allows, amongst others, to register East-West components of the gravity field. The accelerometer stability plays here a much smaller role towards the solution quality than in case of a single-pair mission. Considering the optimistic scenario of \textit{case 4} and an EA corner frequency of 1 mHz, the performance gain is now only limited to degrees below $n=20$. Nevertheless, it allows for improvements of up to one order of magnitude in the very low degrees up to $n=5$ which is still substantial with regards to time-variable gravity applications.\\
\end{paracol}
\begin{paracol}{2}
\switchcolumn
\end{paracol}
\begin{paracol}{2}
\switchcolumn
It is important to note that the retrieval performance gain that could be achieved through hybridization is not limited to the frequency bands specified in Figure~\ref{fig:Hybrid_Noise_param}, but instead extends into the high-degree spectrum of the gravity field solution (cf. Figure~\ref{fig:Simu_cf_GRACE}, Figure~\ref{fig:Simu_cf_Bender} and Figure~\ref{fig:Simu_Stripes}). This behaviour arises from the fact that a one-to-one relation between the instrument's frequency spectrum, related to time domain of the orbit track, and the spherical harmonic spectrum, related to the space domain of the sphere only holds for zonal coefficients, while coefficients of high orders contain significant amounts of low-frequency information \citep{Rummel_1993}. Therefore, if the quality of low-frequency observation increases, then, consequently, the retrieval quality of sectorial and tesseral coefficients is also improved. Figure~\ref{fig:Simu_Stripes}-a) and -c) show the relative improvements made in analyzing the formal error triangles of \textit{case 3} and \textit{case 4} with respect to the reference EA scenario. We can see that the hybridization may significantly influence high degree sectorial coefficients as well as high-order tesseral coefficients. This finding is of particular interest for a GRACE-type mission, as this behaviour can contribute to the reduction of the typical striping pattern observed in Figure~\ref{fig:Simu_Stripes}-c), -a') and -b') which, aside from temporal aliasing, occurs due to the low retrieval quality of (near-)sectorial coefficients, which in turn stems from a lack of observations in East-West direction. In case of a Bender-type mission (cf. Figure~\ref{fig:Simu_cf_Bender}), the accelerometer’s noise impact on the retrieval of the (near-)sectorial coefficient groups is far more limited in comparison to a single polar pair, because the overall observation geometry is already significantly improved through the additional observation components in East-West directions provided by the inclined pair.
\begin{figure}[H]
\centering
\includegraphics[scale=0.25]{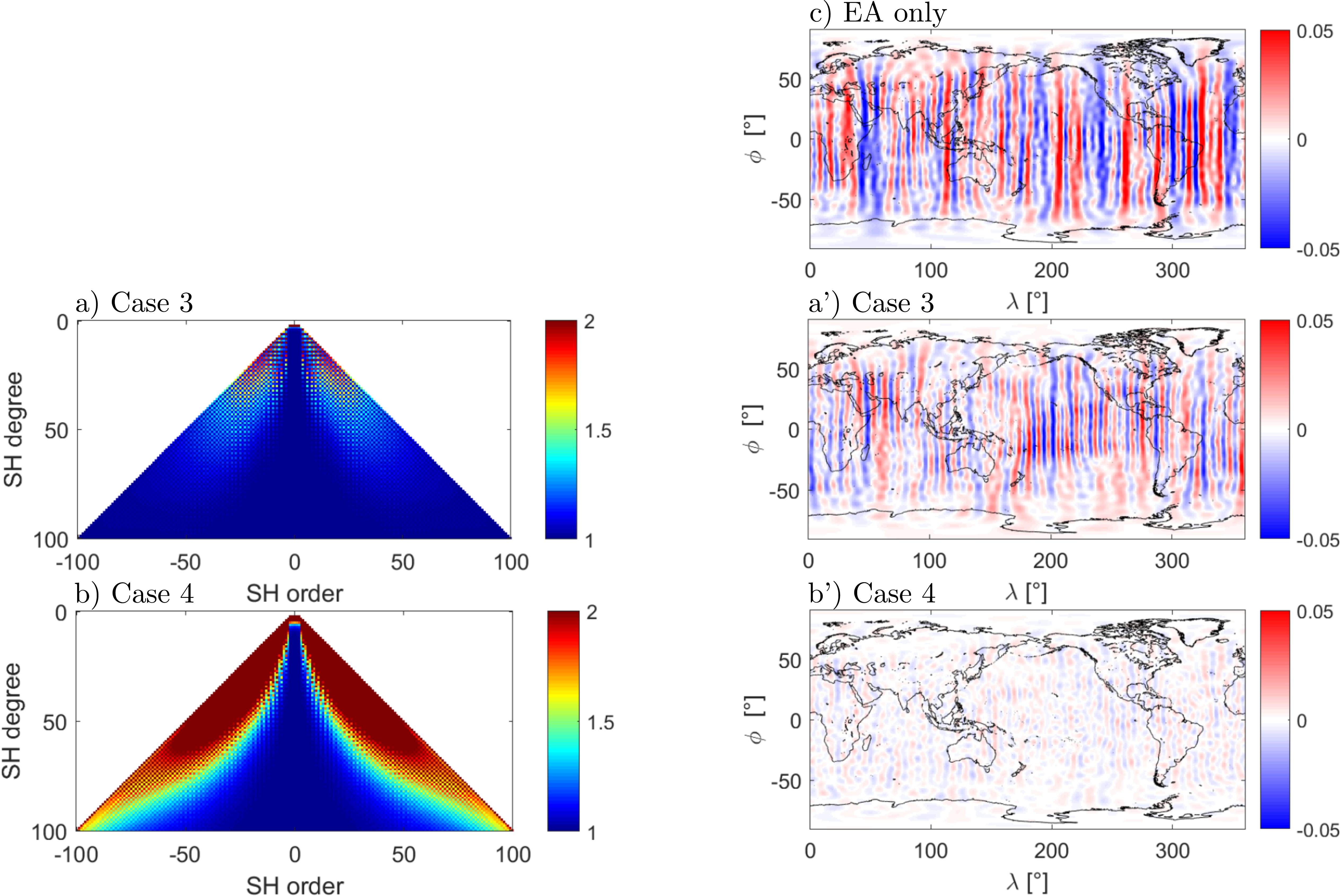}
\caption{Relative improvement of the formal errors for \textit{case 3} (a)  and \textit{case 4} (b) towards the EA scenario for a \textbf{GRACE-type mission} under consideration of the static gravity field signal and a hybrid accelerometer with a $1/f^3$ EA noise slope as well as an EA corner frequency of 1 mHz. Global geoid height errors are shown for EA-only (c), \textit{case 3} (a') and \textit{case 4} (b').}
\label{fig:Simu_Stripes}
\end{figure}
%

\subsubsection{Temporal variation of gravity field}
A series of simulations have been conducted to investigate the hybrid accelerometer’s impact when temporal variations of the gravity field due to tidal and non-tidal effects (atmosphere (A), ocean (O), continental hydrology (H), land ice mass (I), and solid earth (S)) are taken into account. All the results are detailed in \citep{abrykosov_impact_2019}. Introducing the different components of the temporal signal induces temporal aliasing which degrades the gravity retrieval performance. The different components of the temporal signal are introduced subsequently. For the non-tidal signal components, the updated ESA Earth System Model is used while the ocean tides are introduced by means of the models FES2004 (used as “truth”) and EOT08a (used as a priori model for de-aliasing). 
In either simulation case, the retrieval error is significantly increased in comparison to a scenario assuming only instrument errors due to the now introduced temporal aliasing. The differences between \textit{cases 1-4} have been greatly diminished, meaning that the aliasing constitutes the predominant error source for gravity retrieval – especially if AO (atmosphere and ocean) and the ocean tides are considered. Nevertheless, the retrieval errors of \textit{cases 1-4} obtained from simulations based on HIS (land hydrology, land ice mass and solid Earth) variations (i.e. where AO and ocean tides are disregarded) still show notable variations primarily in the spectrum below $n=10$ and therefore suggest that the accelerometer performance is not entirely masked by the aliasing errors. In case of additional AO, some variations can still be found in the low-degree spectrum, but they are far less pronounced than in case of HIS alone. Once the ocean tides are included in the simulations, the accelerometer performance becomes fully subordinate to the temporal aliasing. These findings clearly indicate the need for high-quality de-aliasing models resp. techniques to co-estimate AO and ocean tides in order to see the hybrid accelerometer’s additional value for gravity field retrieval in the context of a real satellite gravity mission. Consequently, this finding strongly advocates the implementation of a Bender-type mission (or a multi-pair mission in general), as such a configuration enables the co-estimation of high-frequency mass variations for the low-degree coefficients (i.e. reduces the temporal aliasing significantly).
\subsubsection{Bias and scale factor impact}
The impact of accelerometer bias and particularly scale factor uncertainty on the gravity field retrieval has been investigated. First, it has been concluded that accelerometer measurements suffering from a constant observation offset would only affect the retrieval of the term $C_{00}$ and, consequently, some further (primarily zonal) coefficient groups. However, since $C_{00}$ is fixed and linked to the total Earth's mass, the gravity solution is scaled accordingly, thus fully removing any effects resulting from the bias.
The following is dedicated to evaluate the impact of an improved knowledge or stability of the hybrid accelerometer's scale factor with respect to the one of a stand-alone EA. We will consider that the retrieval error $\Delta$ has two dominant contributions one coming from the measured signal impacted by the scale factor uncertainty and the other coming from the instrument noise and the objective is to determine which one dominates in terms of ASD of acceleration noise. For this analysis, we define the relative scale factor knowledge as $2\times 10^{-3}$ for the EA and $10^{-5}$ for the hybrid measurements, assumed to be invariant over the observation period.
We consider first a mission flying in drag-free mode as specified in NGGM requirements \citep{massotti_next_2021} where the thrusters compensate the drag signal at a certain level leaving the so-called residual drag to be measured by the accelerometer. In this case, the contribution of the scale factor error to $\Delta$ is fully masked by the instrument noise in both cases of a stand-alone EA and a hybrid instrument.\\
We consider now a mission where no drag compensation is applied at all, with a level of drag acceleration signal given in Figure~\ref{fig:Drag_signal}, extrapolated from \citep{knabe_hybridization_2020}, here slightly more conservative especially with regards to the cross-track and radial signals' spectra. 
\begin{figure}[H]
  \centering
  \includegraphics[scale=1]{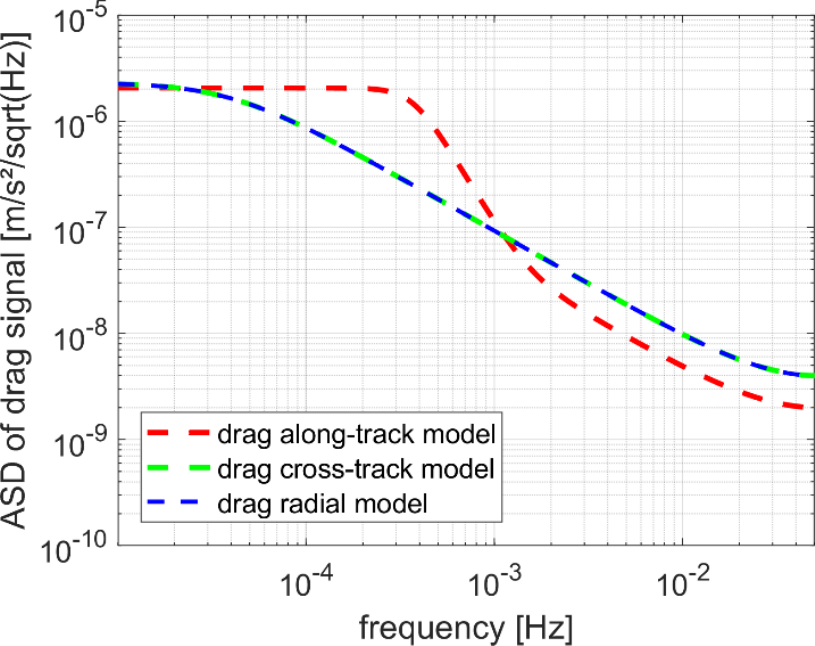}
  \caption{Drag signal model at 361 km altitude, extrapolated from \citep{knabe_hybridization_2020}.}
  \label{fig:Drag_signal}
\end{figure}
It can be seen on Figure~\ref{fig:SF_EA}-a, where a stand-alone EA is considered, that the total retrieval error is dominated clearly by the scale factor error from 0.2 mHz to 10 mHz leading to an increased retrieval error (cf. Figure~\ref{fig:SF_EA}-b).

\begin{figure}[H]
  \centering
  \includegraphics[scale=0.8]{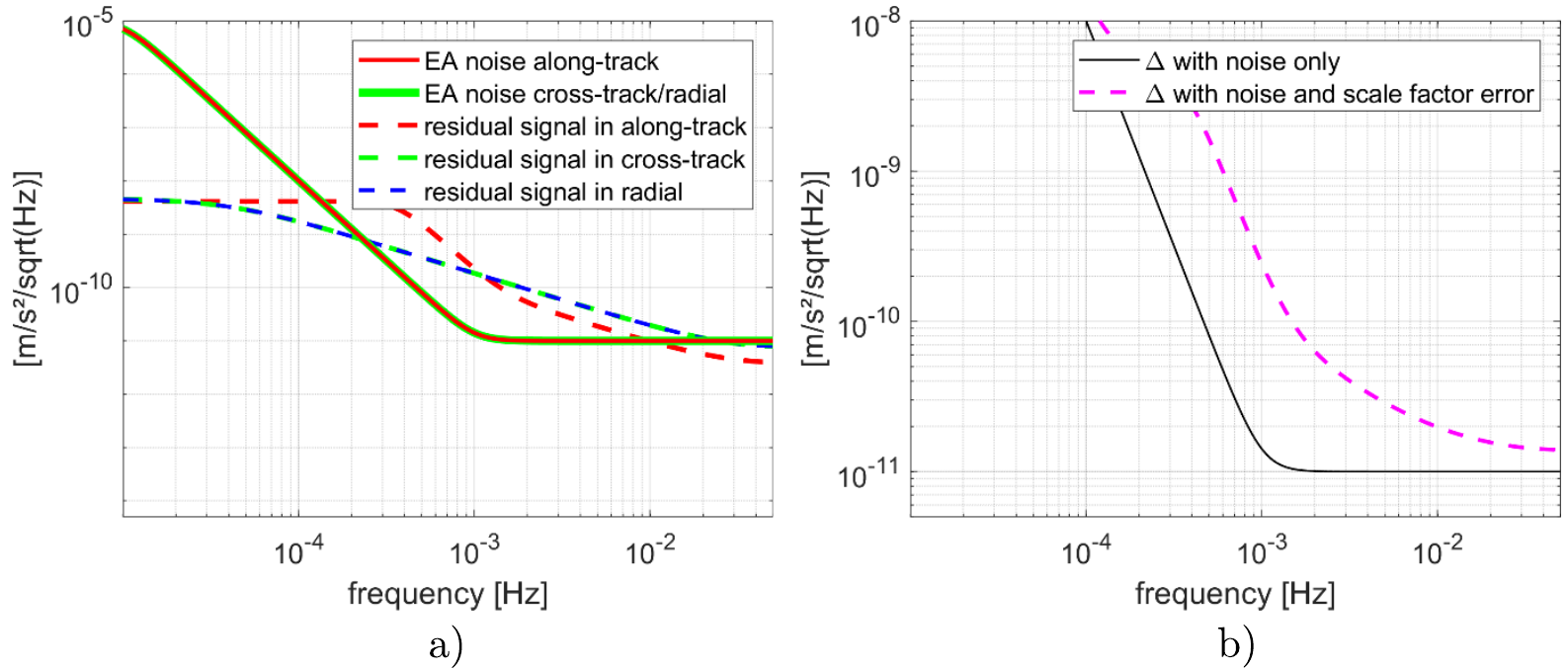}
  \caption{Retrieval error with scale factor and instrument noise contributors,  considering a stand-alone EA (a) and the corresponding LOS projection (b).}
  \label{fig:SF_EA}
\end{figure}
When a hybrid instrument is now considered (cf. Figure~\ref{fig:SF_Hyb} where a \textit{case 2} 1D-hybridization is analyzed), the impact of the scale factor error is not any more significant relatively to the instrument noise. If we consider particularly a 1D-hybridization along-track and a \textit{case 2} CAI noise level, the impact of the scale factor error is not any more significant relatively to the instrument noise. We can see in Figure~\ref{fig:SF_Hyb}-a that the residual signal in cross-track and radial direction features partially higher amplitudes than the instrument noise level but this effect is fully mitigated once the Line-Of-Sight (LOS) projection is computed. This performance already poses a significant improvement compared to a stand-alone EA.
\begin{figure}[H]
  \centering
  \includegraphics[scale=0.8]{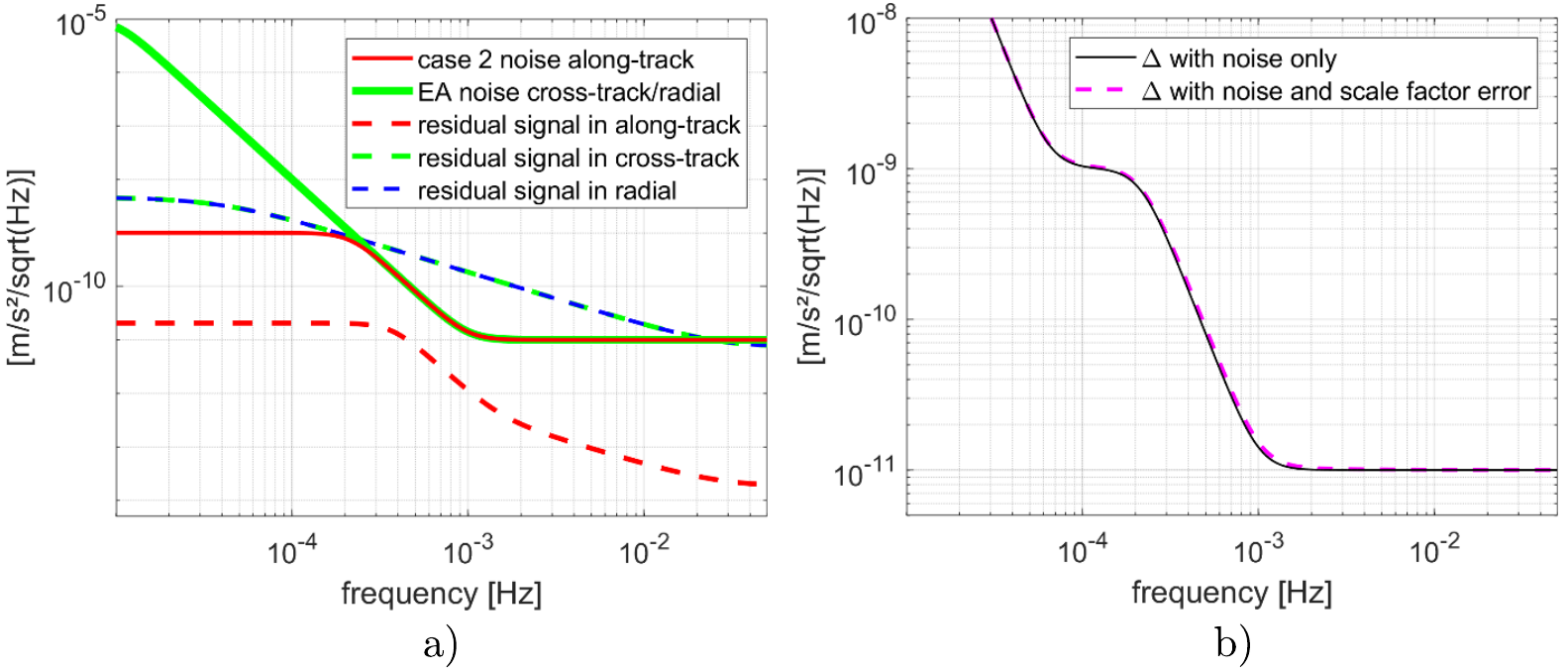}
  \caption{Retrieval error contributors with scale factor and instrument noise  contributors, considering a \textit{case 2} 1D-hybridized instrument (a) and the corresponding LOS projection (b).}
  \label{fig:SF_Hyb}
\end{figure}

\subsection{Experimental demonstrations}
This section is dedicated to the presentation of the main experimental results aiming to investigate in a lab, on ground, the concept of a space hybrid instrument.
\subsubsection{Hybridization demonstration}
A hybridization algorithm, similar to the one described in section \ref{section_Algo}, has been implemented in the experiment, coupling both the atom accelerometer and the electrostatic accelerometer. The EA is used to determine the CAI fringe index for each cycle. Also for each cycle, the EA bias is evaluated thanks to the CAI. We can see on Figure~\ref{fig:Exp_Hyb-signals} such experimental implementation of this hybridization algorithm for a CAI interrogation time $T = 20$ ms that illustrates qualitatively, in a clear way, the gain that could be reach in a future space mission.
These results show that the hybridized system takes the best part of each EA and atomic instrument, namely the short term sensitivity of the EA and the long term stability of the CAI. The hybridization algorithm shows a correction on the EA signal that begins to be effective from roughly 3 s of integration time. Note that the continuous feature of the EA allows to explore also the sensitivity region below 0.25 s which is not the case for the CAI which is limited by its measurement cycling frequency of 4 Hz.
\begin{figure}[H]
  \centering
  \includegraphics[scale=0.55]{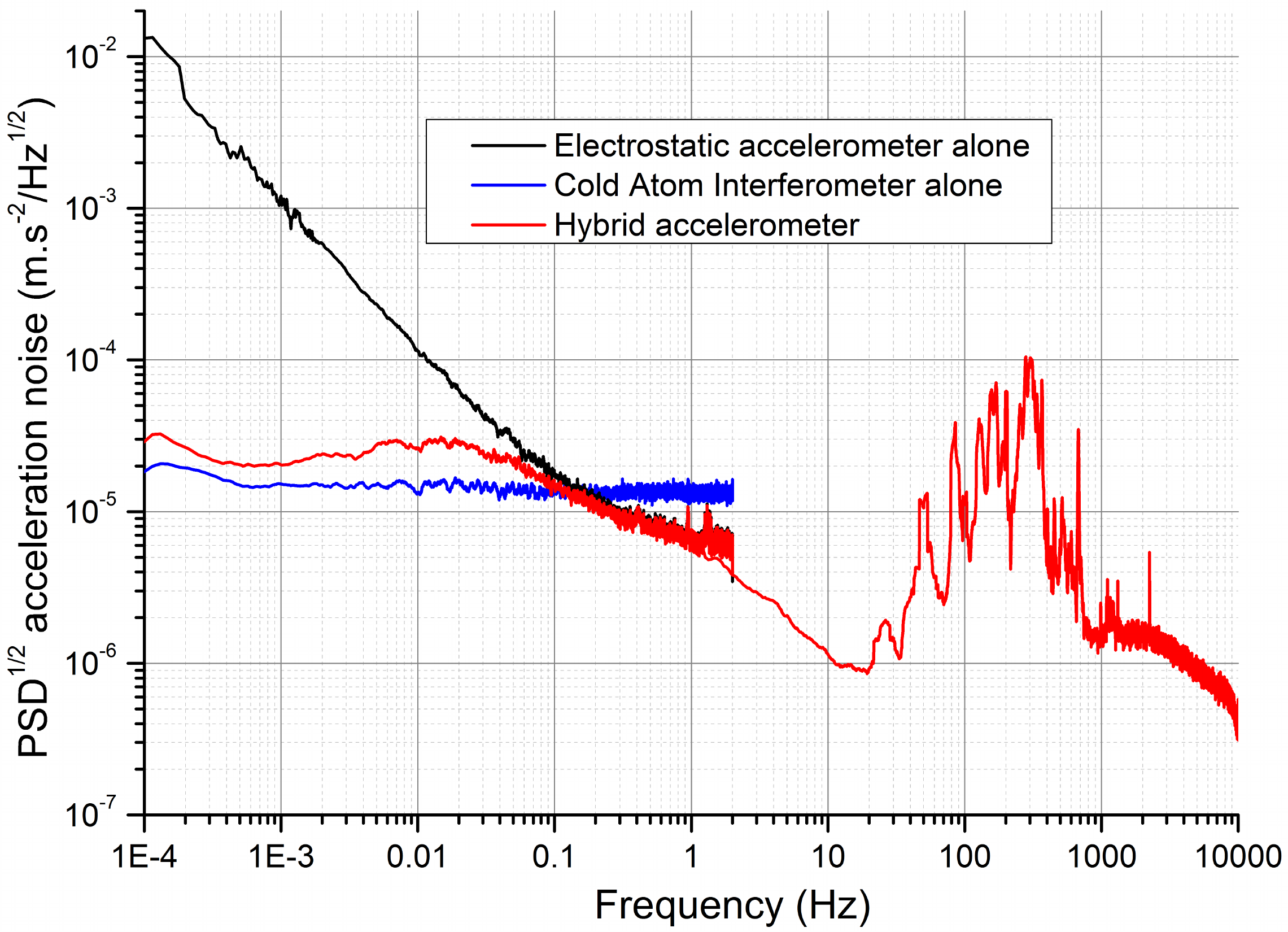}
  \caption{ASD acceleration noise for the EA (black curve), the cold atom gravimeter (blue curve), and the EA corrected by the CAI through a hybridization algorithm (red curve). The interrogation time of the CAI has been set to $T = 20$ ms for highlighting the interest of hybridization and reproducing qualitatively the noise assumptions treated in the numerical gravity simulations. Note that, at high frequency ($\> 1$ Hz), the red curve is completely superimposed on the black one.}
  \label{fig:Exp_Hyb-signals}
\end{figure}
We would like to emphasize that these presented data don’t correspond to what is really expected in term of performance from an EA, a CAI or a hybridized instrument in a future space mission. The EA ground prototype has been for instance designed specifically to sustained the vertical gravity acceleration. The performance of the EA along the vertical axis is therefore degraded due to the high voltage necessary for the proof-mass levitation.

\subsubsection{Satellite rotation compensation}
The characterization of CAI contrast dependency relative to, independently, platform rotation or EA's proof-mass rotation, has been studied in details and will be described in a dedicated forthcoming article \cite{Marquet_Rotation}. We report here the main results.
\paragraph{Proof-mass rotation}
The proof-mass of the EA, acting as the reference mirror for the CAI, has been excited in rotation during CAI measurement cycles. The method of rotation excitation follows a sine function as described in section \ref{Rot-comp_method}. We consider here the specific case of an excitation frequency of $\nu_{exc}=1/2T\approx 10.87$ Hz and study the effect on atomic fringe contrast for different excitation phases. Here the impact causing the contrast drop comes mainly from the Coriolis acceleration and angular acceleration of the proof-mass that act with different weights depending on the phase and frequency of the excitation.

We can see on Figure~\ref{fig:Results_PM_Rotation}-a) the variation of the $\Psi$ angle of the proof-mass around the $Z$ axis for a specific phase excitation of 0 rad  and $\pi/2$ rad. The angle of the proof-mass is monitored in real time during the interferometer sequence by capacitive detection of the EA. These two phase values correspond to excitation configurations where the atomic contrast is at its extrema (see Figure~\ref{fig:Results_PM_Rotation}-b). The excitations represented here correspond to peak-to-peak amplitudes of around 106 $\mu$rad, 7.2 mrad/s and 494 mrad/s$^{2}$. For an excitation frequency of 10.87 Hz and a 0 rad phase, we can see on the black curve of Figure~\ref{fig:Results_PM_Rotation}-a) that, during the three laser pulses, the mirror of the CAI (the EA's proof-mass) is at the same angular position. So for this phase configuration of excitation, the atomic contrast has no reason to be affected by rotation effects. 
The black dashed line on Figure~\ref{fig:Results_PM_Rotation}-b) results from calculations (see section \ref{Rot-comp_method}) of the contrast loss using the expected experimental parameters as inputs. The input parameters for the calculated curve are a transverse atomic temperature of 1.6 $\mu$K and an atomic source of 0.8 mm diameter (corresponding to $4\sigma$ with $\approx 87 \%$ of the atoms) with the assumption of a Gaussian position and velocity distribution. We can see that the calculations are in very good agreement with the experimental data. 
\end{paracol}
\begin{figure*}
\centering
  \includegraphics[scale=0.33]{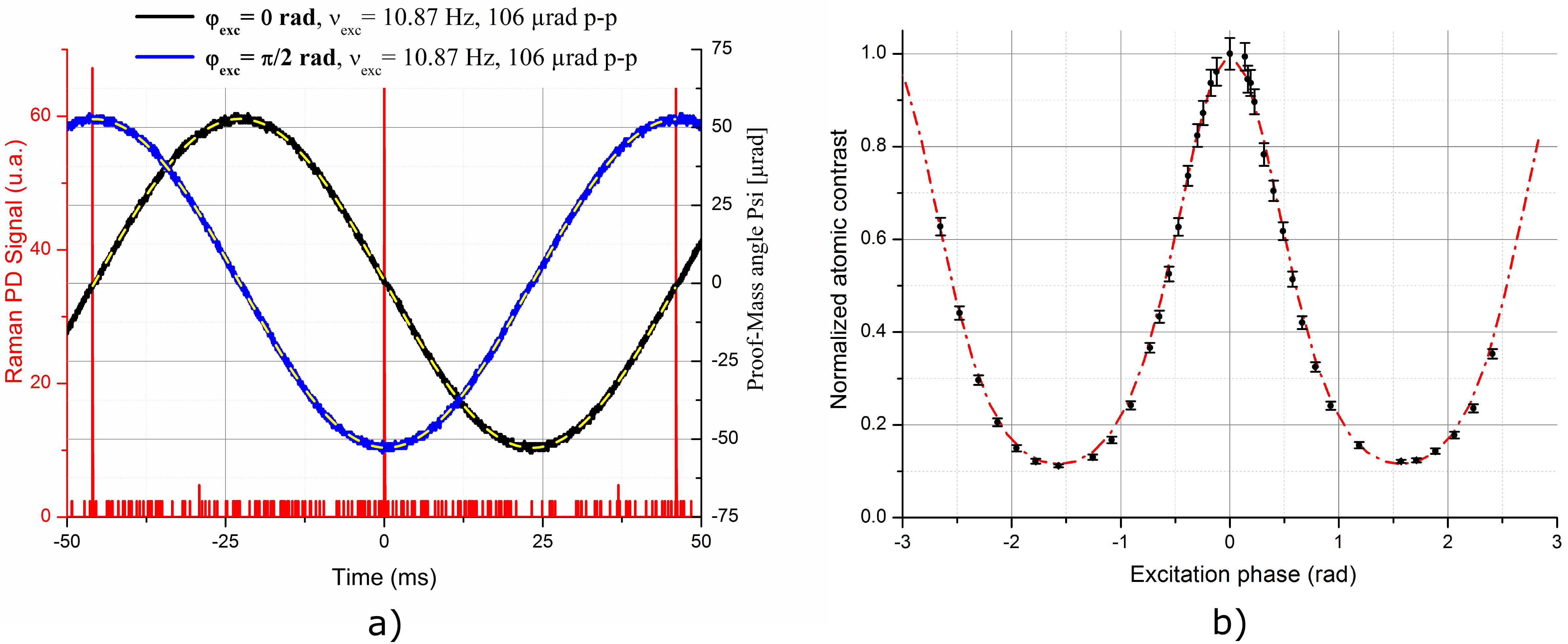}
  \caption{EA's PM rotation and impact on CAI's contrast. a) Proof-mass angle $\Psi$ measurement during the interferometer phase delimited by the three laser pulse detected thanks to a photodiode (red signal). The rotation excitation is made at 10.87 Hz with an amplitutde of 106 $\mu$rad p-p. Two excitation phases are represented, 0 rad (black) and $\pi/2$ rad (blue). The excitation phase is defined relatively to the first Raman laser pulse of the CAI. The dashed yellow lines are sine fitting functions. b) Evolution of the atomic contrast according to the excitation phase for a 10.87 Hz sine excitation of amplitude 106 $\mu$rad p-p. The experimental data are represented in black dots with $\pm 2 \sigma$ error bars.  The red dashed line results from a calculation of contrast loss using as inputs the same experimental excitation parameters. The interrogation time of the CAI is $T=46$ ms.}
	  \label{fig:Results_PM_Rotation}
\end{figure*}
\begin{paracol}{2}
\switchcolumn
Note that the ability to control the CAI mirror as it has been done offers the possibility to characterize the atomic source in terms of position and velocity dispersion. Indeed by analyzing quantitatively the loss of contrast due to rotation due to Coriolis effect or angular acceleration, we should be able to determine precisely these characteristics \cite{Marquet_Rotation}. The estimation of the transverse mean position and velocity of the atomic cloud relatively to the mirror is also achievable by analyzing the variation of the measured acceleration for different rotation excitations \cite{Marquet_Rotation}. This could be of prime interest for space in-flight characterizations of the CAI.
\paragraph{Platform rotation}
We show now the impact of excitation of the whole setup by rotating the platform supporting all the hybrid instrument consisting mainly of the cold atom gravimeter and the electrostatic accelerometer. The excitation in rotation is much less accurate than the one that can be done with the proof-mass. On Figure~\ref{fig:Contraste_Table}, we report the atomic contrast loss due to platform rotation following a sine function. The angular velocity of the table is measured with a gyroscope. This amplitude of excitation allows to fully reduce the atomic contrast.
\begin{figure}[H]
  \centering
  \includegraphics[scale=0.3]{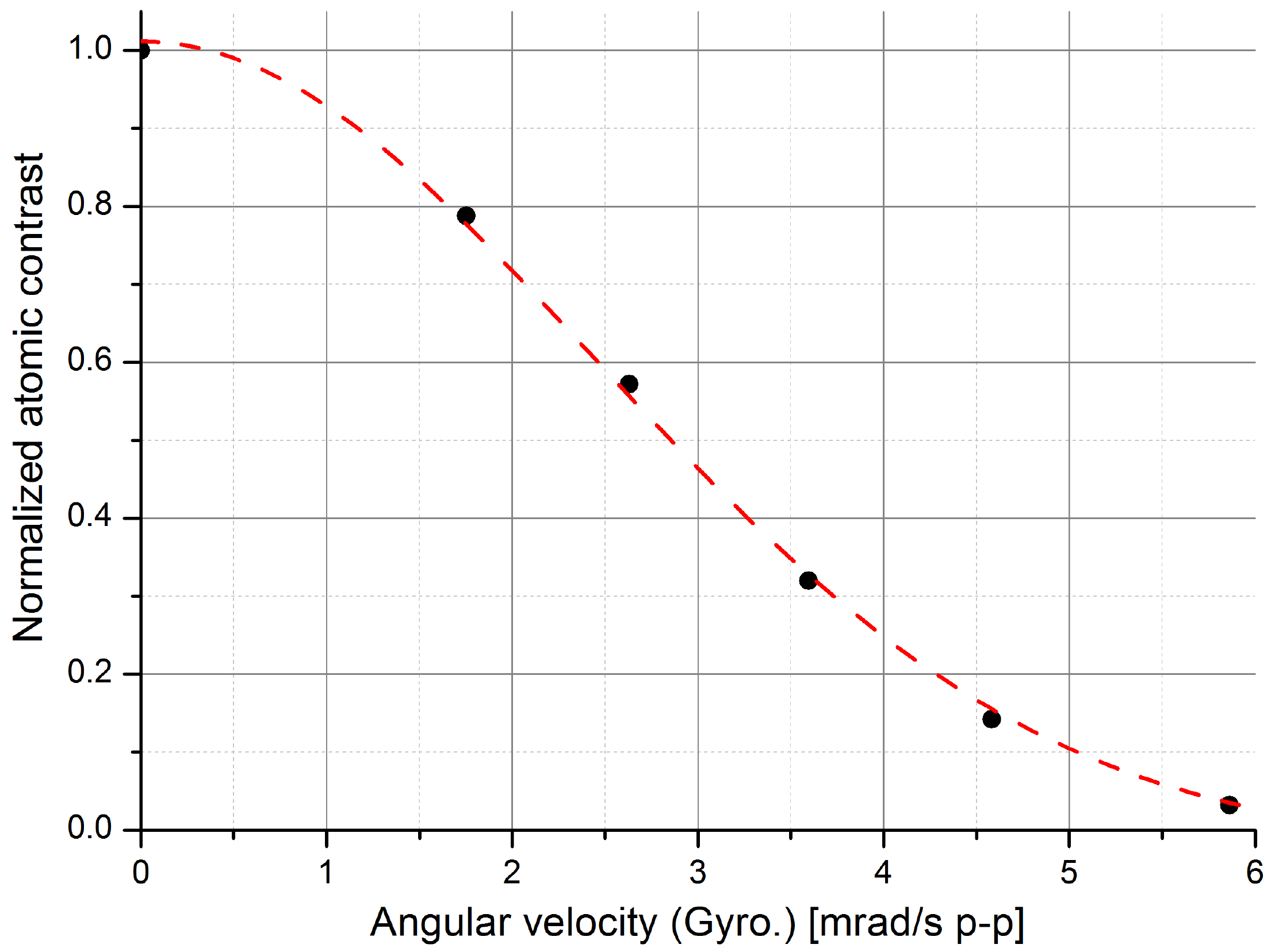}
  \caption{Contrast loss due to platform excitation with a sine actuation at 8.1 Hz around the $Z$ axis. The angular velocity of the platform is measured thanks to a gyroscope. The red dashed line represents a Gaussian fit in agreement with the expected contrast decrease. The interrogation time of the CAI is $T=46$ ms.}
  \label{fig:Contraste_Table}
\end{figure}

In the next paragraph, we present a way to compensate the table rotation by actuating the PM of the EA.
\paragraph{Rotation compensation}
This section is devoted to present a very preliminary demonstration of rotation compensation with the hybrid prototype. Note that the compensation of rotations with constant angular velocities has already been demonstrated with  a reference mirror driven by PZT actuators \cite{lan_influence_2012, hauth_first_2013, dickerson_multiaxis_2013, duan_suppression_2020, zhao_extension_2021}. Here, in our experiment, we consider additionally the impact of angular accelerations. The platform has been excited around the $Z$ axis with a sine function at a frequency of $\approx$ 5.435 Hz corresponding to $1/(4T)$ where $T=46$ ms. The CAI is therefore submitted to all types of acceleration coming from the effect of rotation, i.e. centrifugal, Coriolis and angular accelerations. These accelerations induce a loss of contrast of the atomic signal as it has been seen in the previous paragraphs. The gyroscope signal is integrated to drive the angular control of the proof-mass. After integration, a gain $G_{comp}$ is applied to adapt the amplitude of the feedback that is fed to the secondary entry of the EA controlling the $\psi$ angle of the proof-mass.\\

First results on active rotation compensation have been obtained for an excitation of the platform at 5.435 Hz, with an amplitude of angular velocities of 5 mrad/s p-p and angular accelerations of 170 mrad/s$^{2}$ p-p.
The effect of the implemented compensation scheme on the atomic contrast is visible on Figure~\ref{fig:Rotation-Comp}-a). The variation of the atomic contrast is analyzed relatively to the gain amplitude of the compensation signal. The offset signal of the CAI contrast has been subtracted and then normalized by the atomic contrast value obtained with no rotation. We can see that if no compensation signal is applied (gain = 0), the CAI contrast drops by $\approx 93\%$. The CAI contrast does not fall completely due to the limited excitation amplitude of the platform. For higher excitation amplitudes, perturbations of the EA's proof-mass along the horizontal plan are too strong to allow precise control of the PM.
Increasing the gain on the compensation signal allows to recover the CAI contrast. An optimum gain of $\approx$ -2.6 (x2000) enables up to 85 $\%$ of the CAI contrast to be recovered. Increasing the gain even further leads to an over-compensated regime where the contrast falls again. For gains higher than $\approx$ -5.2 (x 2000), the contrast drops even more than the initial value with no compensation. The higher excitation amplitudes offered by PM control allows this time to decrease the contrast to zero for gains higher than $\approx$ -7 (x 2000).
\end{paracol}
\begin{figure*}
\centering
  \includegraphics[scale=0.35]{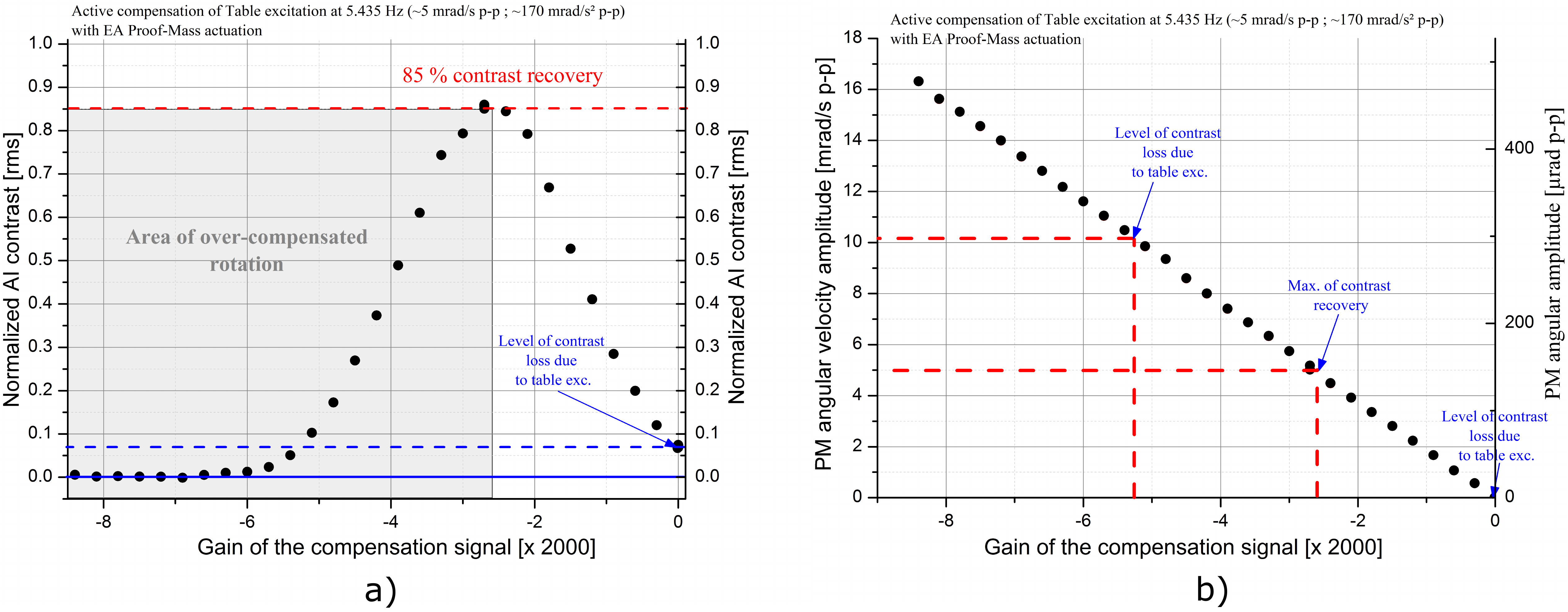}
  \caption{a) Active rotation compensation impact on the CAI contrast for a platform excitation at 5.435 Hz. The variation of the CAI contrast is reported according to the gain amplitude of the compensation signal. b) PM angular velocity amplitude (left axis) and angular amplitude (right axis) according to the compensation gain. For 0 gain, the proof-mass is not moved and the CAI contrast loss is only due to platform rotation. For a gain of $\approx$ -2.6 (x2000), the CAI contrast is retrieved at a maximum of $\approx$ 85 $\%$. For a gain of $\approx$ -5.3(x2000)  the atomic contrast returns to a value similar to that obtained with only the platform excited and the PM motionless.}
 \label{fig:Rotation-Comp}
\end{figure*}
\begin{paracol}{2}
\switchcolumn
Looking to the PM excitation amplitude (cf. Figure~\ref{fig:Rotation-Comp}-b), we can see that we are able to drive rotation amplitudes up to $\approx$ 16 mrad/s p-p, much higher than the platform excitation amplitudes, which explains that the CAI contrast drops efficiently to zero. As expected, the optimal contrast recovery is obtained for the same level of angular velocity that the platform is submitted to, i.e. $\approx$ 5 mrad/s p-p. In this configuration, the PM is motionless in the inertial frame. Increasing the PM amplitude to twice this value, i.e. $\approx$ 10 mrad/s p-p, makes the atomic contrast returning to the value obtained when only the platform is excited which seems logical since the PM is now rotating with an angular velocity of $\approx$ 5 mrad/s p-p in the inertial frame.\\
The non-perfect contrast recovery has not been yet analyzed in details but the main suspicions concern the non-perfect alignment between the rotation axis of the table ($Z$ axis) and the $\Psi$ rotation axis of the PM or generation of parasitic rotation excitations along the $Y$ axis, the second horizontal axis, corresponding to $\Theta$ rotation for the PM. The parasitic rotation excitation has been measured with the 2-axis gyroscope and we evaluated at a level of 25$\%$ of the total excitation the amplitude of the rotation around the $Y$ axis, which has a non-negligible effect on the atomic contrast for excitation amplitudes used in our experiment.\\
Note that it would be possible in principle to use also the angular acceleration output of the EA to benefit from more high frequency information concerning the angular velocity variations at the location of the PM. A compensation scheme merging measurements coming from an external gyroscope and from the angular accelerations measured by the EA seems interesting and should be studied, especially in this specific case of a hybrid instrument.

\subsection{Hybrid instrument simulator}
\subsubsection{EA noise}
The following figure shows the simulated noise of HybridSTAR, the designed EA (see section \ref{Design_EA}) for hybridization, in case of a non-rotating proof-mass:
\begin{figure}[H]
  \centering
  \includegraphics[scale=0.5]{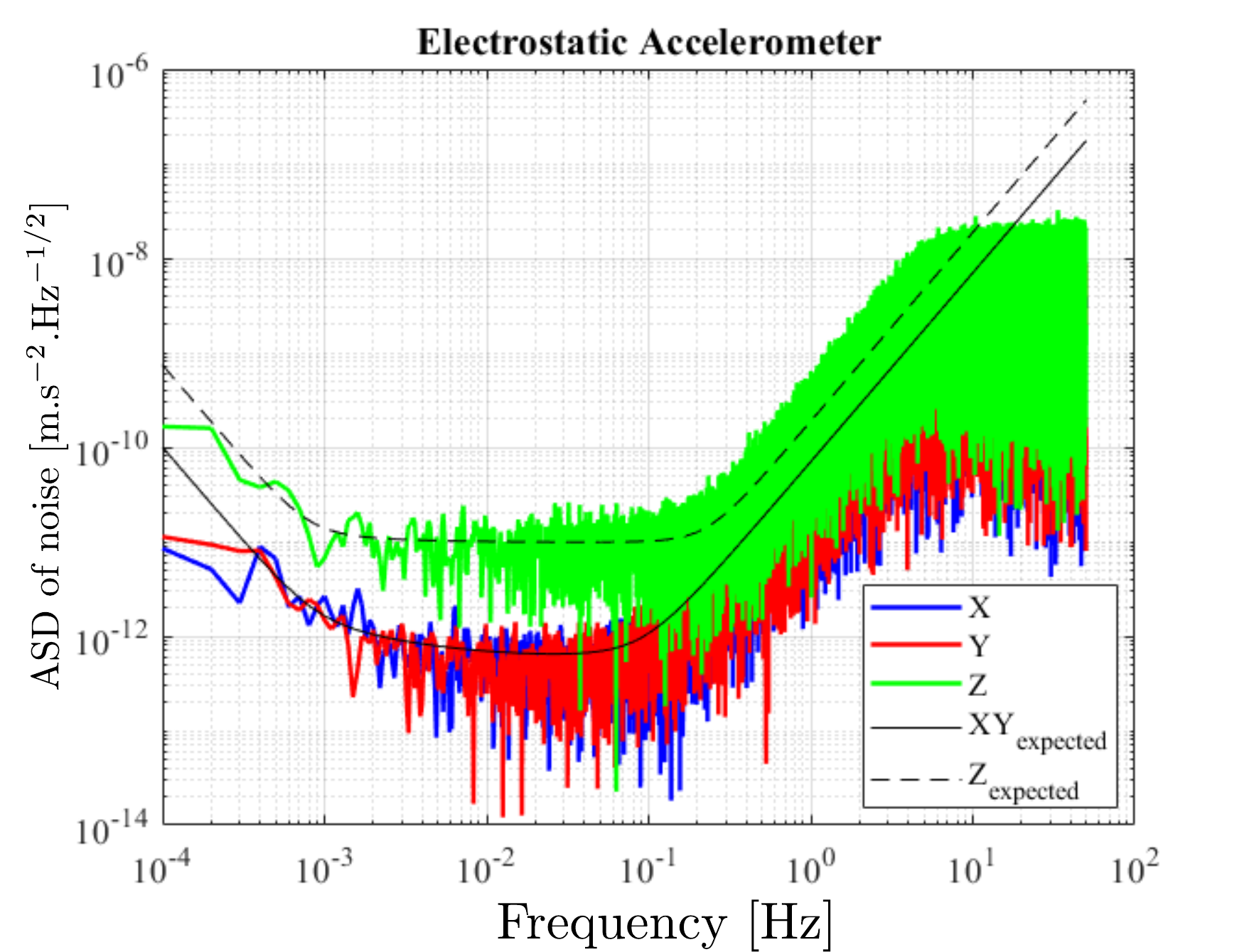}
  \caption{ASD of noise for the EA HybridSTAR given for the three axis in case of a non-rotating proof-mass.}
  \label{fig:Simulator_EA_noise}
\end{figure}
As we can see on Figure~\ref{fig:Simulator_EA_noise}, the Z axis has a higher noise because of the dedicated electrodes design of HybridSTAR that allows to manage the rotation of the proof-mass around the cross-track axis (cf. section \ref{Design_EA}).

\subsubsection{Hybridization signals on ground}
In order to validate the hybridization algorithm implemented in the simulator, the results obtained in term of sensitivity with the on-ground lab-prototype (see Figure~\ref{fig:Exp_Hyb-signals}) are reproduced with the simulation and reported on Figure~\ref{fig:Simu_Hybridization}. In this on-ground experiment, for each measurement cycle, the electrostatic accelerometer is used to determine the interferometric fringe index, and the electrostatic accelerometer bias is evaluated thanks to the atomic interferometer. The results show clearly on the hybrid instrument signal (red curve) the correction of the electrostatic accelerometer drift (black curve) using the long-term stability of the atomic interferometer (blue curve), while the electrostatic accelerometer, with a sampling rate of 1 kHz, allows the exploration of the high frequency region inaccessible to the atomic instrument because of its cycling time.
\begin{figure}[H]
  \centering
  \includegraphics[scale=0.7]{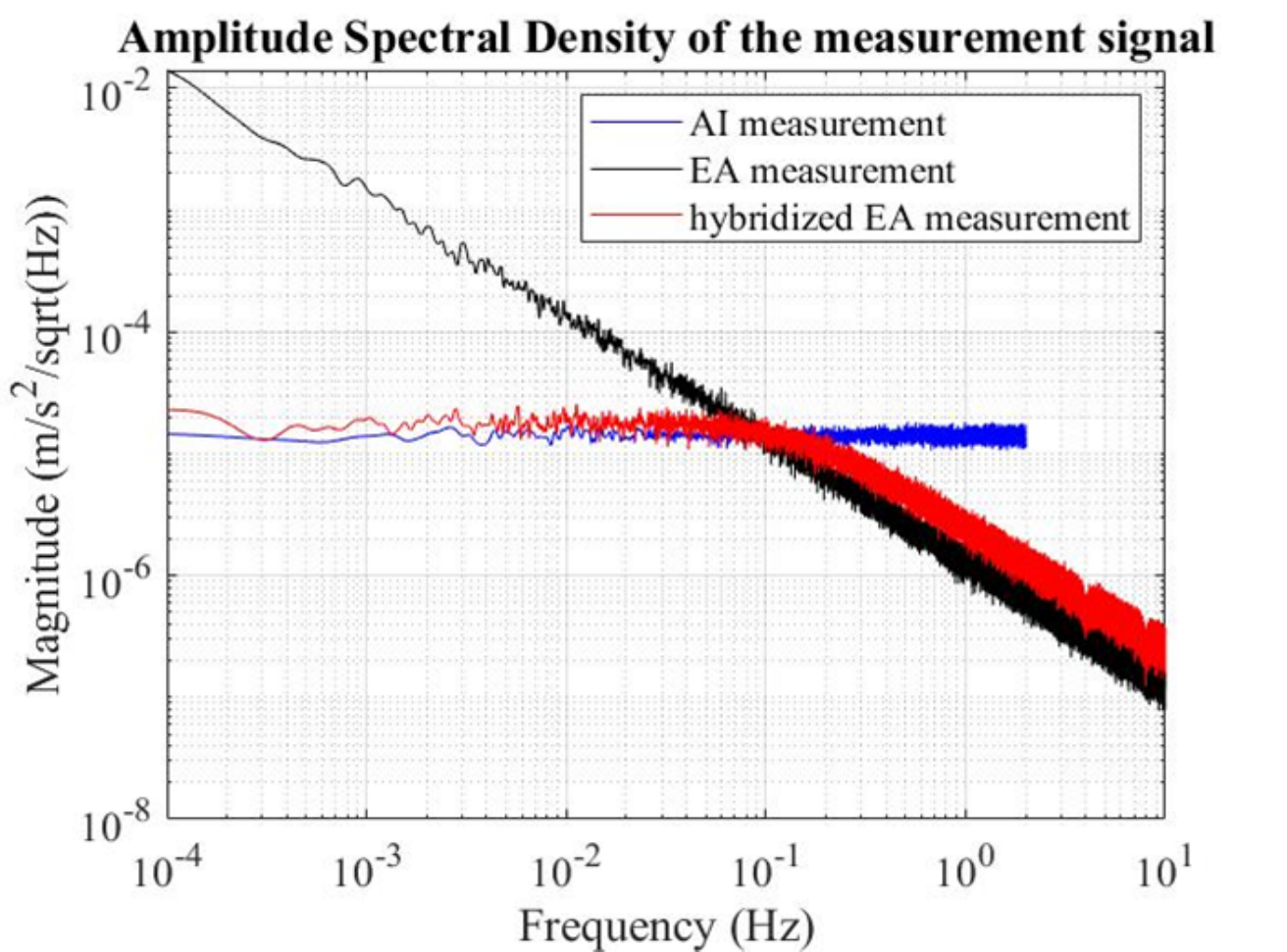}
  \caption{ASD of the acceleration noise, for the electrostatic accelerometer in black, the atom interferometer in blue, and the resulting combination in red. These data are provided by the developed hybrid instrument simulator.}
  \label{fig:Simu_Hybridization}
\end{figure}
\subsubsection{Satellite rotation compensation}
Due to satellite rotation around the Earth, the atoms are submitted to centrifugal and Coriolis accelerations. Because of the initial position and velocity dispersion of the atomic cloud, each atom should provide different outputs in term of acceleration. The resulting interferometric phase difference for each atom induces a contrast loss in the interferometer fringes. This effect is taken into account in the simulator (see Figure~\ref{fig:Simulator_Contrast}) through the random inertial position and velocity affected to the atoms of the cloud. We have considered for the simulation a standard deviation for the position and velocity distribution of respectively 2 mm and 2.5 mm/s. This velocity dispersion corresponds to an effective atomic temperature of 70 nK coming from the velocity selectivity of the foreseen detection scheme (see section \ref{CAI_design}).
\begin{figure}[H]
  \centering
  \includegraphics[scale=0.6]{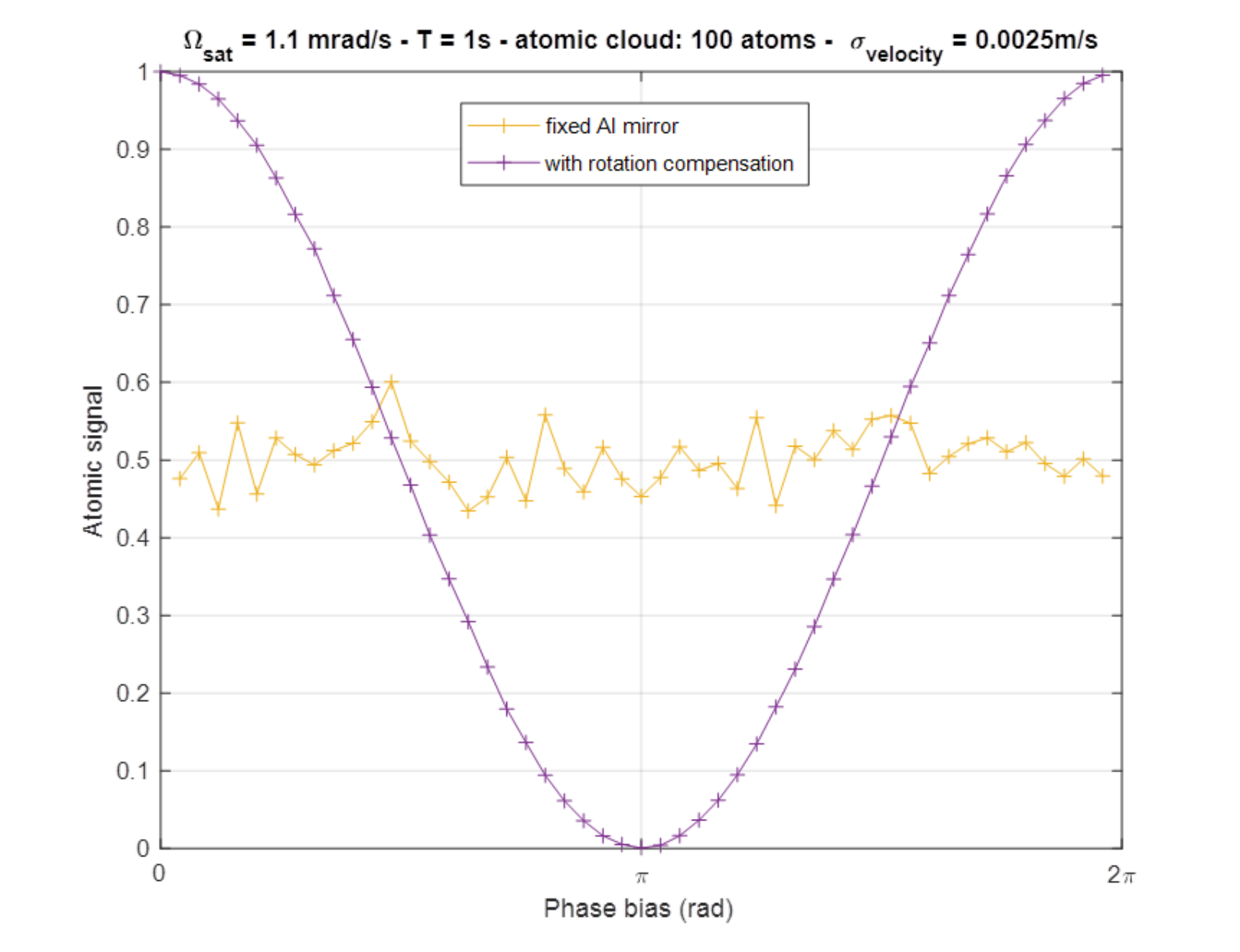}
  \caption{Interferometric signal when using a fixed CAI reference mirror (yellow) and when using the electrostatic proof-mass, acting as the reference mirror, to compensate for the satellite rotation (purple). The simulation is done with a simulated atomic cloud of 100 atoms, a standard deviation of the atomic velocities of 2.5 mm/s, of the initial atomic positions of 2 mm and an interrogation time of $T = 1$ s.}
  \label{fig:Simulator_Contrast}
\end{figure}
With an orbital angular velocity of $\approx$ 1.1 mrad/s and an interrogation time of $T=1 ~\textnormal{s}$, the atomic signal disappears almost completely.\\
It has been proposed to overcome this limitation in CAIs by rotating the CAI reference mirror that acts as the acceleration measurement reference \citep{lan_influence_2012,dickerson_multiaxis_2013} and compensating the detrimental effect of rotation. In our specific hybrid design, the EA's proof-mass is used as the reference mirror for the CAI. Because the atomic interferometer measurement is not affected by the motion of the mirror between the laser pulses whose durations are considered negligible, we choose to move the proof-mass angular position following a sine motion rather than with a constant angular velocity in order to have a smoother control (see Figure~\ref{fig:PM_Comp}). If a saw-tooth signal had been used, the control of the PM would have been challenging in order to be fast enough to allow the repetition of the command for every interferometric cycle while allowing such a large motion range. A PID-type controller is used to control the electrostatic proof-mass. With a large bandwidth of 27 Hz and a high control gain, the PM is able to accurately follow the compensation command, on the condition of correcting the command’s amplitude and time delay due to the control loop. A simulation was run with the computation of the acceleration with 100 atoms, with the electrostatic proof-mass acting as the CAI reference mirror while compensating the satellite rotation. As shown in Figure~\ref{fig:Simulator_Contrast} (purple curve), the contrast of the interferometric fringes, which was completely lost when using a fixed Raman mirror, is fully recovered. For an efficient correction, the real orbital rotation shall be known with an accuracy better than 1$\%$.
\begin{figure}[H]
  \centering
  \includegraphics[scale=0.55]{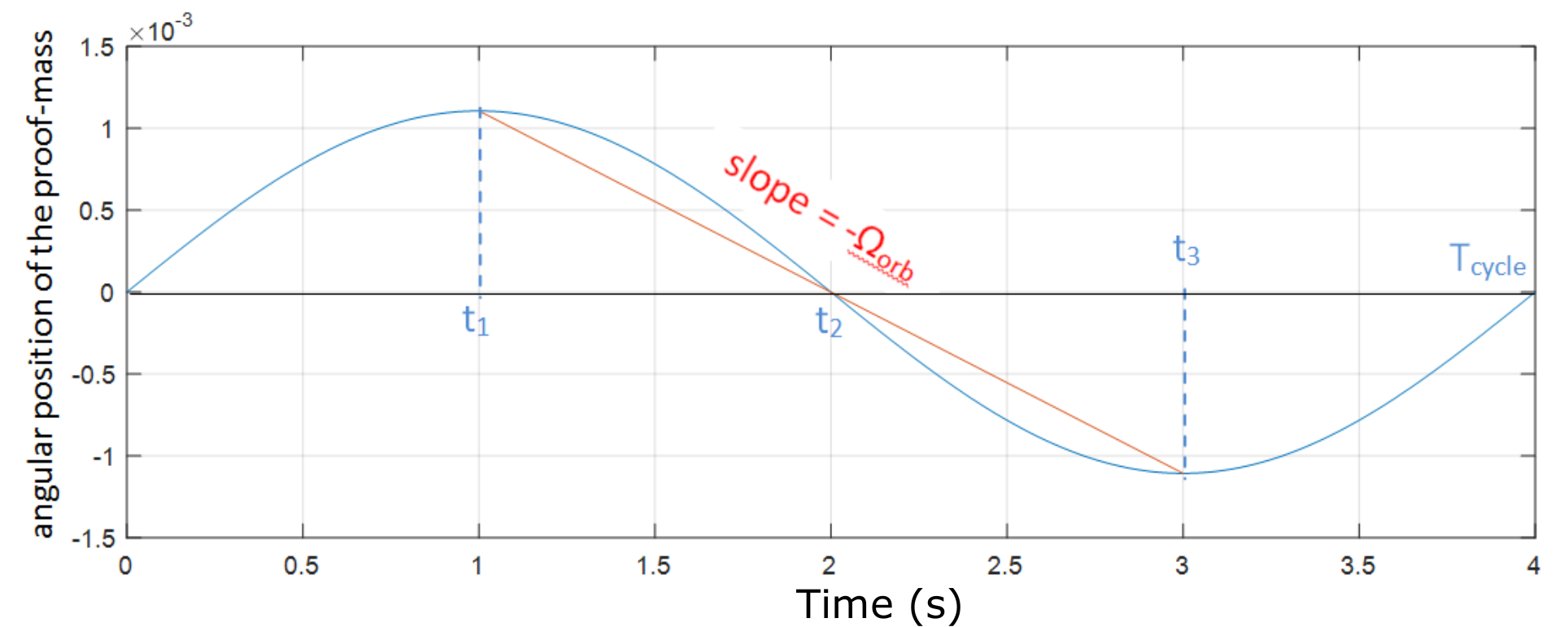}
  \caption{Control of the electrostatic proof-mass angular position during an interferometric cycle. The angular velocity to be compensated between $t_{1}$ (first CAI laser pulse) and $t_{3}$ (last CAI laser pulse) is represented in red.}
  \label{fig:PM_Comp}
\end{figure}
As seen previously (cf. Figure \ref{fig:Simulator_EA_noise}), the ability to rotate the EA proof-mass induces an increase of EA noise on the radial axis due to the common electrodes for the control of translation along Z and the rotation around Y. To decrease this impact, a new design of the electrodes with a separation between linear and angular control could be studied in the future.
For the X axis, along-track, the main impact of the rotation is due to the fact that the measurement is done on the proof-mass reference frame, which is in rotation with respect to the satellite frame. The electrostatic forces exerted by the electrodes on the along-track axis are perpendicular to the proof-mass, as they follow the electric field lines. In consequence, the direction of the electrostatic forces will depend on the rotation of the proof-mass around the cross-track axis. Figure~\ref{fig:Simu_EA-Rot_DSP} shows the noise on each axis with rotation of the proof-mass. Peaks at cycle period appear. The peaks at multiple of the frequency of PM rotation are due to the non-linearities of the electrostatic forces which are proportional for each electrode to the inverse of the square of the distance between the electrode and the proof-mass. Along radial axis Z, the peaks are more visible as there is also the effect of the variable potential applied on the electrodes to rotate the proof-mass.\\
It is necessary to take into account this orientation in order to deduce the acceleration on the along-track axis and on the radial axis. It could be done by using the angular position of the proof-mass given by the capacitive detector. The error in the estimation of the acceleration on the along-track axis due to the rotation of the proof-mass depends on the projection of bias and noise of radial axis as well as the error in the rotation reconstruction (due to the detector error). The rotation noise is directly related to the detector noise level, which is about 10 nrad.Hz$^{-1/2}$. The detector bias is about 10 $\mu$rad.
\end{paracol}
\begin{figure*}
\centering
  \includegraphics[scale=0.4]{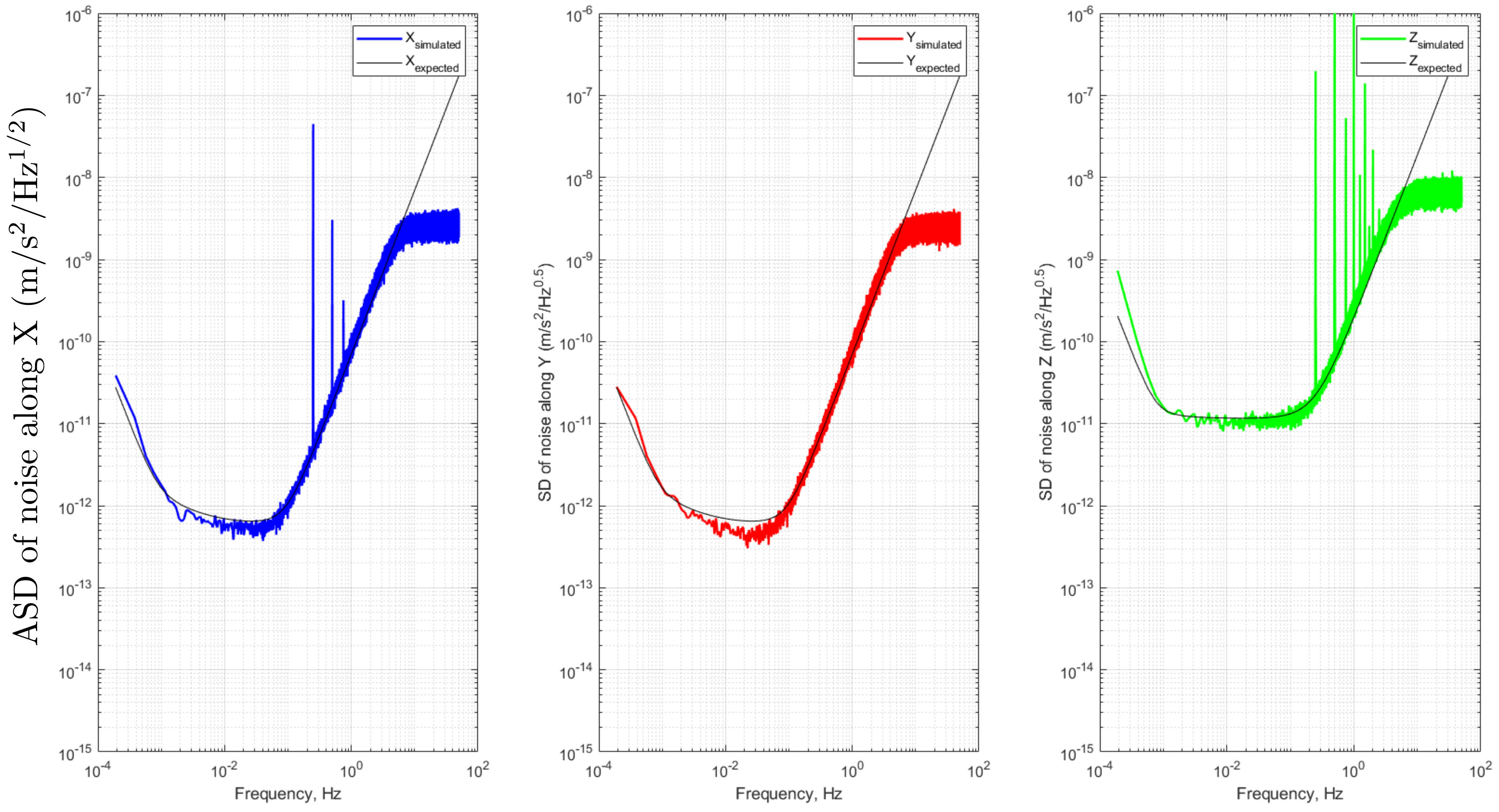}
  \caption{HybridSTAR accelerometer noise performance obtained with the hybrid instrument simulator. The EA's proof-mass is rotated around the $Y$ axis for satellite rotation compensation.}
  \label{fig:Simu_EA-Rot_DSP}
\end{figure*}
\begin{paracol}{2}
\switchcolumn

\subsection{Hybrid instrument design}
In this purpose of investigating this concept of hybrid accelerometer for future space missions, we establish here a preliminary global design of instrument with the related estimation of Size, Weight and electrical Power consumption (SWaP) and give also some performance evaluations. 
\subsubsection{Atom interferometer}
\paragraph{Global architecture}
\label{CAI_design}
For the atomic accelerometer, we considered primarily an atomic source constituted by a molasses source of $^{87}$Rb that benefits from a more mature technology development and that should lead to a more favorable SWaP for a satellite-based mission than an accelerometer based on a BEC source \cite{debs_cold-atom_2011, abend_atom-chip_2016} or evaporatively cooled source \cite{karcher_improving_2018, asenbaum_atom-interferometric_2020}. Nevertheless, the performance in term of accelerometry with a BEC are expected to be better \cite{hensel_inertial_2021}. Indeed, a BEC source benefits from a lower velocity dispersion and a better control/stability of the position and initial velocity of the atomic cloud. A parallel with a BEC-based instrument is given at the end of the design presentation.\\
For the CAI, we consider the standard configuration of a three light pulses Mach-Zehnder interferometer and a Raman double diffraction scheme \citep{leveque_enhancing_2009}, well adapted for microgravity environments, with a laser retro-reflected on the EA's proof-mass.\\
The vaccum chamber (see Figure~\ref{fig:Hybrid_Global-Design}) is based on a glass cell setup connected to a titanium base where the vacuum is maintained thanks to non-evaporable getters and an ionic pump. Two rubidium dispensers are also integrated in the titanium base that allow the atomic cloud to be loaded from a background vapor.\\
The detection of the atomic population is realized thanks to four photodiodes surrounding the atomic cloud and a laser beam selecting spatially the atoms in a region of typically 1 cm x 1 cm x 1 cm, providing a effective velocity selection of the atoms and decreasing artificially the temperature of the detected atoms to $\approx 100$ nK for typical foreseen interrogation times. Two layers of mu-metal magnetic shield are foreseen to lower the impact of external magnetic fields, leading to a theoretical shielding factor of $\approx$ 2000 in the interferometer region.\\
The measurement cycle begins with the cooling and trapping of the atomic cloud in a 3D-MOT (Magneto-Optical Trap). The atoms are loaded for 1 s. A molasses phase of typically 20 ms follows which decreases further the temperature of the atoms down to $\approx 1 \mu$K. A preparation phase aiming to put the atoms in a specific state follows. This phase is a combination of microwave transitions and/or Raman velocity selective transitions and blast pulse to populate a magnetic insensitive atomic state. $2.5 \times 10^{8}$ atoms are expected to participate to the interferometer. The interferometer phase begins with a Raman double diffraction pulse followed by a blast pulse that removes unwanted populated states that would participate to a contrast decrease at the interferometer output. A second Raman pulse occurs followed also by a blast pulse. The interferometer sequence finishes by a third Raman pulse that recombines the two interferometer paths. The population in the two hyperfine atomic states, that reflects the interferometer phase, is measured with an independent detection beam that selects, in conjunction with four photodiodes, the center of the cloud and therefore the colder atoms. The number of detected atoms should be of $\approx 3 \times 10^{5}$ for a typical interrogation time $T=1$ s (see next section \ref{CAI_Perf}).
\paragraph{Performance aspects}
\label{CAI_Perf}
In the following, we give a rough estimation of some important effects that could limit the performance of the cold atom interferometer in space. This constitutes a non-exhaustive analysis with estimations given for a CAI based on a molasses source.
\begin{itemize}
\item	Fundamental limit:\\
The fundamental limit in term of sensitivity is given by the so-called Quantum Projection Noise (QPN) \cite{itano_quantum_1993, gauguet_characterization_2009}. This noise per shot at the output of a double diffraction atomic accelerometer is given by:
\begin{eqnarray}
\sigma^{QPN}_{a}=\frac{1}{2 k_{eff}T^{2}}\frac{1}{C\sqrt{N_{det}}}
\end{eqnarray}
\indent where $N_{det}$ is the number of detected atoms contributing to the interferometer signal. This formula assumes to work at mid-fringe. $C$ refers to the interferometer contrast. $k_{eff}\approx 4\pi/\lambda_{Rb} \approx 1.6 \times 10^{7}$ where $\lambda_{Rb}$ is the laser wavelength associated to the rubidium atomic transition.\\
The QPN sensitivity of the CAI, relatively to the interrogation time $T$, is reported on Figure~\ref{fig:QPN_sens}-a). The contrast C is calculated taking into account an imperfect contrast without rotation of 0.5 coming for instance from imperfect Raman beamsplitters and mirrors.
\begin{figure}[H]
  \centering
  \includegraphics[scale=0.32]{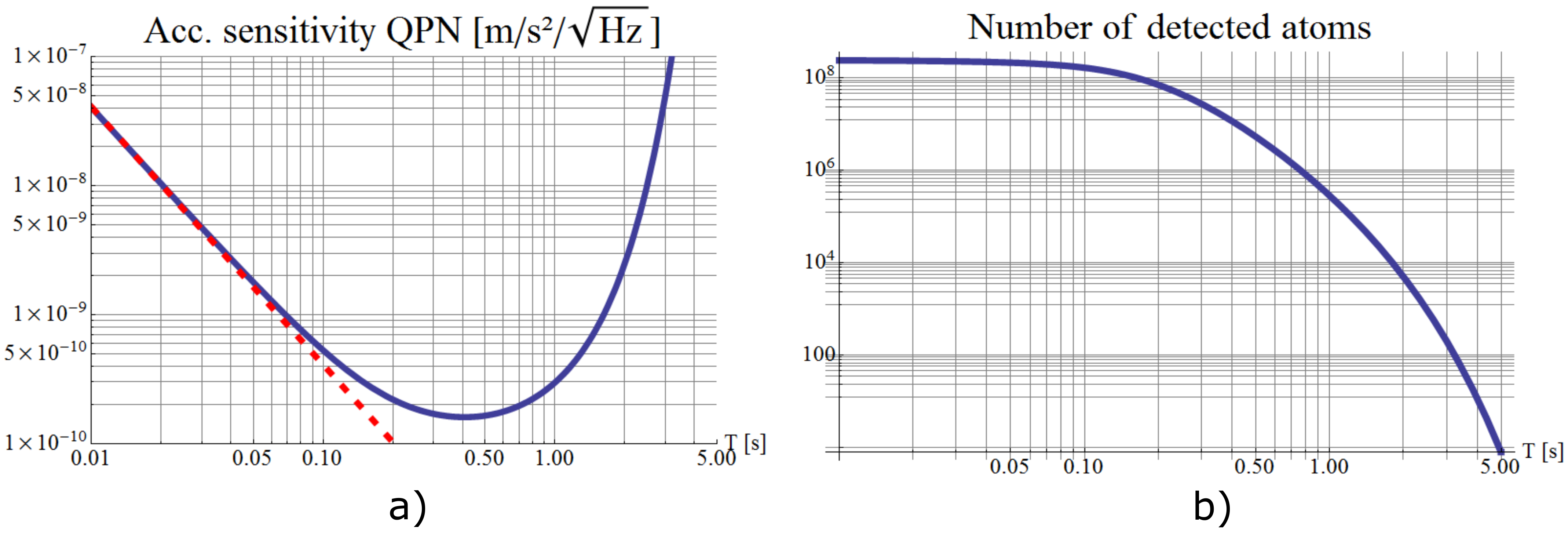}
  \caption{a) Dependence of the acceleration sensitivity according to the interrogation time $T$. The dashed red line shows the sensitivity decrease, scaling as $1/T^{2}$. b) Dependence of the number of detected atoms according to the interrogation time $T$.}
  \label{fig:QPN_sens}
\end{figure}
We can see that increasing the interrogation time does not improve continuously the sensitivity which is a consequence of atoms losses due to collisions during free-fall with residual background gas, of contrast loss due to satellite rotation, and finally, of the spatial expansion of the atomic cloud relative to the finite size of the detection area that conducts to a loss of detected atoms and an increase of the quantum projection noise. This effect is illustrated on Figure~\ref{fig:QPN_sens}-b)  that represents the evolution of the number of detected atoms according to the interrogation time. The input parameters lead to a number of 2.5 x 10$^{8}$ atoms entering the interferometer. We can see, as explained, that the number of detected atoms decreases quickly with the interrogation time reaching for $T = 0.4$ s approximately 10$^{7}$ detected atoms. The assumption of an accelerometer limited by quantum projection noise should not stand for a number of detected atoms lower than 10$^{3}$-10$^{4}$, where other types of noise should dominate for the detection.
So, typically in our case, considering $3\times10^{5}$ detected atoms, a contrast $C=0.34$ (beamsplitter efficiency + compensated satellite rotation), an interrogation time $T=1$ s and a cycling time $T_{c}=3$ s, we find a quantum projection acceleration noise of $\sigma^{QPN}_{a} = 3\times10^{-10}$ m.s$^{-2}$.Hz$^{-1/2}$.
\item	Detection noise:\\
Considering a molasses atomic source allowing $3\times10^{5}$ detected atoms, a contrast $C=0.34$ (beamsplitter efficiency + compensated satellite rotation), with an interrogation time $T=1$ s and a cycling time $T_{c}=3$ s, a near state-of-the-art detection noise of $\sigma_{P}=10^{-3}$, and an interferometer operating at mid-fringe, we find an acceleration noise contribution of the detection of $3\times10^{-10}$ m.s$^{-2}$.Hz$^{-1/2}$, similar to the quantum projection noise, constituting the fundamental limit of sensitivity performance.
\item	CAI contrast:\\
Any imperfections of the beamsplitters or mirrors forming the interferometer reduce the number of atoms involved in the interferometer and lead to parasitic paths that could impact the contrast of the interferometer fringes. The finite position dispersion of the atomic cloud in conjunction with the inhomogeneous intensity profile of the Raman laser reduce also the global probability transition and induce a loss of contrast. For this contrast estimation, we consider a finite velocity distribution of the atoms with an effective atomic temperature of 0.1 $\mu$K thanks to velocity selection and double diffraction transition with an effective Rabi frequency of $2\pi \times 100$ kHz, leading to an estimation of $C\approx0.5$.\\
Satellite rotation is also responsible of contrast loss through centrifugal acceleration, Coriolis acceleration or angular acceleration. To estimate the impact of satellite rotation on the CAI contrast (see Equation \ref{Contrast_Decay_Formula}), we assume an angular acceleration $\dot{\Omega}_{sat}=2\times10^{-6}$ rad.s$^{-2}$, corresponding to the maximum angular acceleration measured in GOCE, an angular velocity $\Omega_{sat}=1.16$ mrad.s$^{-1}$, an atomic velocity dispersion of $\approx$ 1 cm/s ($\approx$ 1 $\mu$K) and an initial atomic position dispersion of 2 mm. Figure~\ref{fig:Contrast_loss_MOT} illustrates that the dominant contribution is by far the Coriolis acceleration that limits drastically the possibility to increase the interrogation time $T$. The two other terms are nevertheless non-negligible for $T$ approaching 1 s.
\begin{figure}[H]
  \centering
  \includegraphics[scale=0.8]{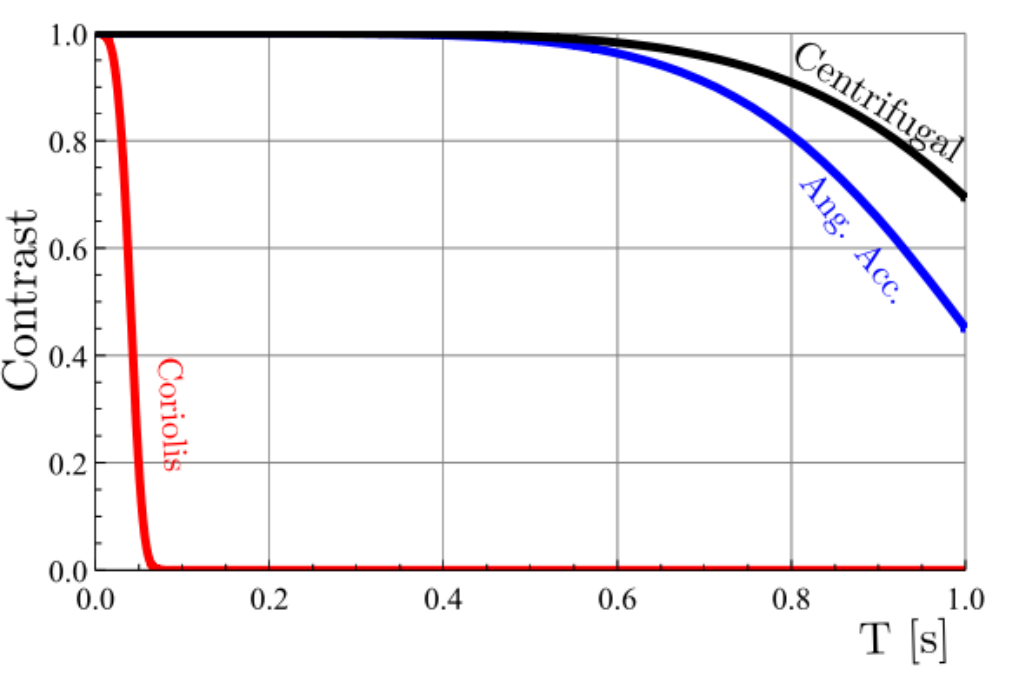}
  \caption{Contrast loss due to the different rotation terms for a satellite in orbit.}
  \label{fig:Contrast_loss_MOT}
\end{figure}
To allow measurements with the CAI, it appears fundamental to reduce the apparent satellite rotation. The contrast drop is at a level of $\approx$ 40$\%$ by reducing the effective angular velocity down to a level of 5 $\mu$rad/s and at the same time by reducing the effective atomic temperature to 0.1 $\mu$K, which seems possible with velocity selection during the detection phase and/or with a preliminary Raman velocity selective pulse before the interferometer phase. Without atomic velocity selection, the residual angular velocity should be reduced down to a level of 1 $\mu$rad/s.
\item	Scale factor:\\
Considering CAI parameters given in Table \ref{CAI_Param_SF}, the relative scale factor uncertainty has been estimated at a level of $4\times 10^{-7}$. Considering linear drag accelerations with amplitudes lower than $10^{-6}$ m.s$^{-2}$ and a noise in the measurement bandwidth (1-100 mHz) lower than $5\times10^{-9}$ m.s$^{-2}$ \citep{massotti_next_2021}, the scale factor uncertainty should impact the acceleration measurement at a level of $4\times10^{-13}$ m.s$^{-2}$ with a noise impact of $2\times10^{-15}$ m.s$^{-2}$. Note that in some specific mode of operation where the CAI is forced to operated at mid-fringe, the scale factor uncertainty could be even further reduced to $<10^{-9}$.
\begin{specialtable}[H]
\small
\caption {CAI parameters for scale factor uncertainty estimations. $\sigma_{i}$ refers to the uncertainty on the parameter $i$.}
\centering
\label{CAI_Param_SF}
\begin{tabular}{c|c|c|c|c}
\hline\hline
Parameters & $k_{eff}\approx1.6 \times 10^{7}$ m$^{-1}$ & $\tau_{R}$ = 50 $\mu$s & $T=1$ s & $\Omega_{eff}=\pi/(2\tau_{R})$ \\
\hline
Uncertainty & $\sigma_{k_{eff}}/k_{eff}=10^{-9}$ & $\sigma_{\tau_{R}} = 10$ ns & $\sigma_{T}=10$ ns & $\sigma_{\Omega_{eff}}/\Omega_{eff}=10^{-2}$ \\
\hline\hline
\end{tabular}
\end{specialtable}
\item Magnetic fields:\\
Even if the atoms are in one of the $m_{F}=0$ first order insensitive magnetic state, they are still affected by the second order Zeeman effect \cite{peters_high-precision_2001}. Considering a uniformity of the magnetic field on the order of $0.4\%$ over $\approx$ 1 cm and assuming a bias magnetic field during the interferometer phase of 10 mG, we can anticipate a parasitic gradient magnetic field of typically $\approx$ 40 $\mu$G/cm. For an interrogation time of $T = 1$ s, this gradient would lead to an acceleration error of $\approx 10^{-10}$ m.s$^{-2}$.
The rotating Earth magnetic field on the order of $\approx 0.4$ G, that creates magnetic field rate variations up to 0.5 mG/s, should not have a significant impact on the measurement error thanks to the double diffraction scheme but should be kept significantly lower than the bias magnetic filed to keep the atom spins aligned along the quantization axis during the interferometric phase. The foreseen double layer of magnetic shield should allow this objective.
\item Light shifts:\\
The effect of one- and two-photon light shifts \citep{peters_high-precision_2001,gauguet_off-resonant_2008} occurs during each Raman light pulse and could be responsible of an energetic displacement of the two involved hyperfine ground states that has an impact on the phase at the output of a single diffraction Mach-Zehnder interferometer. Here a double diffraction scheme should be implemented and in principle no impact of light shifts is expected, at least for differential AC Stark shifts. The average AC Stark shift should be the remaining contribution if we consider the spatial arm separation of the interferometer that should be around 2 cm for an interrogation time of 1 s. Considering a Raman beam waist of 1.3 cm and a half angular divergence of 0.1 mrad \citep{weiss_precision_1994}, the relative intensity difference between the extreme points on the interferometer should be of $\approx 2\times10^{-4}$ and the impact on the acceleration systematics of few $10^{-11}$ m.s$^{-2}$. Note that this estimation does not take into account steeper intensity gradients such as laser speckle.
\item Misalignements:\\
The measurement axis of the atomic instrument is given by the direction of $\vec{k}_{eff}$. Any misalignments between this direction and the satellite measurement axis should lead to errors coming from projection of accelerations. Let’s consider first that the Raman mirror is fixed relatively to the satellite. In the atomic instrument, the Raman laser is retro-reflected on the mirror and therefore the effective wavevector $\vec{k}_{eff}$ is always orthogonal to the mirror plane. A defect in the retro-reflection will only impact at first approximation the scale factor through a modification of $\left\|\vec{k}_{eff}\right\|$. A misalignment of the mirror will induce therefore a misalignment of the measurement axis by the same angle. The Raman mirror will constitute the reference for the atomic measurement and therefore has to be precisely integrated in the satellite to avoid important misalignment errors. Misalignment errors of the mirror referential relatively to the satellite referential should be of the same order of magnitude than the one foreseen for the electrostatic accelerometer in NGGM, typically of $\approx 100$  $\mu$rad. Angular errors in the mirror mechanical structure itself should be significantly lower, estimated at first sight of  $\approx$ 10 $\mu$rad. The main contribution comes from the projection of transverse drag accelerations. We consider for all axes residual linear drag accelerations with amplitudes lower than $10^{-6}$ m.s$^{-2}$ and a noise in the measurement bandwidth (1-100 mHz) lower than $5\times10^{-9}$ m.s$^{-2}$.Hz$^{-1/2}$ \citep{massotti_next_2021}. This leads to acceleration errors with a maximum amplitude of $10^{-10}$ m.s$^{-2}$ and a noise of $5\times10^{-13}$ m.s$^{-2}$.Hz$^{-1/2}$ in the measurement bandwidth.
\item Raman wavefront aberrations:\\
Wavefront aberrations of the Raman laser are identified as a main limitation in cold atom inertial sensors \citep{wicht_phase_2005,louchet-chauvet_influence_2011,schkolnik_effect_2015,zhou_observing_2016,trimeche_active_2017,karcher_improving_2018}. This is due to the fact that the measurement is derived from the displacement of the free-falling atoms relatively the Raman laser equiphase surfaces. Any distortions of the equiphase surfaces introduce therefore parasitic phase shifts at the interferometer output that translate into acceleration biases and long term drifts due to atom kinematics variations (initial atomic position/velocity and temperature). For an atom whose trajectory is not perfectly aligned with the Raman laser, the phases of the three pulses are different, leading to a systematic shift.
The resulting phase error at the output of the interferometer depends therefore on the initial density distribution of the atomic cloud, its expansion during the free-fall and the averaging over the final spatial distribution during the detection process.
State-of-the-art determination on the impact of wavefront aberrations on acceleration measurements has been achieved using an evaporatively-cooled atomic source, with temperature down to 50 nK, reaching an uncertainty on the effect of $1.3\times10^{-8}$ m.s$^{-2}$\citep{karcher_improving_2018}.\\
Since the error phase shifts due to wavefront aberrations result from the convolution between the distribution of atomic trajectories and the Raman beam wavefronts, and consequently, depends on parameters such as atomic temperature, atoms initial position/velocity, and Raman laser beam shape, it is important to consider the related properties of the atomic source. If we simplify the problem and assume only a quadratic dependency of the wavefronts on the transverse position of the atoms, with an atomic cloud centered on the mirror and with no mean velocity, we can give here a rough estimation of acceleration stability. Considering a high quality mirror of flatness ($\lambda/300$) over a diameter of 20 mm, corresponding to a radius of curvature of $\approx$ 19 km, and a temperature variation of 100 nK, this leads to acceleration variations of $5\times10^{-10}$ m.s$^{-2}$. Note that this model constitutes a huge simplification of the problematic of wavefront aberration and that it does not reflect totally the achievable performances. It has to be taken as a best case scenario. To evaluate more rigorously the stability of the acceleration measurement, we have to take into account the position and velocity stability of the atomic source along with the temperature stability. This leads to a complex analysis that should be treated with dedicated simulations \citep{trimeche_concept_2019}.
\item Coupling with satellite rotation:\\
All the acceleration terms coming from the satellite rotation, fundamentally linked to the centrifugal acceleration, to Coriolis acceleration and to angular acceleration could contribute to the measurement error of the atomic accelerometer. We give in Table \ref{Rot_Grad_err} an estimation of the dominant errors terms. We consider here that the CAI, and more particularly the atomic cloud, is located by construction at the position of the satellite centre-of-mass. Table \ref{Molasse_param} lists the assumptions concerning the atomic source allowing the numerical evaluation of acceleration errors gathered in Table \ref{Rot_Grad_err}. We can see that the Coriolis term, involving the transversal atomic velocity, has a largely dominant contribution to the instrument stability. Assuming a compensated satellite rotation around both the Y and Z axis at a level of 100 nrad/s, we can see that we reach a level of $10^{-10}$ m.s$^{-2}$ of acceleration stability. The error is dominated by the Coriolis acceleration and the angular acceleration terms.
\begin{specialtable}[H]
\small
\caption {Estimation of main errors coming from the coupling of the atomic source kinematics with the satellite rotation and the Earth gravity gradient. The estimations with rotation compensation are given considering a satellite rotation compensation with a residual at a level of 100 nrad/s.}
\centering
\label{Rot_Grad_err}
\begin{tabular}{c||c||c|c}
               & Acc. stab.                & Acc. stab.      & Acc.  Stab \\
Inertial terms & without rot. comp.        & with rot. comp. & with rot. comp. \\
               &                           & around $Y$      & around $Y$ and $Z$\\ 
							 & $\left[\pm \textnormal{m.s}^{-2}\right]$  & $\left[\pm \textnormal{m.s}^{-2}\right]$ & $\left[\pm \textnormal{m.s}^{-2}\right]$\\
\hline
$2 \cdot \Omega_{y} \cdot v_{z}$ & $2.3\times10^{-6}$ & $2\times10^{-10}$ & $2\times10^{-10}$\\
$2 \cdot \Omega_{z} \cdot v_{y}$ & $2\times10^{-7}$ & $2\times10^{-7}$ & $2\times10^{-10}$\\
$\Omega^{2}_{y} \cdot x$ & $2.7\times10^{-10}$ & $2\times10^{-18}$ & $2\times10^{-18}$\\
$\dot{\Omega}_{y} \cdot z$ & $2\times10^{-10}$ & $2\times10^{-10}$ & $2\times10^{-10}$\\
$\dot{\Omega}_{z} \cdot y$ & $2\times10^{-10}$ & $2\times10^{-10}$ & $2\times10^{-10}$\\
\hline
$\Gamma_{xx} \cdot v_{x} \cdot T$ & $3\times10^{-9}$ & \multicolumn {2}{c}{$3\times10^{-9}$}\\
$\Gamma_{xx} \cdot x$ & $6\times10^{-10}$ & \multicolumn {2}{c}{$6\times10^{-10}$}\\

\end{tabular}
\end{specialtable}
\begin{specialtable}[H]
\small
\caption {Assumptions on CAI atomic source parameters and satellite rotation parameters. $\delta_{i}$ refers to the amplitude of variation impacting parameter $i$.}
\centering
\label{Molasse_param}
\begin{tabular}{c|c|c|c}
\hline\hline
$x,y,z = 0$ & $v_{x,y,z} = 0$ & $\Omega_{y} = 1.16$ mrad/s & $\Omega_{x,z} = 0.1$ mrad/s (max)\\
\hline
$\delta_{x,y,z} = \pm~200~\mu$m & $\delta_{v_{x},v_{y},v_{z}} = \pm$ 1 mm/s & \multicolumn {2}{c}{$\dot{\Omega} = 10^{-6}$ rad.s$^{-2}$ (max, all axes)}\\
\hline\hline
\end{tabular}
\end{specialtable}
\item Coupling with Earth gravity gradient:\\
To evaluate the impact of the gravity gradient on the CAI acceleration measurement stability, we consider the atomic source parameters reported in Table \ref{Molasse_param} and a rough value of the Earth gravity gradient $\Gamma_{xx} = \Gamma_{zz} = 3 \times 10^{-6}$ s$^{-2}$. Two types of terms have been considered. The first ones, which have a dominant contribution, depend only on the gravity gradient value and on the position of the atomic cloud. The second ones, which are negligible, are crossed terms depending also on the satellite rotation. Concerning the dominant errors terms, which are reported in Table \ref{Rot_Grad_err}, their impact should be largely compensated thanks to an adequate frequency jump during the second laser Raman pulse \citep{roura_circumventing_2017,damico_canceling_2017,overstreet_effective_2018}.
\end{itemize}  
\subsubsection{Electrostatic accelerometer}
\label{Design_EA}
\paragraph{Global architecture}
In the frame of NGGM \citep{massotti_next_2021}, a new design of electrostatic accelerometer, MicroSTAR, with a cubic proof-mass and numeric control loop has been proposed. The main advantage of this proof-mass shape is to have the same performance along the 3 axes, and to provide 3 adapted angular acceleration measurements which can be used for attitude control, as it was done in the MICROSCOPE mission. HybridSTAR accelerometer (cf. Figure \ref{fig:EA_design}-a), the EA foreseen for the hybrid instrument, is a variation of MicroSTAR accelerometer adapted for the needs of the cold atom sensor. With a satellite in an Earth pointing attitude control mode, the atomic sensor has a loss of contrast in case of a satellite-fixed mirror for the Raman laser beam, which can lead to the impossibility to provide a long interrogation time measurement. To compensate this effect, HybridSTAR is designed in order to use the proof-mass as a mirror for the Raman laser beam as well as to allow a rotation of the proof-mass around the cross-track axis to compensate for the satellite rotation. Following this objective, the proof-mass size has been increased and consequently the electrode plates have been changed compared to MicroSTAR.\\
The main differences with the other previous EA accelerometers are the shape of the proof-mass, and the dissociation of the stops from the ULE plates. A window has been also added on the housing to allow the passage of the Raman laser towards the proof-mass. The dimension of the cubic proof-mass has been increased and adapted to the size of the Raman laser ($\phi = 25$ mm). Figure \ref{fig:EA_design}-b) shows the presence of an additional big hole at the center of the electrodes for the passage of the Raman laser beam. The 6 electrodes plates are identical to have a symmetrical core around the proof-mass, and for redundancy manufacturing aspects.
\begin{figure}[H]
  \centering
  \includegraphics[scale=0.8]{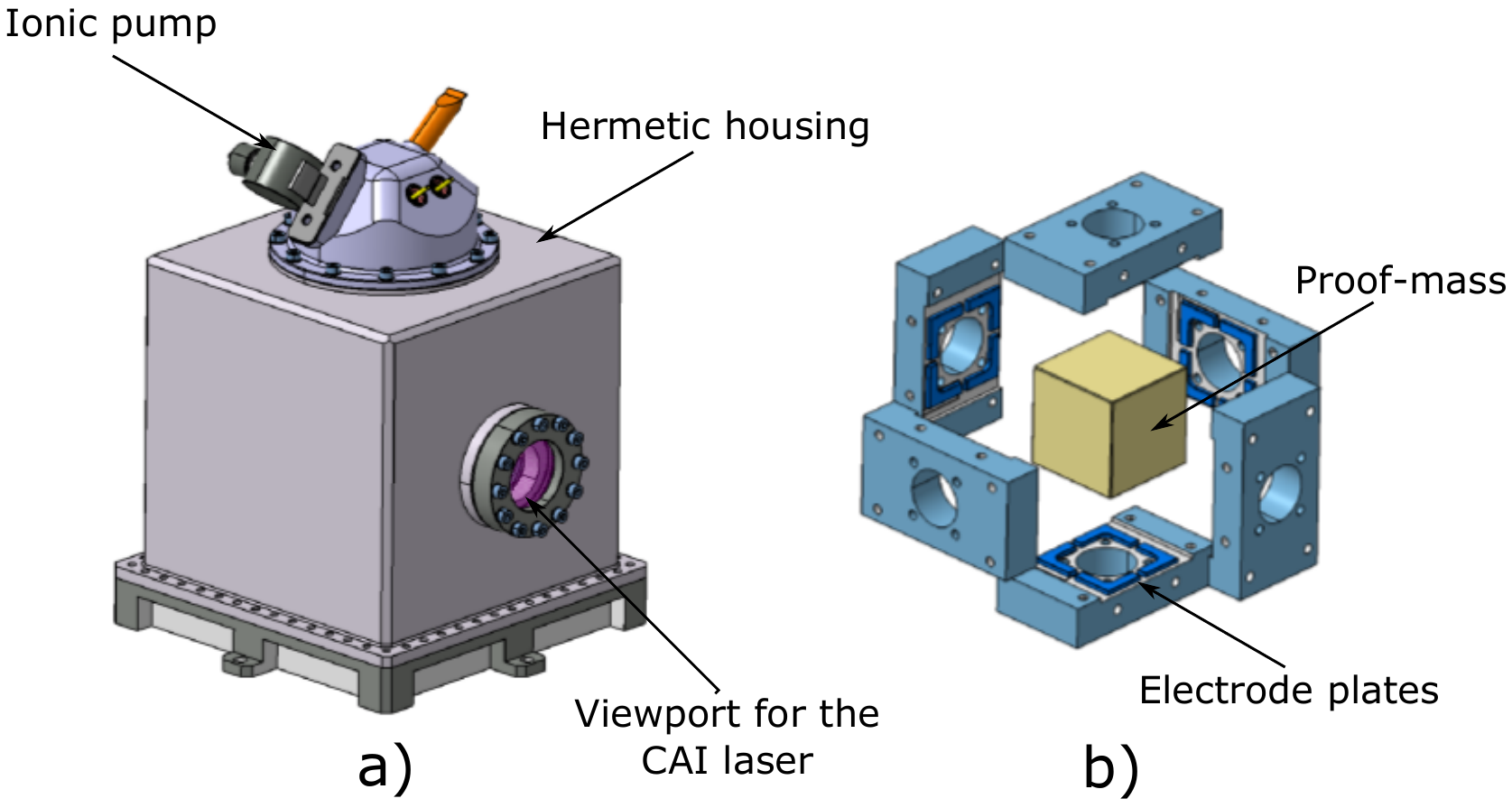}
  \caption{a) HybridSTAR electrostatic accelerometer. b) HybridSTAR electrode plates with the cubic proof-mass.}
  \label{fig:EA_design}
\end{figure}
\noindent HybridSTAR accelerometer is composed of the following sub-assemblies:
\begin{itemize}
\item[-]	A sole plate, which receives the core, the housing and hermetic feedthroughs, to ensure the mechanical and electrical interfaces;
\item[-] A hermetic housing with the pumping system, to maintain and ensure the vacuum inside the core;
\item[-] A core, with a cubic proof-mass polarized with a 5 $\mu$m diameter Platinum wire and 6 identical electrode plates, providing the same sensitive measurements for the 3 linear and 3 angular accelerations;
\item[-] A cage, independent of the core, which receives supports with the stops  limiting the free motion of the proof-mass, preventing any contact with the electrode plates.
\end{itemize}  

\paragraph{Performance aspects}
The electrostatic accelerometer is designed to fulfill the requirements for NGGM \citep{massotti_next_2021}. In order to have the capability to rotate the proof-mass around the cross-track axis, the design is slightly different for radial axis, which will be in charge of this rotation. In particular, the detectors along this axis shall be able to see a great motion of the proof-mass. The polarization wire introduces also a different contribution depending on its direction. This wire is implemented along the radial axis as this axis is already degraded due to the rotation control of the proof-mass. Figure \ref{fig:EA_DSP}-a) provides the ASD of noise of the accelerometer for the along-track axis as well as the cross-track axis. The polarization voltage is at 10 V and the detection voltage at 5 Vrms. The gap between electrode and proof-mass is 300 $\mu$m. The mass of the PM is about 394 g.
Above 0.1 Hz, the main contributor is the detector noise. In the measurement bandwidth $\lbrack$1-100 mHz$\rbrack$, the main contributor is due to the contact potential difference, which is a worst case based on estimation from previous missions. Below 1 mHz, the main contributor is the thermal sensitivity of the bias, considering the same stability than in GOCE mission. A better thermal stability will decrease this noise. 
In this configuration, the measurement range is $\pm$ 7.5 10$^{-6}$ m/s$^{2}$ and the control range is $\pm$ 9.4 10$^{-5}$ m/s$^{2}$ (the control range is the maximal acceleration before losing control of the proof-mass).\\
Figure \ref{fig:EA_DSP}-b) shows the performance on radial axis. The performance is degraded with respect to the other axes due to the control of the rotation with the same electrodes that control also the translation:
\begin{itemize}
\item[-] Above 0.1 Hz due to the increased detector noise, consequence of the decrease of the detector gain in order to be able to measure the proof-mass rotation when it compensates the satellite rotation of the Raman laser beam;
\item[-] in the measurement bandwidth, the performance is degraded, first, due to the higher voltage to be measured, consequence of the control of the rotation, and secondly, due to the increase of the electrostatic stiffness caused by the higher voltage on the electrodes (impact on the detector noise contribution);
\item[-] below 1 mHz, the thermal sensitivity of the bias is increased due to the thermal sensitivity of the wire stiffness. 
\end{itemize}
\end{paracol}
\begin{figure*}
\centering
  \includegraphics[scale=0.55]{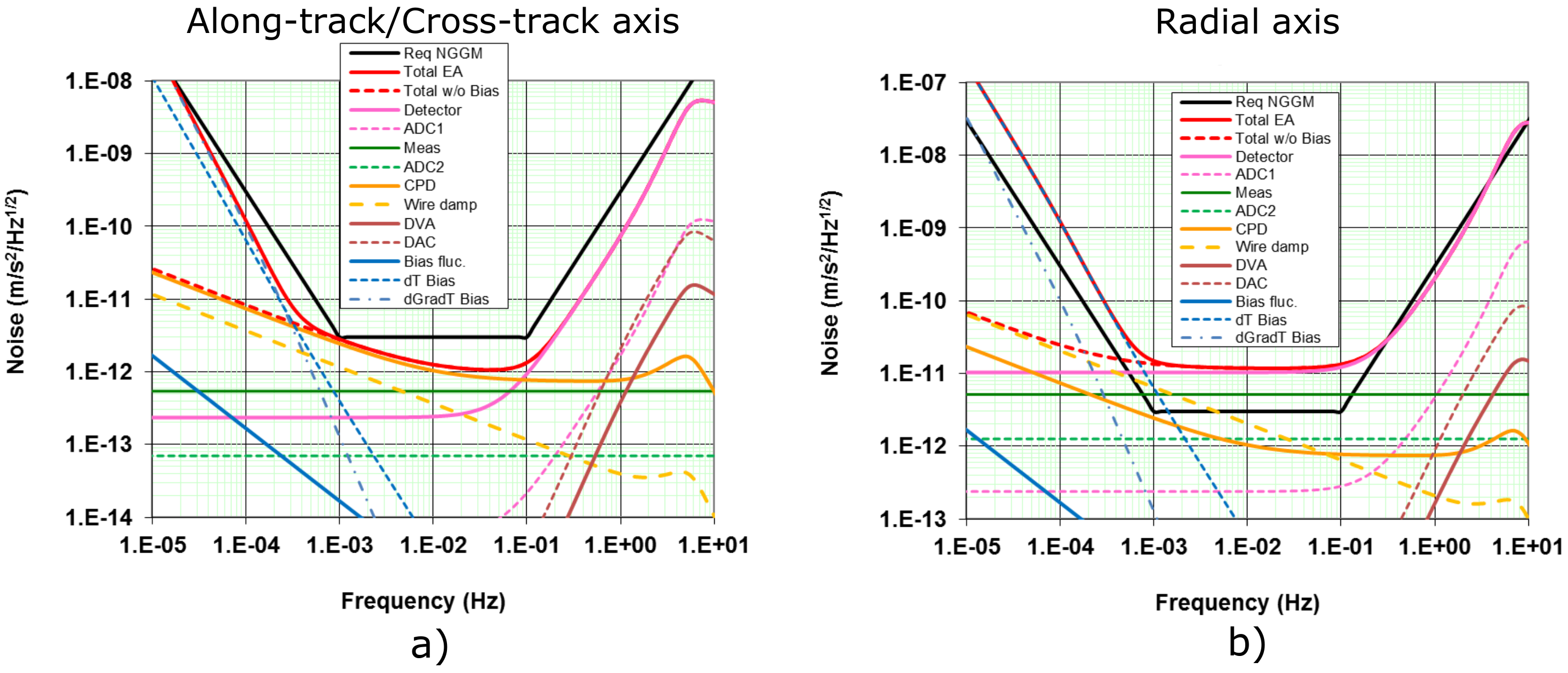}
  \caption{a) HybridSTAR accelerometer performance on along-track and cross-track axis. b) HybridSTAR accelerometer performance on radial axis.}
  \label{fig:EA_DSP}
\end{figure*}
\begin{paracol}{2}
\switchcolumn
Note that an improvement of the radial axis performance is feasible by changing the design of the EA electrodes in a way to separate the angular control of the PM from its linear control.
We can also notice that the levels of noise estimated in Figure~\ref{fig:EA_DSP} are compliant with the results obtained with the simulator (cf. Figure~\ref{fig:Simulator_EA_noise}).

\subsubsection{Global hybrid accelerometer design}
Several architectures could be considered, with more or less complex design, and technological risks vs global performances of the hybrid accelerometer. The more risky architecture would be the one where the atomic cloud is at the center of gravity of the proof-mass with a common vacuum chamber, to have the two accelerometers at the center of gravity of the spacecraft. The main interest is to remove gravity gradient/angular acceleration effects between the CAI and EA, but this architecture would have complex impact on the EA design. This solution would require a dedicated study to analyze its feasibility.
To limit technological risks, the global architecture of hybrid accelerometer is designed to allow nominal work of EA, regardless of the operation conditions of the CAI. The center of gravity of the proof-mass and the one of the atomic cloud are at a distance of few centimeters, but the EA is completely independent of the CAI.\\
We present on Figure~\ref{fig:Hybrid_Global-Design} the global architecture of the hybrid instrument.\\
\end{paracol}
\begin{figure*}
\centering
  \includegraphics[scale=0.27]{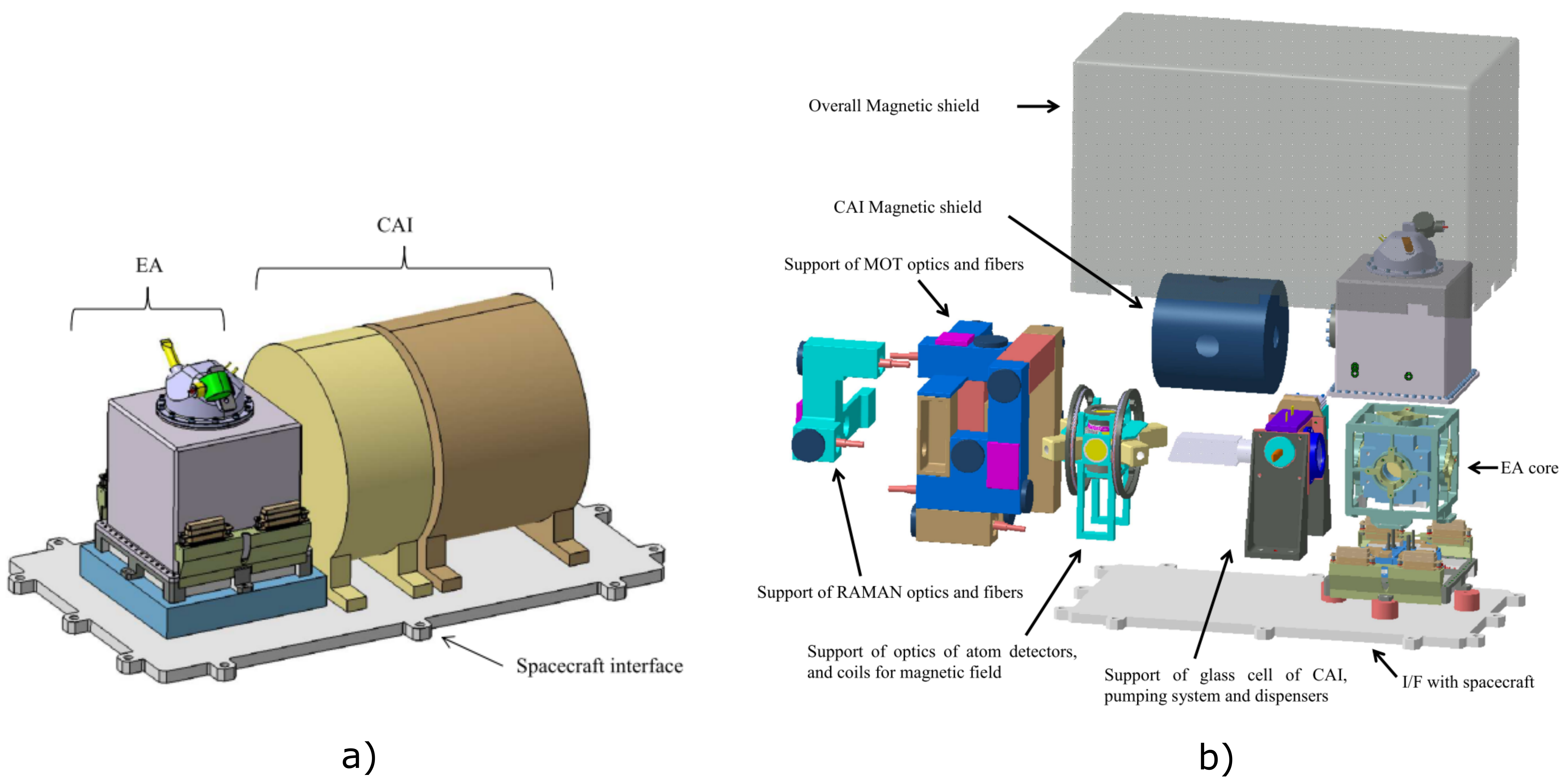}
  \caption{a) Global hybrid instrument architecture. b) Exploded view of the hybrid architecture.}
  \label{fig:Hybrid_Global-Design}
\end{figure*}
\begin{paracol}{2}
\switchcolumn
\noindent This architecture is composed of the following sub-assemblies:
\begin{itemize}
\item[-] The interface with the spacecraft;
\item[-] The EA, fully functionally independent;
\item[-] The support of the glass cell of the CAI, with pumping system and dispenser;
\item[-] The support of optics of atom detectors, and the coils for magnetic field;
\item[-] The support of MOT fibers and optics;
\item[-] The support of fiber and optics of Raman laser;
\item[-] The two magnetic shields, at CAI level and overall instrument.
\end{itemize}
All those sub-assemblies are mounted on the spacecraft interface.\\

The MOT is positioned at 240 mm from the center of gravity of the proof-mass (217.8 mm to the mirror face PM 44.4 x 44.4 x 44.4 mm). The main structural parts of the CAI will be manufactured in titanium alloys, as much as possible, for magnetic and weight aspects. If necessary for thermal stability, Invar alloy will be used (as EA with magnetic thermal treatment). The spacecraft interface could be manufactured in honeycomb sandwich with carbon panel.
\subsubsection{SWaP}
The elaborated budget for the hybrid instrument is given in the following table:
\begin{specialtable}[H]
\small
\caption {Size Weight and Power budget for the hybrid instrument. Note that the CAI budget for the electronics, laser and microwave system has been extrapolated by taking into account the PHARAO's flight model budget \citep{laurent_acespharao_2015,leveque_pharao_2015}.}
\centering
\label{SWaP_Hybrid}
\begin{tabular}{c|cc|ccc|ccc}
\hline\hline
 & \multicolumn {2}{c|}{Physics Package} & \multicolumn {3}{c|}{Electonics/laser/MW} & \multicolumn {3}{c}{Whole Instrument}\\
& Size & Weight & Size & Weight & Power & Size & Weight & Power\\
\cline{1-9}
EA & 7 L & 6 kg & 9 L & 6 kg & 45 W & 16 L & 12 kg & 45 W\\
\hline
CAI & 36 L & 24 kg & 57 L & 39 kg & 100 W & 93 L & 63 kg & 100 W\\
\hline
Hybrid & 43 L & 30 kg & 66 L & 45 kg & 145 W & \textbf{109 L} & \textbf{75 kg} & \textbf{145 W}\\
\hline\hline
\end{tabular}
\end{specialtable}
We reach therefore a total budget of 109 L, 75 kg and 145 W for the hybrid instrument based on an electrostatic accelerometer and a molasses-based CAI.\\

In a more prospective way, the molasses source for the CAI could be upgraded to a BEC source \citep{muntinga_interferometry_2013, rudolph_high-flux_2015, schuldt_design_2015, becker_space-borne_2018, elliott_nasas_2018, aveline_observation_2020, frye_bose-einstein_2021}. In such an approach, the budget of a hybrid instrument has been evaluated to the less favorable SWaP of 156 L, 102 kg and 245 W (contribution of the EA: 16 L, 12 kg, 45 W) but this should lead to better performances with respect to acceleration measurements. This is mainly due to a better control of the kinematics of the atomic cloud, with a position/velocity stability that should be better than $\pm$ 10 $\mu$m/$\pm$ 100 $\mu$m/s \cite{karcher_improving_2018} and should lead to an expected gain of roughly one order of magnitude regarding acceleration stability.\\
Note that concerning the potential implementation of a CAI in a satellite, it seems that the electrical power consumption could constitute a severe limit. One possible mitigation strategy could be the implementation of an intermittent mode of operation where the CAI alternates between a standard mode of operation with a full power consumption (245 W) and a standby mode characterized by a reduced power consumption (95 W) (cf. Figure~\ref{fig:intermittent_mode}). A preliminary analysis has shown under some assumptions (e.g. a CAI limited by the QPN, a BEC atomic source, $T=2$ s, $T_{c}=6$ s...) that the average electrical power consumption could be lowered to around 110 W considering an effective measurement cycling time of 42 s ($n=7$). This mode of operation seems compatible in term of noise with an objective of hybridization at a level of $10^{-10}$ m.s$^{-2}$.Hz$^{-1/2}$ (see Figure~\ref{fig:PSD_Intermittent}).
\end{paracol}
\begin{figure}[h]
\centering
  \includegraphics[scale=0.32]{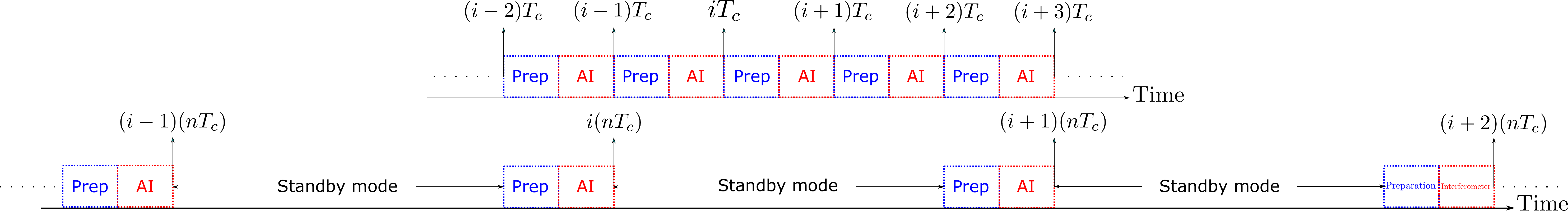}
  \caption{Temporal sequence describing the intermittent mode. The standard mode (up) allows acceleration measurements each $T_{c}$. In the intermittent mode (bottom), the measurements are now only each $n \cdot T_{c}$. In this example, we take $n=4$. In between, the CAI is put in a low power consumption mode (Standby mode) allowing to lower the average electrical power consumption.}
   \label{fig:intermittent_mode}
\end{figure}
\begin{paracol}{2}
\switchcolumn
\begin{figure}[H]
  \centering
  \includegraphics[scale=0.35]{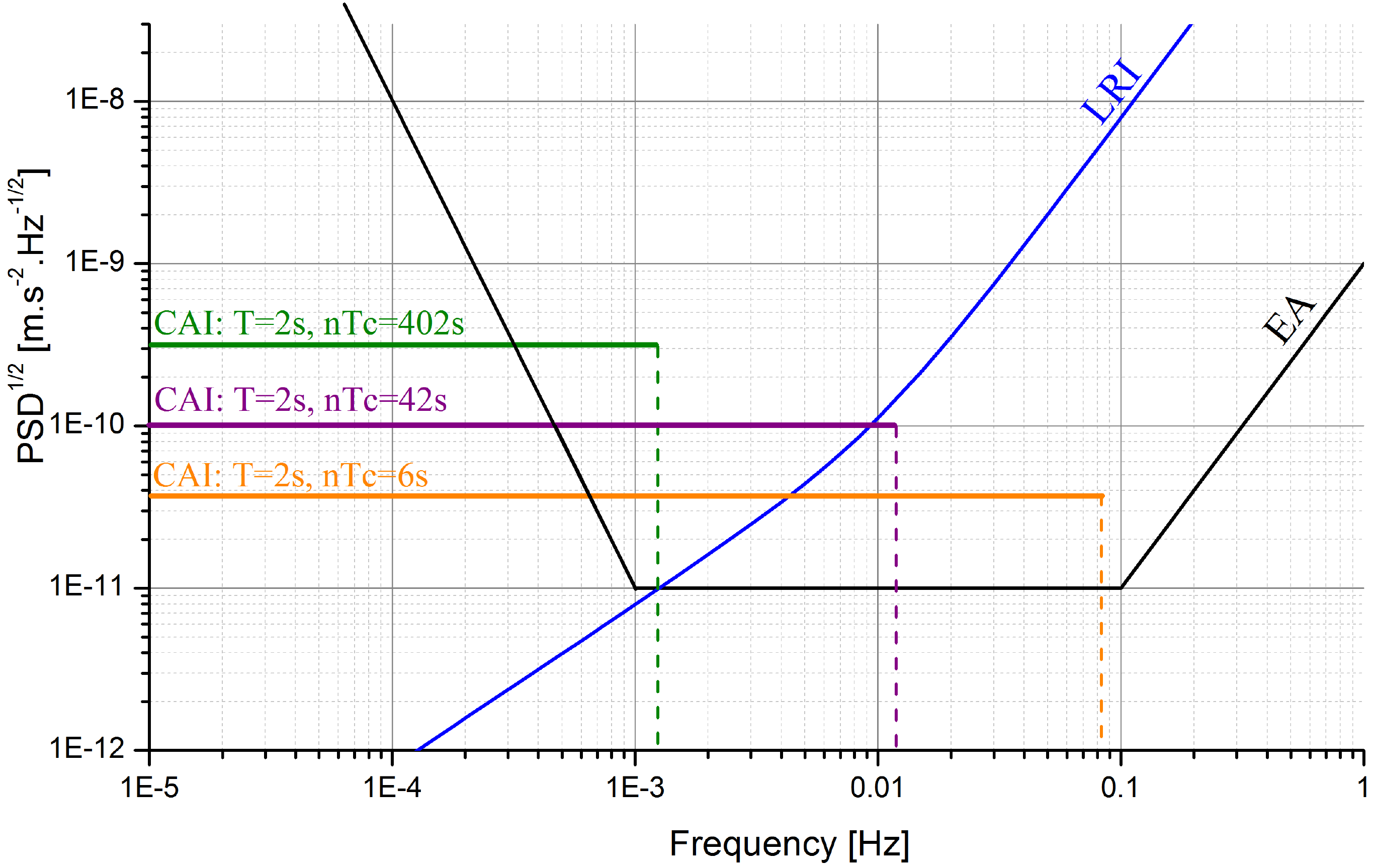}
  \caption{Evolution of the CAI noise (QPN) due to intermitent operation for an interrogation time $T = 2$ s and a cycling time $T_{c}=6$ s in a standard mode. The orange, purple and green curves correspond respectively to an effective measurement cycling time of 6 s ($n=1$), 42 s ($n=7$) and 402 s ($n=67$).}
  \label{fig:PSD_Intermittent}
\end{figure}
%
\section{Conclusions}
In this article, a hybrid accelerometer concept, combining an atom interferometer and an electrostatic accelerometer, has been studied for use in future space gravity missions, especially in a Satellite-to-Satellite Tracking scenario similar to GRACE, GRACE-FO or future NGGM. To assess the potential of such instrument, numerical simulations were conducted to evaluate the associated benefits in term of gravity retrieval. These simulations showed that improved gravity retrieval could be achieved for some threshold low frequency atomic instrument performance but that this beneficial impact vanishes once temporal variations of the gravity field are taken into account, because of temporal aliasing effects. Improved de-aliasing models and/or more sophisticated parametrization schemes which allow for a reduction of temporal aliasing effects (or, preferrably, a Bender-type constellation employing both) are then required in order to maximize the impact of accelerometer performance gain. The hybrid instrument, allowing a better knowledge of the scale factor accelerometer, should also lead to relaxation of the drag-free requirements for the satellite formation.
Some assumptions of acceleration performance levels were made concerning the hybrid instrument. To evaluate their validity, a simulator modeling the hybrid sensor has been developed with the objective to estimate ultimately the achievable acceleration performance of such instrument in orbit. The hybrid instrument simulator has been tested successfully by reproducing noise characteristics of a hybrid accelerometer prototype operating on ground. The detrimental impact of satellite rotation on the CAI contrast has been also studied. While the contrast is completely lost in-orbit when a fixed CAI reference mirror is used, it is possible to compensate for the orbital rotation with the proof-mass of the electrostatic accelerometer acting as the reference mirror. In this case, we observe a total recovery of the interferometric contrast. Although the simulator could give a quite good overview of the hybrid instrument behavior, it should be further developed to take into account more effects that could have significant impact on the final performance of the accelerometer. The development of this simulator will be of prime importance in the future to extrapolate on-ground performance towards space environment.
To further validate the feasibility of such hybridized configuration, some experimental demonstrations have been conducted with a lab hybrid prototype constituted of a cold atom gravimeter and a dedicated ground electrostatic accelerometer. Acceleration measurements have been obtained following a hybridization scheme, highlighting the interest of EA hybridization with a CAI. The potential of the EA instrument to cancel satellite rotation impact on the CAI contrast has been demonstrated by controlling the electrostatic proof-mass acting as the CAI reference mirror. Experimental demonstration of rotation compensation with a 85$\%$ efficiency of contrast recovery was performed. Further experimental activities are necessary to explain the non-complete contrast recovery and to study the impact on the acceleration error budget. Finally, the ability for the EA to compensate the satellite rotation and to simultaneously allow a precise acceleration measurement should be studied experimentally in a next step.
To evaluate the potential of integration of a hybrid atomic-electrostatic accelerometer in a satellite, a preliminary design has been elaborated. The SWaP of such design has been given along with an associated error budget. All these estimations should be further consolidated with the elaboration of a more detailed architecture and by addressing in a more exhaustive way all the effects that could have a detrimental impact on the CAI output. More particularly, the impact of wavefront aberrations and methods for satellite rotation compensation should be further analyzed, whether with simulation tools or with experimental demonstrations. Considering a CAI based on a molasses source, a hybrid instrument of 109 L, 75 kg and 145 W should be feasible with a level of performance in the range of 10$^{-10}$ m.s$^{-2}$, the impact of wavefront aberrations assumed here mitigated. 
This reported study constitutes a preliminary analysis of an original accelerometer design for future space gravity missions. This configuration based on the emerging cold atom technology seems promising for performance improvements and requires further investigations thanks to simulations or experimental activities to consolidate all the potential envisioned for this technology.

\funding{This research was performed in the framework of the study \textquotedblleft Hybrid Atom Electrostatic System Follow-On for Satellite Geodesy \textquotedblright, Contract No.4000122290/17/NL/FF/mg, funded by the European Space Agency (ESA). The PhD study of N.M. is co-funded by ESA and ONERA.}


\acknowledgments{We would like to thank Bernard Foulon (ONERA) for his early participation to the realization of the EA ground prototype and Fran\c{c}oise Liorzou (ONERA) for her assistance during the experimental activities. We thank also Luca Massotti (ESA), Roger Haagmans (ESA) and Pierluigi Silvestrin (ESA) for fruitful discussions during the course of the study.}




\abbreviations{Abbreviations}{
The following abbreviations are used in this manuscript:\\

\noindent 
\begin{tabular}{@{}ll}
ADC & Analog-to-Digital Converter\\
AOCS & Attitude and Orbit Control System\\
AOHIS & Atmosphere Ocean Hydrology Ice Solid-Earth\\
ASD & Amplitude Spectral Density\\
BEC & Bose-Einstein Condensate\\
CAI & Cold Atom Interferometer\\
CF & Corner Frequency\\
COM & Center-Of-Mass\\
DVA & Drive Voltage Amplifier\\
EA & Electrostatic Accelerometer\\
ESA & European Space Agency\\
IAPG & Institute of Astronomical and Physical Geodesy\\
ICU & Interface and Control Unit\\
ISS & International Space Station\\
LOS & Line-Of-Sight\\
LRI & Laser Ranging Interferometer\\
LSA & Less Sensitive Axis\\
MBW & Measurement BandWidth\\
MOT & Magneto-Optical Trap\\
PIP & Passive Isolation Platform\\
PM & Proof-Mass\\
PSD & Power Spectral Density\\
QPN & Quantum Projection Noise\\
QSG & Quantum Spaceborne Gravimetry\\
RMS & Root Mean Square\\
SH & Spherical Harmonics\\
SST-LL & Low-Low Satellite-to-Satellite Tracking\\
USA & Ultra Sensitive Axis
\end{tabular}}

%
%
%
\end{paracol}
\reftitle{References}


\externalbibliography{yes}
\bibliography{Biblio_Hybrid2021}

%


\end{document}